\pdfoutput=1
\documentclass[british]{aa}
\usepackage{array}
\usepackage{multirow}
\usepackage{subfig}

\makeatletter

\providecommand{\tabularnewline}{\\}

\makeatother

\begin{document}

\title{Radiative levitation in carbon-enhanced metal-poor stars with s-process enrichment}\titlerunning{Radiative levitation in CEMP stars with \emph{s}-process enrichment}

\author{E. Matrozis\thanks{Member of the International Max Planck Research School (IMPRS) for Astronomy and Astrophysics at the Universities of Bonn and Cologne}\and R. J. Stancliffe}

\institute{Argelander-Institut für Astronomie (AIfA), University of Bonn, Auf dem Hügel 71, DE-53121, Bonn, Germany\\
\email{elvijs@astro.uni-bonn.de}}

\date{Received ; accepted}

\abstract{A significant fraction of all metal-poor stars are carbon-rich. Most of these carbon-enhanced metal-poor (CEMP) stars also show enhancement in elements produced mainly by the \emph{s}-process (CEMP-\emph{s} stars) and evidence suggests that the origin of these non-standard abundances can be traced to mass transfer from a binary asymptotic giant branch (AGB) companion. Thus, observations of CEMP-\emph{s} stars are commonly used to infer the nucleosynthesis output of low-metallicity AGB stars. A crucial step in this exercise is understanding what happens to the accreted material after mass transfer ceases. Here we present models of the post-mass-transfer evolution of CEMP-\emph{s} stars considering the physics of thermohaline mixing and atomic diffusion, including radiative levitation. We find that stars with typical CEMP-\emph{s} star masses, $M\approx0.85~\mathrm{M}_{\sun}$, have very shallow convective envelopes ($M_{\mathrm{env}}\lesssim10^{-7}~\mathrm{M}_{\sun}$). Hence, the surface abundance variations arising from the competition between gravitational settling and radiative levitation should be orders of magnitude larger than observed (e.g. $[\mathrm{C/Fe}]<-1$ or $[\mathrm{C/Fe}]>+4$). Lower-mass stars ($M\approx0.80~\mathrm{M}_{\sun}$) retain thicker convective envelopes and thus show variations more in line with observations but are generally too unevolved ($\log g>4$) when they reach the age of the Universe. We are therefore unable to reproduce the spread in the observed abundances with these models and conclude that some other physical process must largely suppress atomic diffusion in the outer layers of CEMP-\emph{s} stars. We demonstrate that this could be achieved by some additional (turbulent) mixing process operating at the base of the convective envelope, as found by other authors. Alternatively, mass-loss rates around $10^{-13}~\mathrm{M}_{\sun}\thinspace\text{yr}^{-1}$ could also negate most of the abundance variations by eroding the surface layers and forcing the base of the convective envelope to move inwards in mass. Since atomic diffusion cannot have a substantial effect on the surface abundances of CEMP-\emph{s} stars, the dilution of the accreted material, while variable in degree from one star to the next, is most likely the same for all elements.  }


\maketitle

\section{Introduction}

Currently observable metal-poor stars ($[\mathrm{Fe}/\mathrm{H}]\footnotemark<-2$)\footnotetext{The relative abundance of element A with respect to element B is $\text{[A/B]}=\log\left(C_\text{A}/C_\text{B}\right)-\log\left(C_\text{A}/C_\text{B}\right)_\sun$ where $C$ is the number or mass fraction.} are relatively unevolved low-mass objects that formed within the first couple of billion years after the Big Bang. Because these stars have not modified their surface composition by internal nucleosynthesis, they are expected to carry the signature of the chemical evolution of these early epochs providing us with the means to study this long-gone era. The number of known metal-poor stars in our Galaxy has exploded thanks to large-scale photometric and spectroscopic surveys such as the HK survey \citep{1985AJ.....90.2089B,1992AJ....103.1987B}, and more recently the Hamburg/ESO survey \citep{2001A&A...375..366C,2008A&A...484..721C}, the Sloan Digital Sky Survey \citep[SDSS; e.g.][]{2000AJ....120.1579Y,2012ApJS..203...21A,2014ApJS..211...17A} and its sub-survey, the Sloan Extension for Galactic Understanding and Exploration \citep[SEGUE;][]{2009AJ....137.4377Y}. A common finding of these surveys is that a substantial fraction of all metal-poor stars are relatively carbon-rich with $[\mathrm{C}/\mathrm{Fe}]\gtrsim1.0$. This fraction is around 10\% at $[\mathrm{Fe}/\mathrm{H}]\approx-2$ and increasing towards lower metallicities \citep[e.g.][]{2006ApJ...652L..37L,2012ApJ...744..195C,2013AJ....146..132L,2014ApJ...797...21P}.

These so-called carbon-enhanced metal-poor (CEMP) stars are further classified into CEMP -\emph{no}, -\emph{s}, -\emph{r}, and -\emph{r}/\emph{s} sub-classes depending on whether they show enhancements of elements produced by slow (\emph{s}) or rapid (\emph{r}) neutron-capture nucleosynthesis. Most CEMP stars with $[\mathrm{Fe}/\mathrm{H}]>-3$ display significant \emph{s}-process element enrichment and are classified either as CEMP-\emph{s} ($[\mathrm{Ba}/\mathrm{Fe}]>1$ and $[\mathrm{Ba}/\mathrm{Eu}]>0.5$) or CEMP-\emph{r}/\emph{s} ($[\mathrm{Ba}/\mathrm{Fe}]>1$ and $0<[\mathrm{Ba}/\mathrm{Eu}]<0.5$) stars \citep{2005ARA&A..43..531B}.\footnote{Slightly different distinctions between CEMP stars enriched in \emph{s}-process elements have also been proposed \citep{2006A&A...451..651J,2010A&A...509A..93M,2012A&A...548A..34A}.} While the abundance patterns of these stars have been linked to nucleosynthesis occurring in asymptotic giant branch (AGB) stars \citep{2005ARA&A..43..435H,2008ARA&A..46..241S,2010MNRAS.404.1529B,2011MNRAS.418..284B,2012ApJ...747....2L}, most \emph{s}-process-rich CEMP stars are not luminous enough to be AGB stars. However, results from radial velocity monitoring are consistent with all of them being in binaries which is not the case for the other sub-classes \citep{2005ApJ...625..825L,2014MNRAS.441.1217S,2015A&A...583A..49H,2016A&A...586A.160H,2016A&A...588A...3H}. Therefore, these stars are generally thought to be a product of mass transfer from an AGB companion that later became a white dwarf.\footnote{\citet{2016A&A...588A...3H} have argued that four stars in their observed sample of 22 CEMP-\emph{s} stars appear to be single. Regardless, the mass transfer scenario should apply to most CEMP-\emph{s} stars.}As noted by \citet{2005ApJ...625..825L} this would make CEMP-\emph{s} stars the low-metallicity analogues of CH and Ba stars \citep{1990ApJ...352..709M,2016A&A...586A.158J}. If this is the case, observations of the secondary can in principle be used to infer the nucleosynthesis of low-metallicity AGB stars with initial masses around solar and above, which, although important for galactic and globular cluster chemical evolution, no longer exist in the Local Universe \citep[e.g.][]{1999ApJ...521..691T,2001ApJ...549..346T,2009MNRAS.397.1661V,2011MNRAS.414.3231K,2014ApJ...787...10B,2010MNRAS.407..854D,2014MNRAS.437.3274V}.

A comparison of AGB nucleosynthesis models with abundances of CEMP-\emph{s} stars is straightforward only if the accreted material remains on the surface of the star. This will certainly not be the case once it evolves off the main sequence and develops a deep convective envelope. But as demonstrated by \citet{2007A&A...464L..57S}, this is also unlikely on the main sequence because the higher mean molecular weight of the accreted material should trigger thermohaline mixing \citep{1972ApJ...172..165U,1980A&A....91..175K}. Furthermore, gravitational settling of heavier elements could both modify the extent of this mixing and the subsequent evolution of the secondary \citep{2008MNRAS.389.1828S,2008ApJ...677..556T,2009MNRAS.394.1051S}. If the overall effect is to dilute the accreted material by uniformly mixing it throughout some portion of the secondary, a comparison between AGB nucleosynthesis models and abundances of CEMP-\emph{s} stars is still possible provided this amount of dilution can be estimated \citep[e.g. as attempted by][]{2011MNRAS.418..284B,2012MNRAS.422..849B}. However, the rather impartial (leading to similar dilution of all elements) process of settling will be counteracted by the highly selective process of radiative levitation (also known as radiative accelerations), a process in which the ions in the stellar plasma gain a net outward momentum from absorption of the diffusing photons. Metal-poor stars with masses around $\mathrm{0.8~M}_{\sun}$ have very shallow convective envelopes and, in absence of any counteracting processes, large abundance anomalies can result \citep[e.g.][]{2002ApJ...580.1100R,2002ApJ...568..979R}. If radiative levitation is important during the post-mass-transfer evolution of CEMP-\emph{s} stars, the interpretation of their abundances in the context of AGB nucleosynthesis gets considerably more complicated. In this paper we model the main sequence evolution of CEMP-\emph{s} stars including the effect of radiative levitation to investigate whether this is the case.

We focus primarily on the evolution of carbon and iron surface abundances. Abundances of \emph{s}-process elements are not modelled. However, levitation is expected to have a much greater impact on iron than on carbon \citep{1995A&A...297..223G,1997MNRAS.289..700S,2007MNRAS.382..245S}. We therefore attempt to constrain its overall importance for CEMP-\emph{s} stars by investigating these two elements.

\section{Methods\label{sec:Methods}}

We use the stellar evolution code STARS originally written by \citet{1971MNRAS.151..351E,1972MNRAS.156..361E,1973A&A....23..325E} and since improved by many authors \citep[e.g.][]{1995MNRAS.274..964P,2009MNRAS.396.1699S}. The version used in this work tracks the abundances of the nuclear species $^{1}\mathrm{H}$, $^{3}\mathrm{He}$, $^{4}\mathrm{He}$, $\mathrm{^{12}\mathrm{C}}$, $^{14}\mathrm{N}$, $^{16}\mathrm{O}$, $^{20}\mathrm{Ne}$, $^{24}\mathrm{Mg}$, $^{28}\mathrm{Si}$, and $^{56}\mathrm{Fe}$, the last three of which were previously not tracked in detail. The mass fraction $X_{i}$ of each species \emph{i} is governed by an advection-diffusion equation:
\begin{equation}
\frac{\mathrm{d}X_{i}}{\mathrm{d}t}=\frac{\partial}{\partial m}\left[\left(4\pi r^{2}\rho\right)^2 D_{\mathrm{mix}}\frac{\partial X_{i}}{\partial m}\right]-\frac{\partial}{\partial m}\left(4\pi r^{2}\rho X_{i}w_{i}\right)+R_{i},\label{eq:dxdt}
\end{equation}
where the first term on the right-hand side accounts for convective mixing, thermohaline mixing, and concentration diffusion ($D_{\mathrm{mix}}$ is the sum of the individual diffusion coefficients $D_{\mathrm{conv}}$, $D_{\mu}$, and $D_{i}$, respectively), the second term describes the net effect from atomic diffusion, and the last term, $R_{i}$, accounts for nuclear processing.\footnote{Other symbols in Eq.~\eqref{eq:dxdt} have their usual meaning, namely: $t$ is time; $\rho$ is the mass density; $r$ and $m$ are the radial and mass coordinate, respectively. The diffusion velocity $w_{i}$ is defined in Eq.~\eqref{eq:wi}.} Inside convective regions $D_{\mathrm{conv}}$ is obtained from the mixing length theory (MLT; \citealt{1958ZA.....46..108B}) using a solar-calibrated value of $\alpha_{\mathrm{MLT}}=2.0$ that is fully consistent with the models presented in this paper (see Section~\ref{subsec:uncertainties} for details). Near the convective boundaries the convective mixing coefficient takes the form from \citet{1972MNRAS.156..361E} for numerical stability reasons. The diffusion coefficient for thermohaline mixing is taken from \citet{2010ApJ...723..563D} assuming a finger length-to-diameter ratio of 0.5 as constrained by their numerical simulations. This assumption of more blob-like than finger-like structures is in accord with \citet{1980A&A....91..175K} and results in relatively inefficient thermohaline mixing.

Following \citet{2008MNRAS.389.1828S} we treat atomic diffusion in a trace approximation which allows the diffusion velocity of elements other than hydrogen to be written as \citep{2008EAS....32...81T}
\begin{equation}
w_{i}=\frac{D_{i}}{kT}\left[g\left(\mu-\mu_{i}\right)+\mu_{i}g_{\mathrm{r},i}\right]-D_{i}\alpha_{\mathrm{T},i}\frac{\partial\ln T}{\partial r},\label{eq:wi}
\end{equation}
where $g$ and $g_{\mathrm{r}}$ are the gravitational and radiative acceleration, respectively; $k$ is the Boltzmann constant; $T$ is temperature; $\mu$ is the mean molecular weight of the stellar plasma; $\mu_{i}=m_{i}/\left(1+\bar{Z}_{i}\right)$ is the molecular weight of element $i$ with an atomic mass $m_{i}$ and a mean charge $\bar{Z}_{i}$; and $\alpha_{\mathrm{T},i}$ is the thermal diffusion coefficient. The velocity of hydrogen follows from mass conservation: $X_{\mathrm{H}}w_{\mathrm{H}}=-\sum_{i\ne\mathrm{H}}X_{i}w_{i}$. The diffusion coefficients $D_{i}$ and $\alpha_{\mathrm{T},i}$ are taken from \citet{1986ApJS...61..177P}.

We simulate the accretion of AGB ejecta by adding mass of a given composition to our models. We fix the accretion rate to $10^{-6}~\mathrm{M}_{\sun}\thinspace\mathrm{yr^{-1}}$ following \citet{2007A&A...464L..57S}, and the accreted composition to the average composition of the ejecta from the models of \citet{2012ApJ...747....2L}. These yields together with the zero-age main sequence (ZAMS) abundances \citep{2009ARA&A..47..481A} are given in Table~\ref{tab:xinp}. Mass loss is not included in our models until Section~\ref{subsec:Mass-loss}.

\begin{table*}
\caption{Chemical composition of the secondaries on the zero-age main sequence \citep[ZAMS; abundance distribution from][scaled to $Z=10^{-4}$]{2009ARA&A..47..481A} and of the ejecta from the AGB models of \citet{2012ApJ...747....2L}. The second column lists the age when accretion of the corresponding composition begins ($t_{\text{mt}}$). Mass fractions of all elements other than helium are sums over their isotopes. \label{tab:xinp}}

\begin{tabular*}{1\textwidth}{@{\extracolsep{\fill}}>{\raggedright}p{0.09\columnwidth}>{\raggedright}m{0.07\columnwidth}llllllllllll>{\raggedright}m{1.5cm}}
\hline 
\hline
\multirow{2}{0.09\columnwidth}{{\tiny{}Model}} & \multirow{2}{0.07\columnwidth}{{\tiny{}$t_{\mathrm{mt}}$ (Gyr)}} & \multicolumn{2}{l}{{\tiny{}Mass fraction}} & \multicolumn{8}{l}{{\tiny{}Mass fraction $\times10^{-6}$}} & \multicolumn{2}{l}{{\tiny{}Abundance}} & \multirow{2}{1.5cm}{{\tiny{}Mean mol. weight}}\tabularnewline
\cline{3-14} 
 &  & {\tiny{}H} & {\tiny{}$^{4}\mathrm{He}$} & {\tiny{}$^{3}\mathrm{He}$} & {\tiny{}C} & {\tiny{}N} & {\tiny{}O} & {\tiny{}Ne} & {\tiny{}Mg} & {\tiny{}Si} & {\tiny{}Fe} & {\tiny{}{[}Fe/H{]}} & {\tiny{}{[}C/Fe{]}} & \tabularnewline
\hline 
{\tiny{}ZAMS} & \emph{\tiny{}\ldots} & \emph{\tiny{}$0.75770$} & \emph{\tiny{}$0.24217$} & \emph{\tiny{}$30.30$} & \emph{\tiny{}$17.72$} & \emph{\tiny{}$5.190$} & \emph{\tiny{}$42.95$} & \emph{\tiny{}$9.390$} & \emph{\tiny{}$5.300$} & \emph{\tiny{}$4.980$} & \emph{\tiny{}$9.680$} & \emph{\tiny{}$-2.14$} & \emph{\tiny{}$0.00$} & {\tiny{}$0.5934$}\tabularnewline
\multicolumn{15}{c}{\tiny{}Composition of AGB ejecta}\tabularnewline
\emph{\tiny{}$\mathrm{0.90\thinspace M}_{\sun}$} & \emph{\tiny{}$9.10$} & \emph{\tiny{}$0.73302$} & \emph{\tiny{}$0.26222$} & \emph{\tiny{}$235.8$} & \emph{\tiny{}$3680$} & \emph{\tiny{}$135.1$} & \emph{\tiny{}$217.5$} & \emph{\tiny{}$457.0$} & \emph{\tiny{}$12.77$} & \emph{\tiny{}$4.943$} & \emph{\tiny{}$8.895$} & \emph{\tiny{}$-2.16$} & \emph{\tiny{}$2.35$} & {\tiny{}$0.6046$}\tabularnewline
\emph{\tiny{}$\mathrm{1.00\thinspace M}_{\sun}$} & \emph{\tiny{}$6.30$} & \emph{\tiny{}$0.74907$} & \emph{\tiny{}$0.24956$} & \emph{\tiny{}$261.4$} & \emph{\tiny{}$933.0$} & \emph{\tiny{}$21.47$} & \emph{\tiny{}$92.47$} & \emph{\tiny{}$37.29$} & \emph{\tiny{}$5.395$} & \emph{\tiny{}$4.913$} & \emph{\tiny{}$8.910$} & \emph{\tiny{}$-2.17$} & \emph{\tiny{}$1.76$} & {\tiny{}$0.5972$}\tabularnewline
\emph{\tiny{}$\mathrm{1.25\thinspace M}_{\sun}$} & \emph{\tiny{}$3.06$} & \emph{\tiny{}$0.71670$} & \emph{\tiny{}$0.27604$} & \emph{\tiny{}$228.9$} & \emph{\tiny{}$6032$} & \emph{\tiny{}$42.42$} & \emph{\tiny{}$305.3$} & \emph{\tiny{}$620.9$} & \emph{\tiny{}$14.96$} & \emph{\tiny{}$4.976$} & \emph{\tiny{}$8.869$} & \emph{\tiny{}$-2.15$} & \emph{\tiny{}$2.57$} & {\tiny{}$0.6122$}\tabularnewline
\emph{\tiny{}$\mathrm{1.50\thinspace M}_{\sun}$} & \emph{\tiny{}$1.80$} & \emph{\tiny{}$0.69878$} & \emph{\tiny{}$0.28562$} & \emph{\tiny{}$203.7$} & \emph{\tiny{}$12840$} & \emph{\tiny{}$56.60$} & \emph{\tiny{}$590.2$} & \emph{\tiny{}$1854$} & \emph{\tiny{}$40.11$} & \emph{\tiny{}$5.121$} & \emph{\tiny{}$8.821$} & \emph{\tiny{}$-2.14$} & \emph{\tiny{}$2.90$} & {\tiny{}$0.6212$}\tabularnewline
\hline 
\end{tabular*}
\end{table*}

\subsection{Opacity and radiative accelerations}

We compute the radiative acceleration of each element using the monochromatic data from version 3.3 of the Opacity Project (OP) database \citep{2005MNRAS.360..458B,2007MNRAS.382..245S}. The OP data consists of cross-sections $\sigma_{i}(u\equiv h\nu/kT)$ and electron scattering corrections $a_{i}(u)$ for 17 chemical elements between H and Ni in a temperature range between $\log_{10}T=3.5$ and $\log_{10}T=8.0$. The monochromatic data are used to compute the Rosseland mean opacity $\kappa_{\mathrm{R}}$ and, for each element $i$, a factor $\gamma_{i}$ which is proportional to the radiative acceleration of the respective element:
\begin{equation}
\frac{1}{\kappa_{\mathrm{R}}}=\sum_{j}N_{j}m_{j}\int\frac{1}{\sum_{i}N_{i}\sigma_{i}(v)}\mathrm{d}v,\label{eq:kappar}
\end{equation}
\begin{equation}
\gamma_{i}=\int\frac{\sigma_{i}(u)\left[1-\exp\left(-u\right)\right]-a_{i}(u)}{\sum_{j}N_{j}\sigma_{j}(u)}\mathrm{d}v.\label{eq:gamma}
\end{equation}
Here $N_{i}$ is the number fraction of element $i$, and $v(u)$ is the OP frequency variable
\begin{equation}
v(u)=\frac{15}{4\pi^{4}}\int_{0}^{u}\frac{u^{4}\exp\left(-u\right)}{\left[1-\exp\left(-u\right)\right]^{3}}\mathrm{d}u.\label{eq:freqv}
\end{equation}
With these quantities the radiative accelerations are given by
\begin{equation}
g_{\mathrm{r,}i}=\frac{l_{\mathrm{r}}\kappa_{\mathrm{R}}}{4\pi cr^{2}}\frac{\gamma_{i}}{m_{i}}\sum_{j}N_{j}m_{j},\label{eq:grad}
\end{equation}
where $l_{\mathrm{r}}$ is the radiative luminosity, and $c$ is the speed of light.

The OP team have created the OPserver module \citep{2007MNRAS.378.1031M} which is intended to facilitate the computation of accelerations in stellar evolution calculations. We have made some changes to this module in coupling it to the STARS code. First, we have made it possible to store multiple opacity (and acceleration) tables in memory at the same time. This requires computing the opacity corresponding to a given chemical composition only once. When this composition is encountered during evolution, one only needs to interpolate to the required temperature and density in the corresponding table. The second modification is the same as made by \citet{2011MNRAS.418..195H} in their incorporation of the OPserver module in a version of the STARS code -- instead of calculating the acceleration for multiple relative abundances of a given element, we calculate the accelerations of all elements in a given mixture. Finally, we have added a routine that computes the mean charge of each element from the OP data. These charges are computed on the same temperature and density grid as the opacity and are used to calculate the molecular weights of the elements.

Calculating both the opacity and the accelerations from the monochromatic OP data makes the models self-consistent in that changes in relative abundances modify the structure of the star through the opacity, which, in turn, changes the accelerations. Unfortunately, the OP opacities do not include any contribution from conduction, which becomes important after the main sequence when the central regions of the star become increasingly degenerate. Since in many cases we follow the evolution all the way up the giant branch, we use the opacity tables introduced in the code by \citet{2004MNRAS.348..201E} for regions hotter than $\log_{\mathrm{10}}T=7.3$. These tables are based on the OPAL opacities \citep{1996ApJ...464..943I} supplemented by the low-temperature opacities of \citet{1994ApJ...437..879A} and the conductive opacities of \citet{1969ApJS...18..297H,1970ApJ...159..641C}. While switching to the tabulated OPAL opacities means that changes in the relative abundances no longer modify the structure at high temperatures, by then the effects of atomic diffusion have already started to disappear because of the first dredge-up (FDU), and none of our results depend on this choice (see Section~\ref{subsec:uncertainties}).

\subsection{Grid selection}

Our simulations cover a range of primary masses $M_{\mathrm{1}}$, accreted masses $\Delta M$, and initial secondary masses $M_{2,\mathrm{i}}$ (or, equivalently, final masses $M_{2,\mathrm{f}}$). In this work we consider those systems that are the most probable in the synthetic populations computed by \citet{2015A&A...581A..62A}. According to their work, typical masses are $M_{1}\simeq0.9\text{\text{--}}1.25~\mathrm{M}_{\sun}$, $M_{2,\mathrm{f}}\simeq0.8\text{--}0.9~\mathrm{M}_{\sun}$, and $\Delta M\simeq0.05\text{--}0.2~\mathrm{M}_{\sun}$. These accreted masses and final masses of the secondaries are larger than considered in a related earlier study by \citet{2008MNRAS.389.1828S}. Therefore, we also consider some systems with smaller $\Delta M$ values, namely $0.001$ and $0.01~\mathrm{M}_{\sun}$.

In summary, we evolve stellar models with initial masses of 0.60, 0.65, 0.70, 0.75, and $0.80~\mathrm{M}_{\sun}$ and metallicity $Z=10^{-4}$ ($\mathrm{[Fe/H]}=-2.14$) starting from the pre-main-sequence. At the ages listed in Table~\ref{tab:xinp}, somewhere between $0.001$ and $0.2~\mathrm{M}_{\sun}$ of material of the corresponding composition is added to the models at a rate of $10^{-6}~\mathrm{M}_{\sun}\thinspace\mathrm{yr^{-1}}$ yielding CEMP-\emph{s} stellar models with masses between $0.8$ and $0.95~\mathrm{M}_{\sun}$. These models are evolved up to the core helium flash or an age of 16~Gyr, whichever comes first.

\section{\label{sec:Results}Results}

Two sets of models were initially evolved: in one set only thermohaline mixing, gravitational settling, and thermal diffusion were active; in the other, radiative levitation was also included. Table~\ref{tab:Results_main} lists some properties of these systems, including the {[}C/Fe{]} ratio at the surface at key points of the evolution: after thermohaline mixing, at the point where the convective envelope is smallest in mass (near the turn-off), and after first dredge-up.

\subsection{An illustrative model sequence\label{subsec:A-typical-model}}

\begin{figure*}
\subfloat[Hertzsprung-Russell diagram]{\includegraphics[width=1\columnwidth]{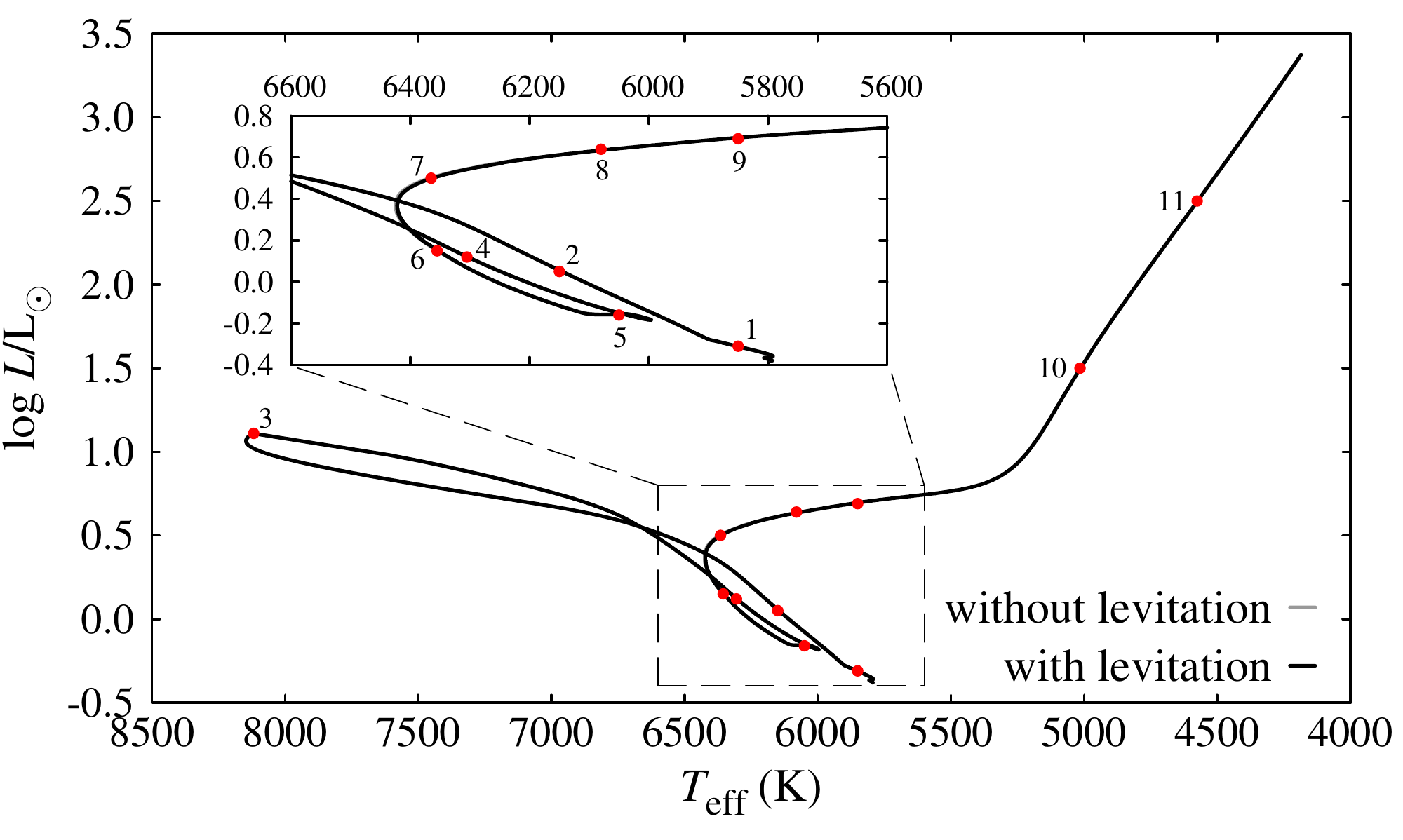}\label{fig:ms0750dm0050mp125-hrd}}\hspace{\columnsep}\subfloat[Evolution of carbon and iron surface mass fractions]{\includegraphics[width=1\columnwidth]{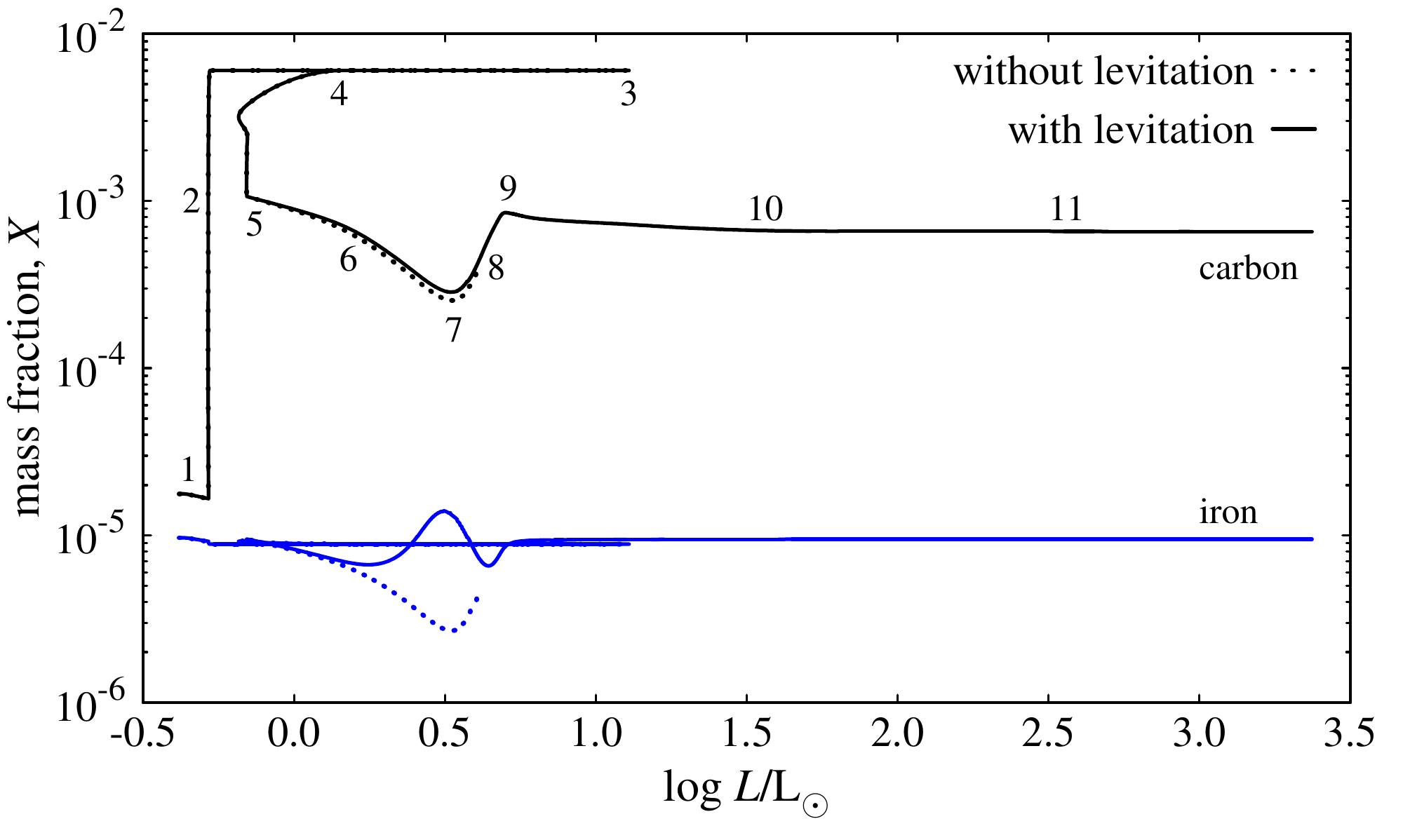}\label{fig:ms0750dm0050mp125-XvsL}}

\subfloat[Interior profiles of carbon]{\includegraphics[width=1\columnwidth]{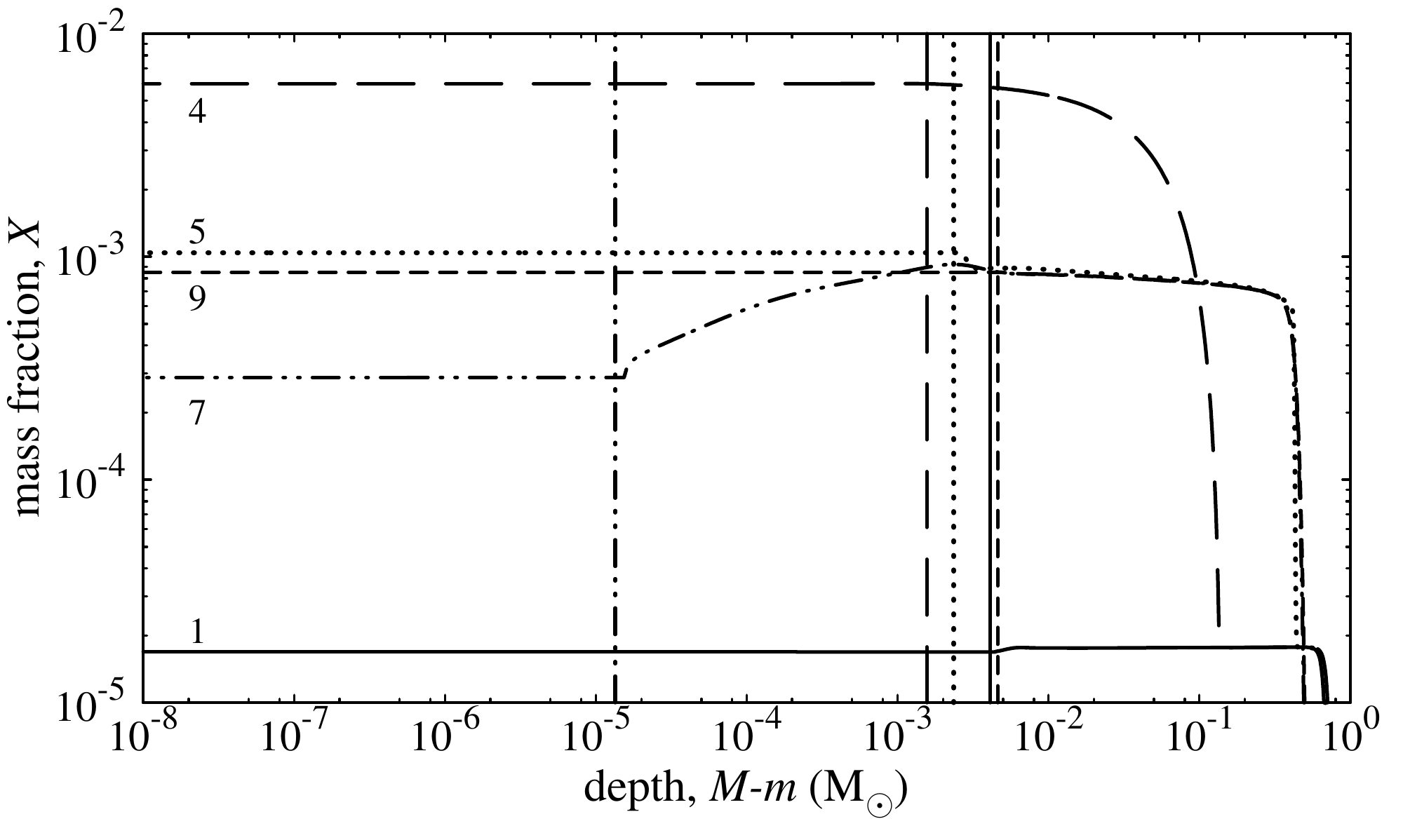}\label{fig:ms0750dm0050mp125-XCvsDpth-gs}}\hspace{\columnsep}\subfloat[Interior profiles of iron]{\includegraphics[width=1\columnwidth]{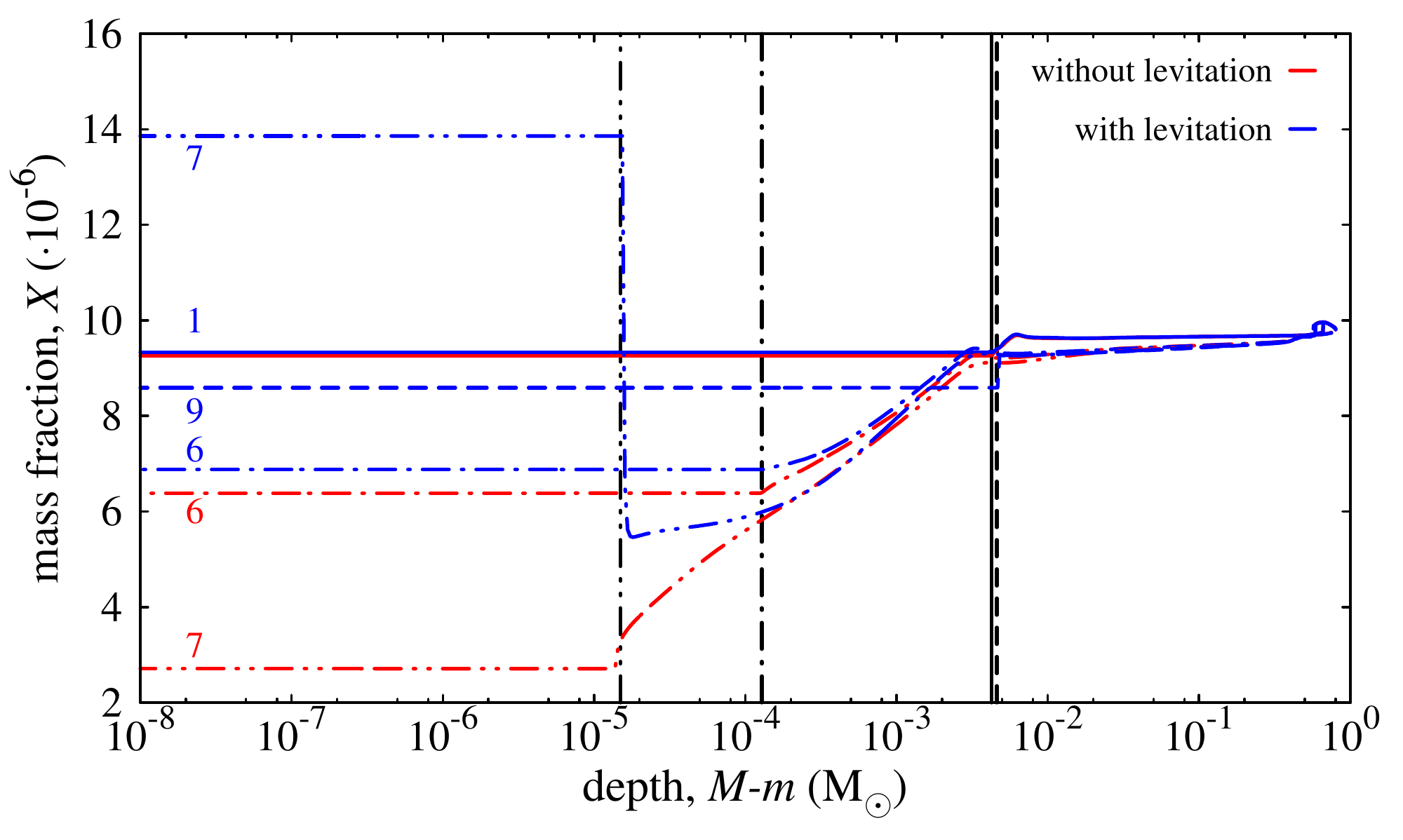}\label{fig:ms0750dm0050mp125-XCFeVSdpth-ra}}\caption{Evolution and abundances of a $M_{2,\mathrm{i}}=0.75~\mathrm{M}_{\sun}$ secondary accreting $\Delta M=0.05~\mathrm{M}_{\sun}$ of material from a $M_{1}=1.25~\mathrm{M}_{\sun}$ primary. The labels highlight specific parts of the evolution discussed in the text. The model sequences with and without radiative levitation overlap at this scale of the HRD. Interior abundance profiles are shown at a few of the stages indicated in the upper panels: before mass transfer (`1', solid); before thermohaline mixing (`4', long-dashed); after thermohaline mixing (`5', dotted); post-mass-transfer main sequence (`6', dot-dashed); minimum of convective envelope mass (`7', dot-dot-dashed); during first dredge-up (`9', short-dashed). The vertical lines in the lower panels indicate the position of the base of the convective envelope at the respective time. The interior profiles of carbon with and without levitation nearly coincide and only the case with levitation is shown.\label{fig:ms0750dm0050mp125}}
\end{figure*}

To understand how the evolution of surface abundances is influenced by the different physical processes included in our simulations, let us consider a particular model sequence in detail. Figure~\ref{fig:ms0750dm0050mp125} illustrates the case of a secondary with an initial mass of $0.75~\mathrm{M}_{\sun}$ that accretes $0.05~\mathrm{M}_{\sun}$ of material from a $1.25~\mathrm{M}_{\sun}$ primary. Multiple stages of evolution can be distinguished.

Prior to mass transfer the secondary slowly evolves as a $0.75~\mathrm{M}_{\sun}$ main sequence star (the part of the evolution labelled `1' in Figs.\,\ref{fig:ms0750dm0050mp125-hrd} and \ref{fig:ms0750dm0050mp125-XvsL}). During this stage gravitational settling dominates and the abundance of every element other than hydrogen decreases at the surface.

At $t=3.06$~Gyr mass transfer begins and the surface composition quickly becomes equal to that of the accreted material (`2'). During the accretion the star becomes hotter and more luminous. This is common for many system configurations in which the Kelvin-Helmholtz timescale of the secondary becomes comparable to the accretion timescale. In some models the effective temperature and luminosity can reach values as high as 9500~K and $30~\mathrm{L}_{\sun}$, respectively. But once accretion stops (`3'), both luminosity and temperature rapidly drop, resulting in loops in the Hertzsprung-Russell diagram (HRD) as the star settles back on the main sequence. Generally, these loops are more characteristic of secondaries with larger initial masses.

Shortly after accretion stops the accreted material starts to mix with the original material of the secondary as a result of the thermohaline instability (`4'). As shown by Fig.~\ref{fig:ms0750dm0050mp125-XCvsDpth-gs}, some of the interior is already mixed by the time the surface abundances change. The mixing takes only a few hundred million years (about $150~\mathrm{Myr}$ in this case) and is over before the star has settled back on the main sequence (`5'). Ultimately, the surface carbon abundance is reduced by about 0.8~dex (regardless of radiative levitation), whereas the iron abundance is barely affected because it is virtually the same in the original and accreted compositions.

Over the rest of the post-mass-transfer main sequence lifetime the abundances are again modified by atomic diffusion (`6'). At first gravitational settling prevails over radiative levitation and the surface becomes increasingly hydrogen-rich as all heavier elements settle out of the surface convection zone. As the star nears the turn-off, this convection zone becomes ever more superficial and radiative effects become increasingly important (Fig.~\ref{fig:ms0750dm0050mp125-XCFeVSdpth-ra}). Once an element's radiative velocity at the base of the convection zone exceeds its settling velocity, the surface abundance of this element increases. This is typically the case with iron. In contrast, if an element's settling velocity is always greater than its radiative velocity, the surface abundance of this element continues to decrease (although less so than in the case when radiative effects are ignored). This is always the case with helium and carbon. The behaviour of other elements is not readily predicted because of the non-monotonic shape of the radiative accelerations (as a function of temperature) and the outward movement of the base of the envelope (decreasing temperature at the base). Therefore, the abundance of most elements alternates between increasing at those times when the radiative velocity exceeds the settling velocity and decreasing at others \citep[e.g. see figure 2 of][]{2002ApJ...568..979R}. 

Eventually, the abundance anomalies, i.e. their values relative to those after thermohaline mixing, reach their maxima (`7'). At this stage the difference between the two sets of models is greatest -- compared to the abundances after thermohaline mixing, in models without levitation the abundances of all elements are reduced (dotted lines in Fig.~\ref{fig:ms0750dm0050mp125-XvsL}), whereas in models with levitation this is not always the case (solid lines in Fig.~\ref{fig:ms0750dm0050mp125-XvsL}; in both cases only carbon and iron are shown for clarity) and readily levitated elements can be over-abundant. The anomalies are maximal shortly after the turn-off when the convective envelope is smallest (Figs.~\ref{fig:ms0750dm0050mp125-XCvsDpth-gs} and \ref{fig:ms0750dm0050mp125-XCFeVSdpth-ra}).\footnote{In fact, the convective envelope has already slightly grown in mass. For a short time the diffusion timescale is shorter than the evolutionary timescale.} This occurs at the same time in models with and without levitation. 

Next, as the convective envelope grows in mass, the material in it is mixed with that of the immediately adjacent, previously radiative layers. In models without diffusion no change in surface abundances would occur until the envelope reached depths where CN cycling had occurred (i.e. at first dredge-up). With diffusion, however, the composition of the envelope is different from that of the radiative layers below, and therefore the effect of the deepening of the envelope is to first undo all the work done by diffusion (`8'). When the envelope mass has reached a few thousandths of a solar mass, all surface evidence of atomic diffusion has been erased and the abundances are similar to those after thermohaline mixing (`9').

First dredge-up homogenizes the composition in the layers above a mass coordinate of $0.3\text{--}0.35~\mathrm{M}_{\sun}$ ($0.34~\mathrm{M}_{\sun}$ in this case). What effect this has on the surface abundances depends on how this depth compares to the depth of thermohaline mixing ($m_{\text{thm}}=0.37~\mathrm{M}_{\sun}$ in this case). If thermohaline mixing is not as deep as the maximum depth reached by the envelope at FDU, the accreted material is further diluted with the original material of the secondary. Otherwise, most abundances do not change. However, some of the accreted carbon will then have been converted into nitrogen. As shown by \citet{2007A&A...464L..57S}, during late FDU (which ends at around $\log L\approx1.5$; `10') this nitrogen is dredged up to the surface.

Finally, after the luminosity bump (`11') $^{3}\mathrm{He}$-burning reduces the mean molecular weight above the hydrogen-burning shell. Thus, a $\mu$-inversion, which is magnified by the settling of $^{4}\text{He}$ \citep{2010A&A...510A.104M}, develops between the shell and the receding convective envelope -- a situation again unstable to thermohaline mixing. This alters the surface abundance of nitrogen by 0.1~dex at most. The much greater carbon abundance remains essentially unchanged. The abundance change after the bump is much smaller than found by \citet{2009MNRAS.396.2313S} because the thermohaline mixing coefficient in this work is about $10^{3}$ times smaller.

This model sequence illustrates the role each physical process plays in all models with atomic diffusion. We see that diffusion modifies the surface composition on the main sequence both before and after mass transfer. This modification is greatest around the turn-off, when the convective envelope is shallowest (point `7' in Fig.~\ref{fig:ms0750dm0050mp125}). We now turn to discussing the expected abundance changes for all CEMP-\emph{s} stars in this evolutionary stage.

\subsection{Abundance anomalies near the turn-off}

\begin{figure}
\centering
\subfloat{\includegraphics[width=0.965\columnwidth]{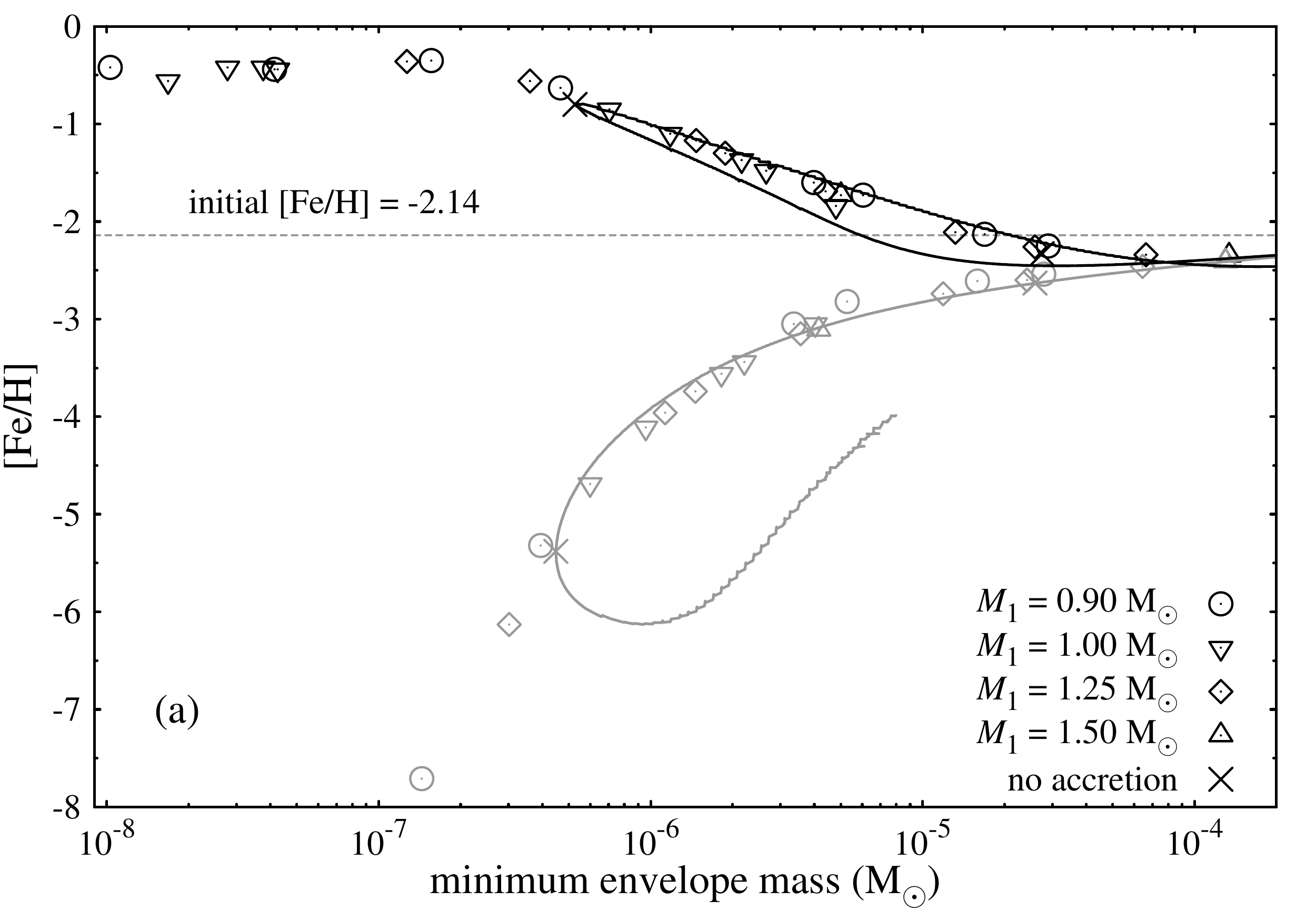}\label{fig:ab_vs_menv-FeH}}

\subfloat{\includegraphics[width=0.965\columnwidth]{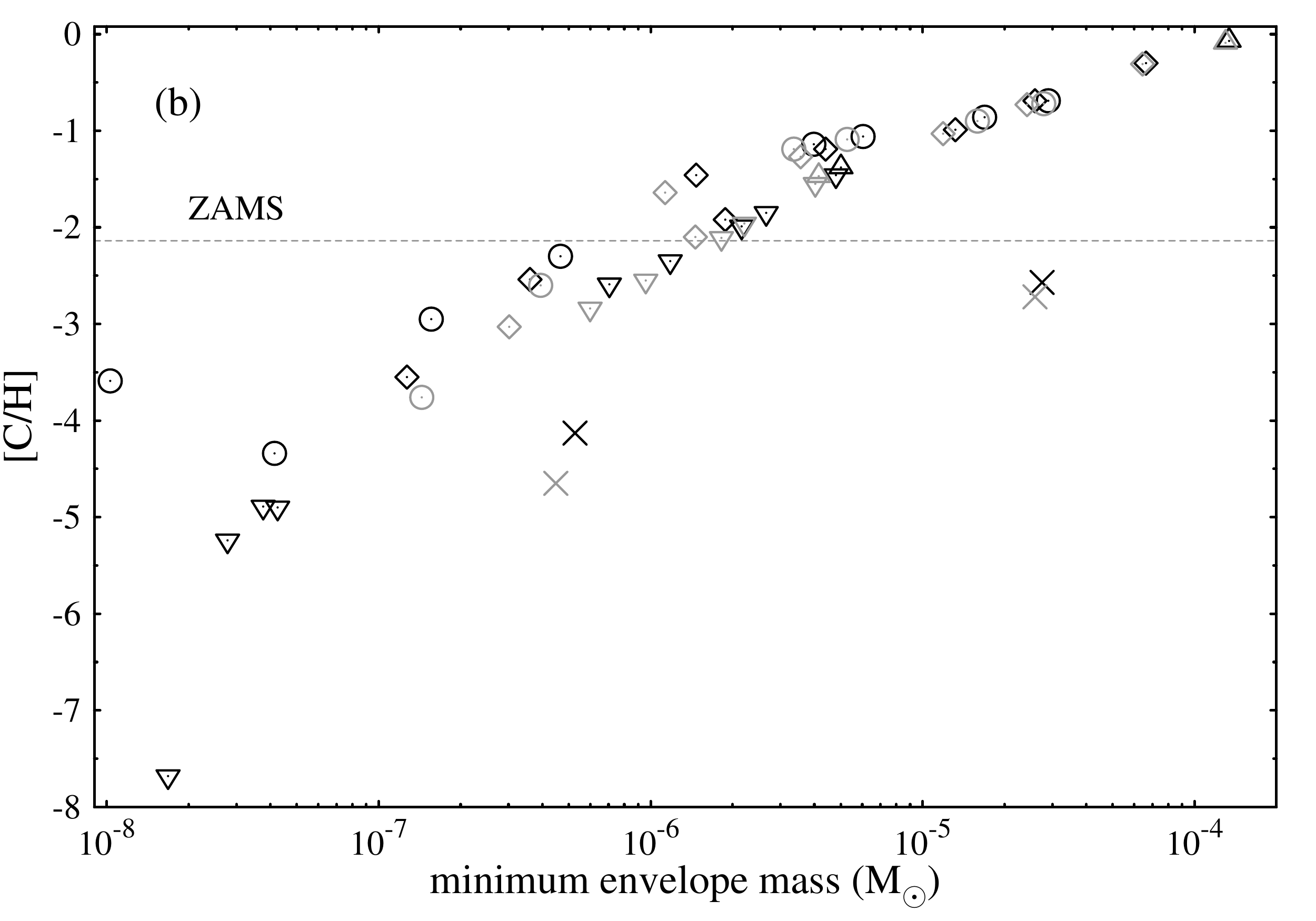}\label{fig:ab_vs_menv-CH}}

\subfloat{\includegraphics[width=0.965\columnwidth]{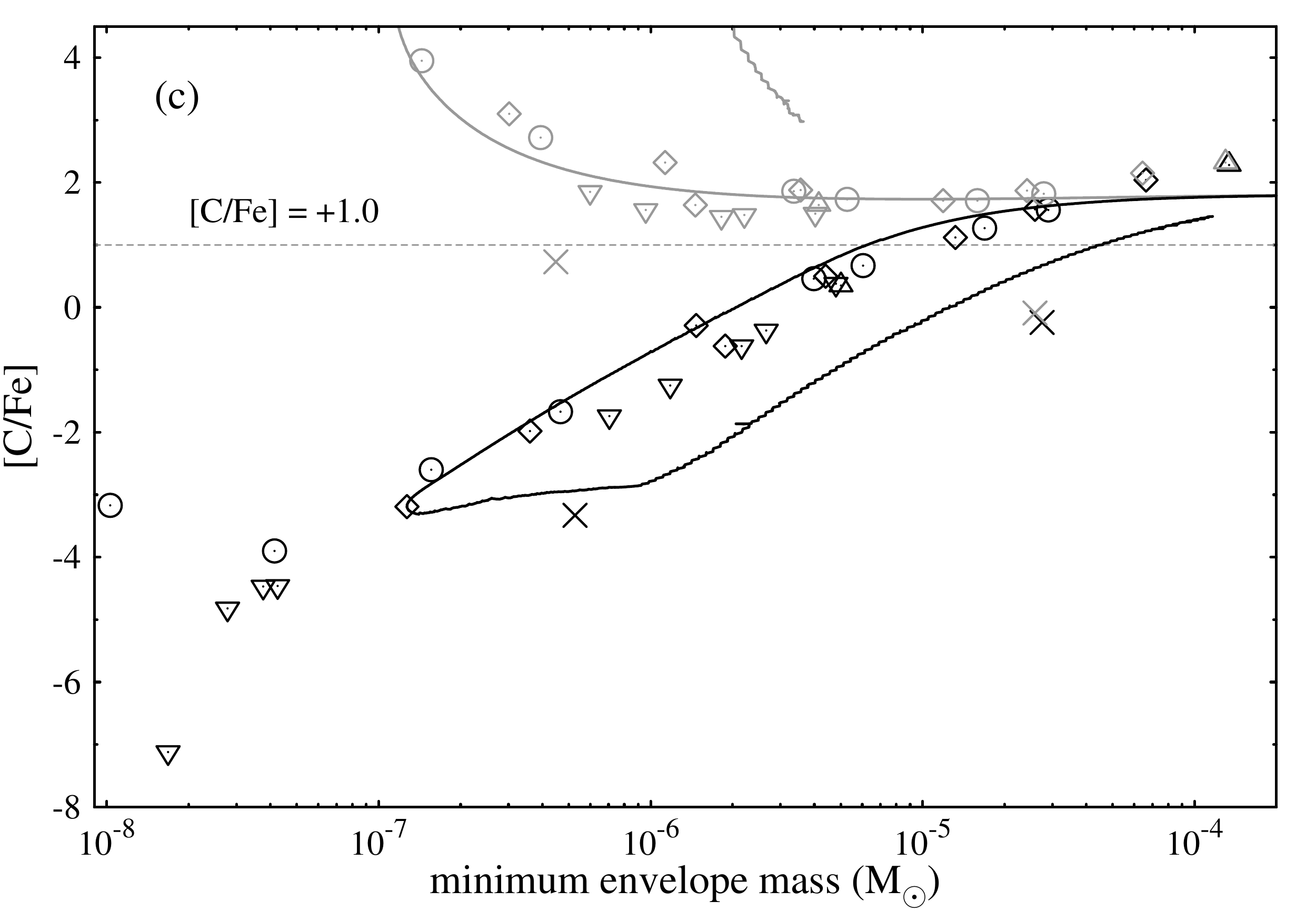}\label{fig:ab_vs_menv-CFe}}

\caption{Symbols show {[}Fe/H{]} (a), {[}C/H{]} (b), and {[}C/Fe{]} (c) in each of the CEMP-\emph{s} models at the point where the mass of the convective envelope is smallest, i.e. just before first dredge-up. Models with and without radiative levitation are plotted with black and grey symbols, respectively. All values for envelope masses below $10^{-7}~\mathrm{M}_{\sun}$ denote upper or lower limits. The {[}Fe/H{]} plot also shows the metallicity evolution in a $0.8~\mathrm{M}_{\sun}$ model sequence with no accretion (solid lines). Similarly, the {[}C/Fe{]} plot shows the evolution in a model corresponding to $M_{1}=1.25~\mathrm{M}_{\sun}$, $M_{2,\text{i}}=0.8~\mathrm{M}_{\sun}$, and $\Delta M=0.05~\mathrm{M}_{\sun}$. Both sequences without levitation stop during the first dredge-up. The criterion for being classified as a carbon-enhanced metal-poor star ($[\mathrm{C/Fe}]\geq1.0$) is from \citet{2005ARA&A..43..531B}.\label{fig:ab_vs_menv}}
\end{figure}

During the main sequence the mass of the convective envelope, $M_{\mathrm{env}}$, of a low-mass star decreases. Therefore, the timescale for atomic diffusion, which is proportional to roughly the square root of $M_{\mathrm{env}}$ \citep{1977Natur.266..433M}, also decreases. In nearly all of our CEMP-\emph{s} models the envelope mass reaches a minimum of less than $10^{-4}~\mathrm{M}_{\sun}$ around the turn-off. The corresponding timescales are short enough compared to the nuclear timescale for atomic diffusion to notably modify the surface composition. Figure~\ref{fig:ab_vs_menv} summarizes the extent of the abundance variations in our models with diffusion (Table~\ref{tab:Results_main}). Specifically, the figure shows the {[}Fe/H{]}, {[}C/H{]}, and {[}C/Fe{]} abundances at the time when the convective envelope mass reaches the minimum in each of the CEMP-\emph{s} models. In models with envelope masses always above about $2\times10^{-5}~\mathrm{M}_{\sun}$ gravitational settling prevails, however, the abundances are decreased only by up to a factor of two from their values after thermohaline mixing. But in models with even smaller $M_{\mathrm{env}}$ at the turn-off, abundances are modified by a factor of ten or more and radiative levitation becomes important (Fig.~\ref{fig:ab_vs_menv-FeH}). The model discussed in Section~\ref{subsec:A-typical-model} is close to this limit -- its envelope mass has a minimum of about $1.2\times10^{-5}~\mathrm{M}_{\sun}$. At this minimum its {[}C/Fe{]} is $1.71$ when levitation is ignored versus $1.12$ when it is included (in both cases down from $1.78$ after thermohaline mixing), and {[}Fe/H{]} is $-2.74$ and $-2.11$, respectively.

The results from many simulations plotted in Figs.~\ref{fig:ab_vs_menv-FeH} and \ref{fig:ab_vs_menv-CFe} form two sequences corresponding to model sets with and without levitation. As shown by the solid lines, to a decent approximation we can interpret these sequences as describing the abundance evolution in a single simulation as the envelope mass changes. For example, as the envelope mass decreases from $10^{-5}$ to $10^{-6}~\mathrm{M}_{\sun}$, {[}C/Fe{]} decreases by about two orders of magnitude in models with levitation because while carbon continues to settle, iron is levitated. On the other hand, in models without levitation {[}C/Fe{]} does not change because both elements settle at similar rates (Figs.~\ref{fig:ab_vs_menv-FeH} and \ref{fig:ab_vs_menv-CH}). At still smaller envelope masses, {[}C/Fe{]} increases because of the low degree of ionization of iron at the base of the envelope. The small mean charge of iron gives a large diffusion coefficient because of the approximately $D\sim\bar{Z}^{-2}$ dependence \citep{1986ApJS...61..177P} and, consequently, a large settling velocity.

Figure~\ref{fig:ab_vs_menv} shows that for given post-thermohaline-mixing abundances the abundance evolution prior to FDU can be parametrized as a function of only $M_{\text{env}}$. But once FDU starts and the envelope deepens, a kind of hysteresis is seen in that the abundances at a given $M_{\text{env}}$ are not the same as they were prior to FDU. This is because diffusion has modified the radiative layers below the envelope in the meantime.

Carbon-enhanced metal-poor stars are distinguished from other metal-poor stars based on their {[}C/Fe{]} value. Assuming that atomic diffusion is correctly predicted and no additional mixing processes operate in the radiative regions below the envelope, Fig.~\ref{fig:ab_vs_menv-CFe} implies that they must have envelope masses larger than $10^{-5}~\mathrm{M}_{\sun}$. Otherwise, they would not be classified as carbon-enhanced. While models without levitation do not have such a limit, the metallicity rapidly decreases below this value because of settling -- at $M_{\mathrm{env}}\approx10^{-6}~\mathrm{M}_{\sun}$ the surface $[\mathrm{Fe/H}]\approx-4$, which is much lower than typical of CEMP-\emph{s} stars.

For a given combination of AGB and CEMP-\emph{s} star masses ($M_{1}$ and $M_{2,\mathrm{f}}$, respectively) the convective envelope is deeper in models with larger accreted mass. For example, a $0.8~\mathrm{M}_{\sun}$ CEMP-\emph{s} star with an initial mass of $0.6~\mathrm{M}_{\sun}$ retains a more massive envelope than one with an initial mass of $0.7~\mathrm{M}_{\sun}$. This is due to the higher average opacity of these stars (more metal-rich stars maintain thicker convective envelopes for the same reason). The difference in $M_{\mathrm{env}}$ can be a factor of 2--10 (depending on $M_{1}$, $M_{2,\mathrm{f}}$, and the range of $M_{2,\mathrm{i}}$) which can lead to substantially different abundances when the envelopes are small (Fig.~\ref{fig:ab_vs_menv}).

As can be seen from Figure~\ref{fig:ab_vs_menv} and Table~\ref{tab:Results_main}, diffusion is extremely efficient in many of our models leading to unrealistic abundance anomalies (e.g. $\mathrm{[C/Fe]}<-1$ or $\mathrm{[C/Fe]}>4$). In most of our more massive CEMP-\emph{s} models ($M_{2,\mathrm{f}}\geq0.85~\mathrm{M}_{\sun}$) diffusion is so efficient that our code is incapable of resolving the steep abundance gradients developing at the base of the envelope and we are forced to stop the computations before the main-sequence turn-off. Such massive CEMP-\emph{s} stars are nevertheless probable according to population synthesis calculations \citep{2015A&A...581A..62A} and would help explain the properties of some CEMP-\emph{s} RR Lyrae stars \citep{2013MNRAS.435..698S}, so we would like to explore their connection to observations. Therefore, we proceed by assuming that diffusion, and possibly thermohaline mixing, is inhibited throughout these stars for one reason or another (leaving open the nature and cause of the underlying mechanism) and evolve two sets of model sequences without atomic diffusion: one set with thermohaline mixing and one without. The results from these simulations are summarized in Table~\ref{tab:Results_massive}.

These models have some key differences in global properties and surface abundances from the model sequences with atomic diffusion. First, the surface abundances do not change prior to mass transfer. More importantly, after thermohaline mixing has reached equilibrium, no further abundance changes occur until FDU (i.e. between the stages labelled `5' and `9' in Figs.~\ref{fig:ms0750dm0050mp125-hrd} and \ref{fig:ms0750dm0050mp125-XvsL}). The importance of FDU still depends on the depth of thermohaline mixing, as in models with diffusion. In models without thermohaline mixing the surface abundances do not change until FDU during which the accreted material is invariably diluted by mixing throughout most of the star (down to a mass coordinate of $0.3\text{--}0.35~\mathrm{M}_{\sun}$).

In agreement with previous studies, models with diffusion are younger (by a few percent) at a given evolutionary stage and have lower effective temperatures at the turn-off than models without diffusion, primarily because of the gravitational settling of helium throughout the star \citep[e.g.][]{1999A&A...344...97C,2002ApJ...571..487V,2012MNRAS.427..127B}. Our non-accreting models with diffusion are about 150~K cooler than models without diffusion (compare Table~\ref{tab:Results_main} and \ref{tab:Results_massive}). Among our CEMP-\emph{s} models, those with thermohaline mixing but without diffusion are generally between 100 to 300~K hotter than models without both. The latter are cooler because of the high opacity of the outer layers owing to the metal-richness of the accreted material. CEMP-\emph{s} models with diffusion likely fall somewhere in between but we cannot do a proper comparison because our massive models with diffusion do not make it to turn-off for numerical reasons.

\subsection{Thermohaline mixing}

\begin{figure}
\includegraphics[width=1\columnwidth]{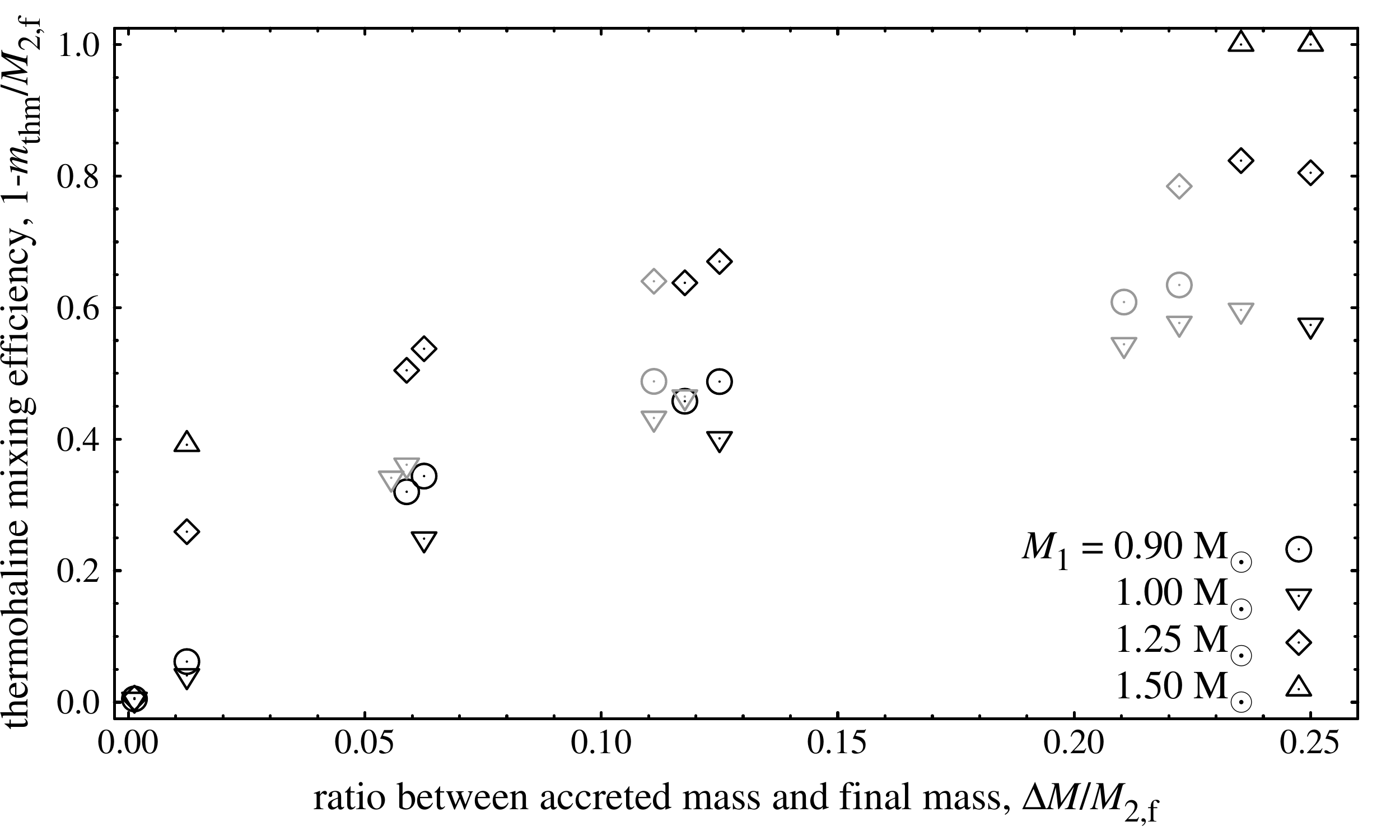}

\caption{Thermohaline mixing efficiency (fraction of the star that is mixed) as a function of the ratio between accreted mass and final mass for different primaries. Black and grey symbols correspond to models with and without diffusion, respectively. Each symbol represents a unique combination of $M_{1}$, $M_{2,\mathrm{f}}$, and $\Delta M$.\label{fig:thmix_eff}}
\end{figure}

The fraction of the star mixed by thermohaline convection, which we shall call the thermohaline mixing efficiency, essentially depends on the mean molecular weight of the accreted material and its total mass compared to the final mass of the star. The more helium- and metal-rich the accreted material, the greater its molecular weight compared to the initial composition, and the greater the portion of the star that gets mixed. From the last column of Table~\ref{tab:xinp} we therefore expect that, for a given amount of accreted material, thermohaline mixing should be most efficient when that material comes from a primary of $1.5~\mathrm{M}_{\sun}$ and least efficient when it comes from a primary of $1.0~\mathrm{M}_{\sun}$, which is indeed the case (Fig.~\ref{fig:thmix_eff}).

\begin{figure}
\includegraphics[width=1\columnwidth]{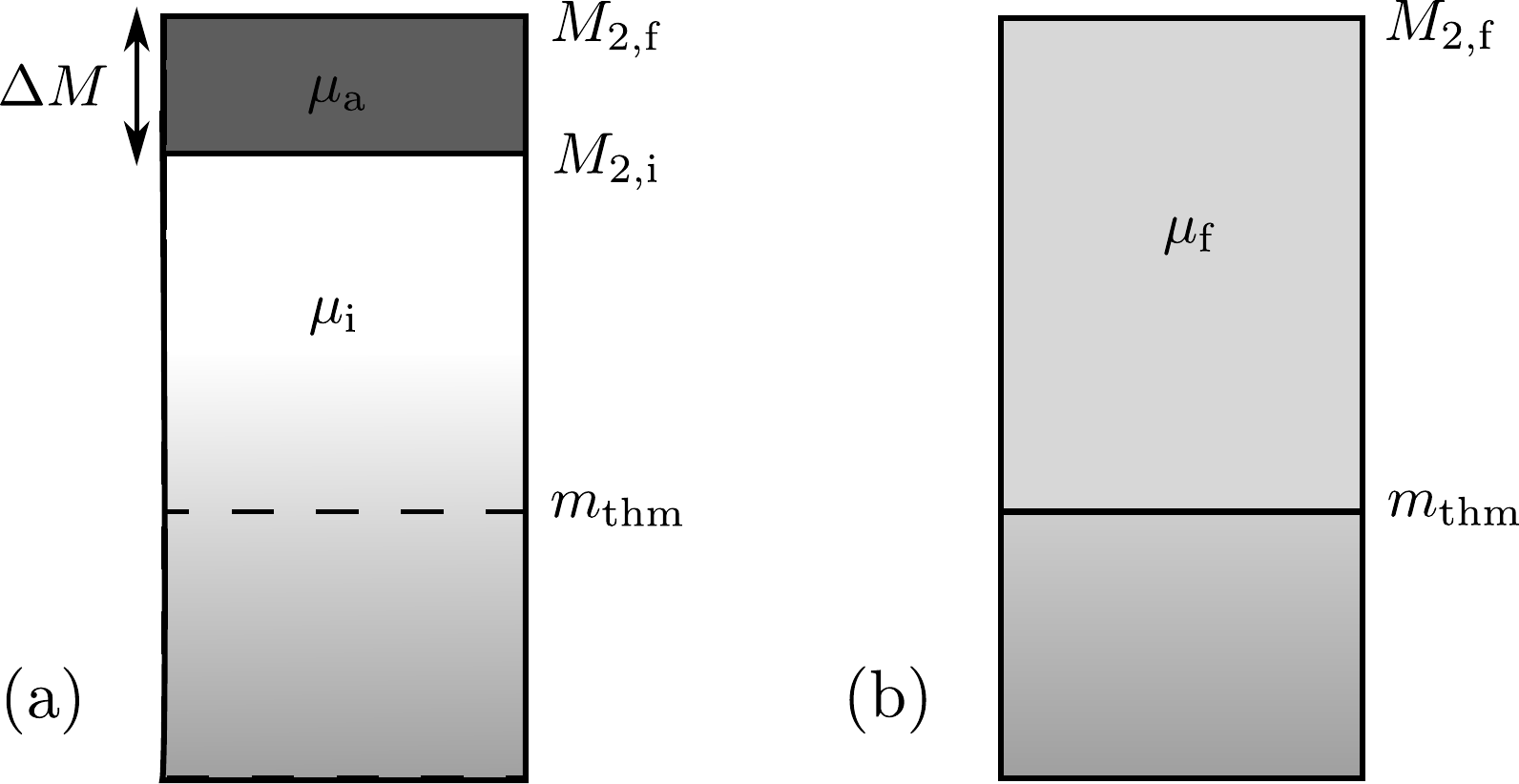}

\caption{Schematic illustration of the structure of the star before (a) and after (b) thermohaline mixing. The shaded area around $m_\text{thm}$ and below indicates a region in which the mean molecular weight has been raised as a result of nuclear processing.\label{fig:thmix_illustration}}
\end{figure}

Furthermore, the greater the amount of the high-$\mu$ material, the deeper the mixing must be for the $\mu$-gradient to be removed. If an amount $\Delta M$ of AGB ejecta with a mean molecular weight $\mu_{\text{a}}$ is mixed with $M_{2,\text{f}}-\Delta M-m_{\text{thm}}$ of the unpolluted material with an average molecular weight $\mu_{\text{i}}$ ($<\mu_{\text{a}}$) before the $\mu$-gradient is removed, a mixed region $M_{2,\text{f}}-m_{\text{thm}}$ with molecular weight $\mu_{\text{f}}$ results (Fig.~\ref{fig:thmix_illustration}). Equating the states before and after mixing, one gets that the removal of the $\mu$-gradient implies a linear relationship between the accreted-to-final mass ratio and mixing efficiency. Indeed, a linear relationship is a reasonable approximation in the range $0.05\lesssim\Delta M/M_{2,\mathrm{f}}\lesssim0.2$ of accreted-to-final mass ratios (Fig.~\ref{fig:thmix_eff}). However, higher and lower ratios of $\Delta M/M_{2,\mathrm{f}}$ require special consideration.

First, a high $\Delta M/M_{2,\mathrm{f}}$ corresponds to a case where a lot of mass is transferred to a low-mass star, which implies that the mass transfer takes place when the secondary is still nearly on the ZAMS. Two outcomes are possible. If the accreted material has a sufficiently high molecular weight compared to the molecular weight throughout the star, thermohaline mixing affects the whole star, and the composition is nearly homogenized. For example, this occurs when a $0.65~\mathrm{M}_{\sun}$ secondary accretes $0.2~\mathrm{M}_{\sun}$ of material from a $1.5~\mathrm{M}_{\sun}$ primary. On the other hand, if the accreted material has a lower molecular weight than some central region of the star, increasing the accreted mass will not lead to a much deeper mixing because of the steepness of the $\mu$-gradient near the center. The mixing efficiency can therefore decrease for higher $\Delta M/M_{2,\mathrm{f}}$ (although at this point almost all of the star will be mixed, anyway).

Second, in models with diffusion low accreted-to-final mass ratios lead to relatively inefficient thermohaline mixing. Contrary to models without diffusion, where the molecular weight is constant outside of nuclear burning regions, in these models there is a stabilizing $\mu$-gradient throughout the star owing to gravitational settling. This presents a ``$\mu$-barrier'' to thermohaline mixing that must be overcome for mixing to happen \citep{2008ApJ...677..556T}. If only a tiny amount of material is accreted ($\Delta M\lesssim0.001~\mathrm{M}_{\sun}$), thermohaline mixing can be almost completely inhibited (left bottom corner of Fig.~\ref{fig:thmix_eff}). Nevertheless, even in these cases the surface carbon content is depleted by a factor of two or more because the mass of the mixed region, $M_{2,\text{f}}-m_{\text{thm}}$, is always greater than $\Delta M$. 

For $\mathrm{\Delta}M>0.01~\mathrm{M}_{\sun}$ the $\mu$-barrier is largely overwhelmed. However, the mixing is nevertheless slightly less efficient than in models without diffusion even for large amounts of accretion. Similar conclusions were reached by \citet{2008MNRAS.389.1828S}. Overall, the surface carbon abundance is typically reduced by somewhere between 0.3 to 1~dex depending on the relative amount of the accreted material and its molecular weight. Since radiative accelerations have almost no influence on the molecular weight profile deep in the star, they have almost no influence on the efficiency of thermohaline mixing.

\section{\label{sec:Discussion}Comparison with observations}

We have presented four sets of models of CEMP-\emph{s} stars. Two of the sets comprise models with thermohaline mixing and atomic diffusion (one set with, one without radiative levitation). The other two sets comprise models without diffusion (one set with, one without thermohaline mixing). We now compare the four sets of models with observations of CEMP stars.

The largest data set of Galactic metal-poor stellar spectra currently available is that from the SDSS and SEGUE surveys. \citet{2013AJ....146..132L} used the spectra collected by SDSS/SEGUE to derive the stellar parameters and carbon abundances ({[}C/Fe{]}) in close to 250\,000 stars, around 10\,000 of which have $\text{[Fe/H]}<-2$. We now use this homogeneous metal-poor sample \citep[priv. comm.]{2013AJ....146..132L} to compare the observed carbon abundances with our models.

\begin{figure*}
\subfloat{\includegraphics[width=1\columnwidth]{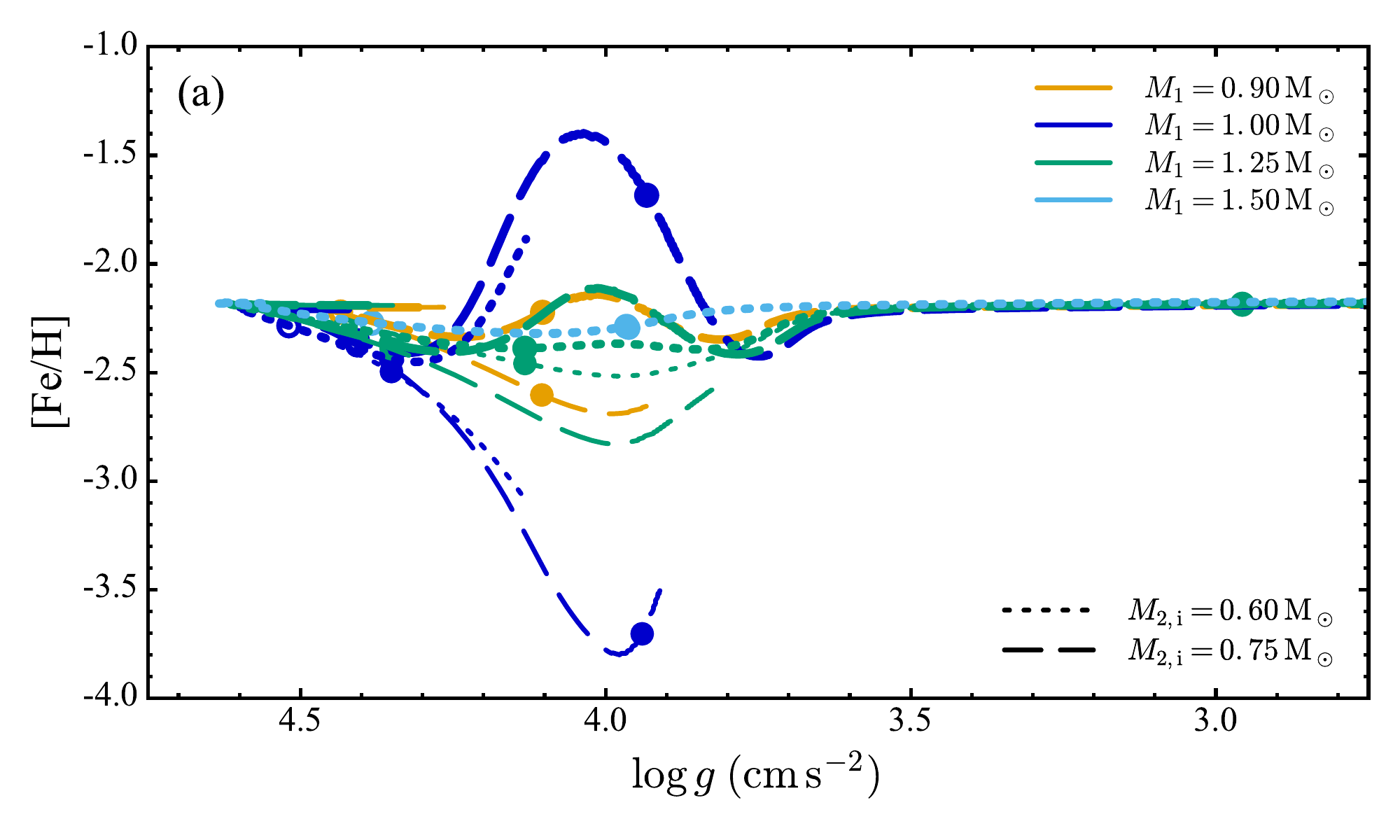}\label{fig:feh_gsra_0.8}}\hspace{\columnsep}\subfloat{\includegraphics[width=1\columnwidth]{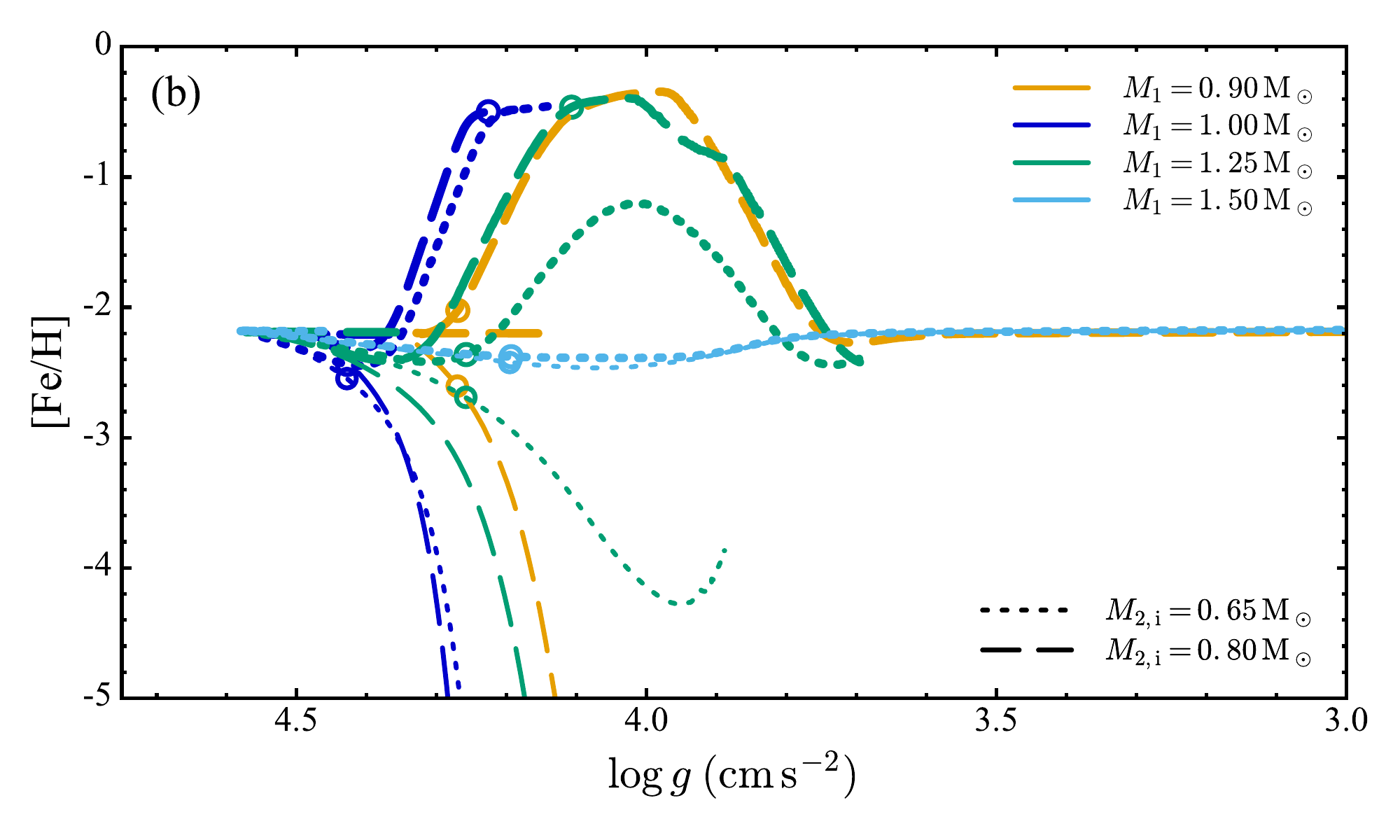}\label{fig:feh_gsra_0.85}}

\subfloat{\includegraphics[width=1\columnwidth]{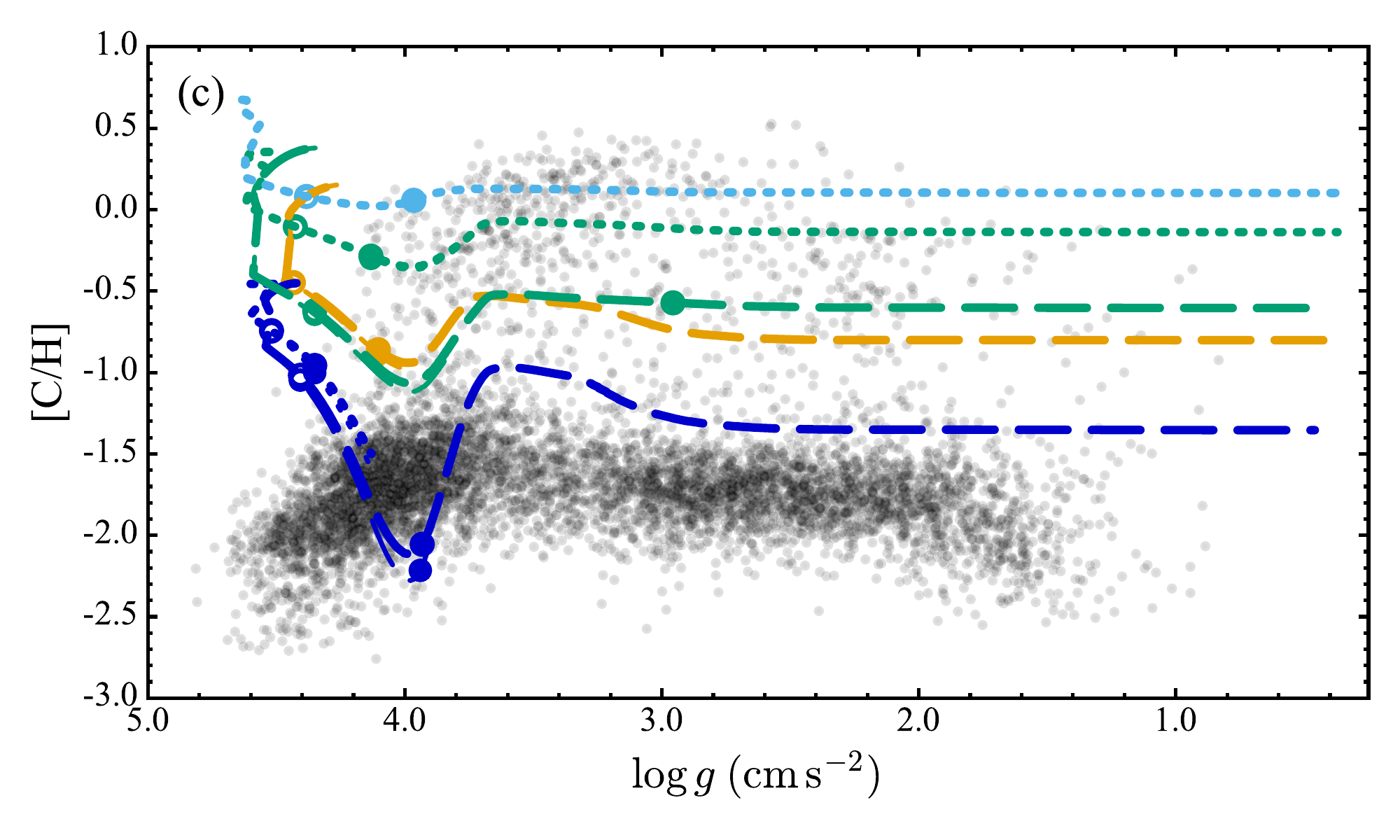}\label{fig:ch_gsra_0.8}}\hspace{\columnsep}\subfloat{\includegraphics[width=1\columnwidth]{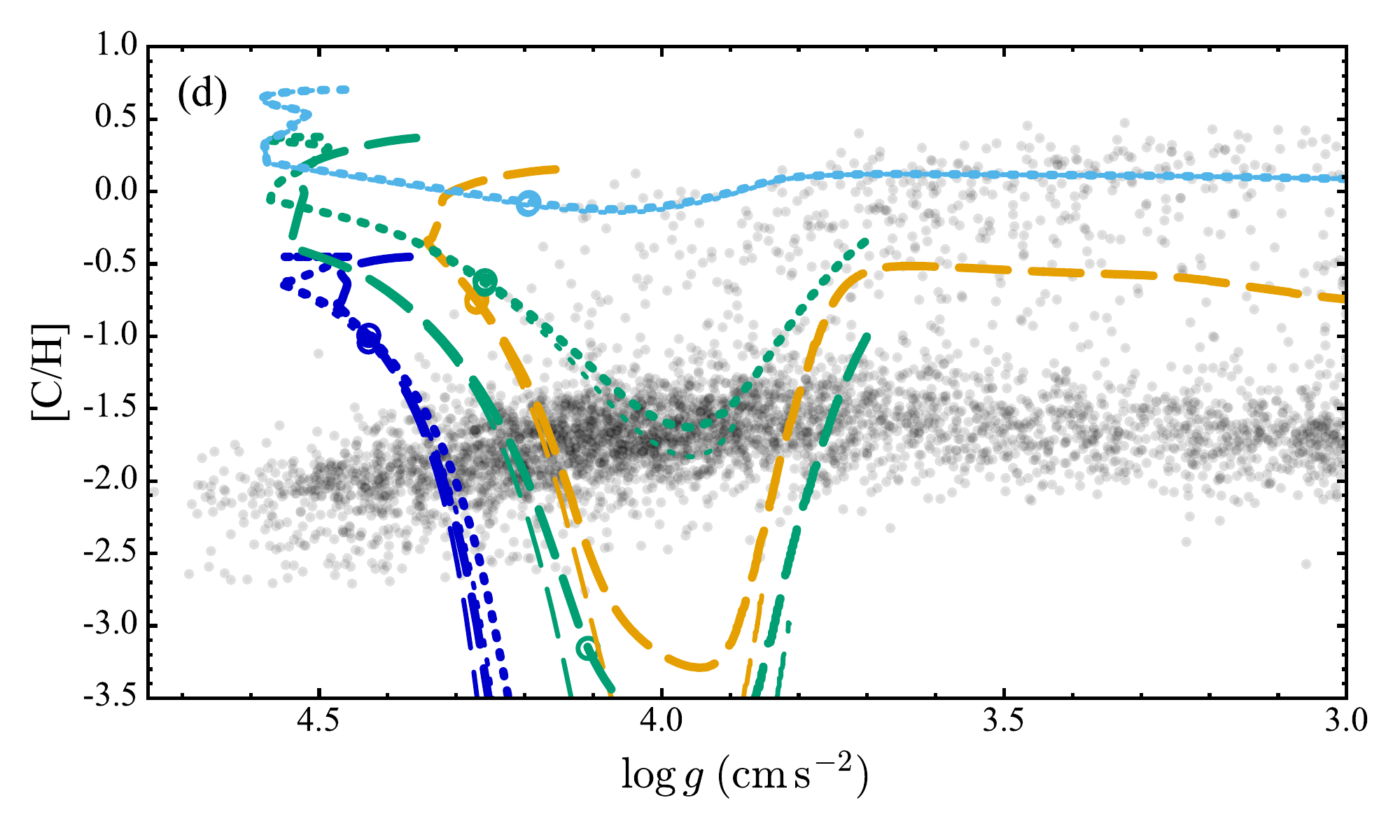}\label{fig:ch_gsra_0.85}}

\subfloat{\includegraphics[width=1\columnwidth]{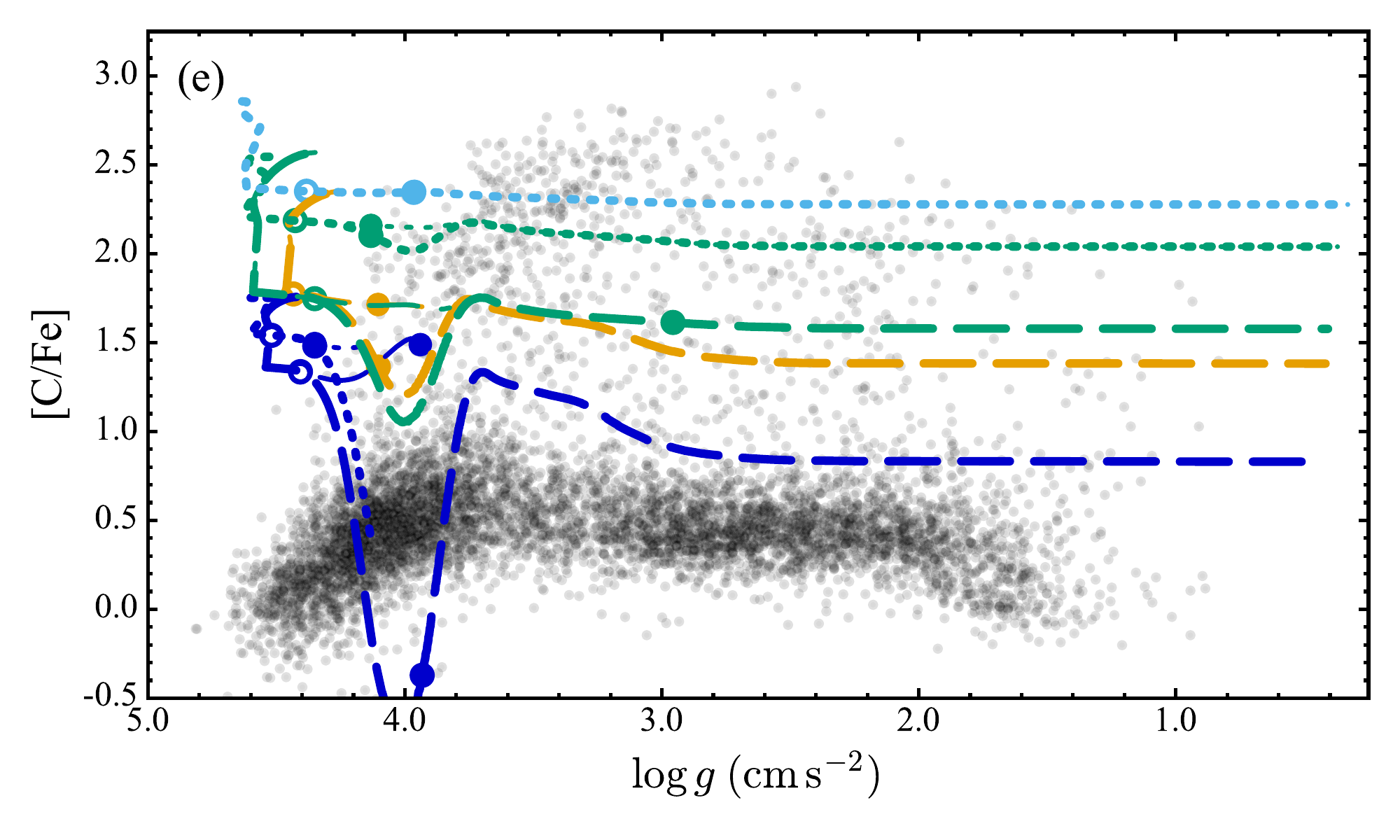}\label{fig:cfe_gsra_0.8}}\hspace{\columnsep}\subfloat{\includegraphics[width=1\columnwidth]{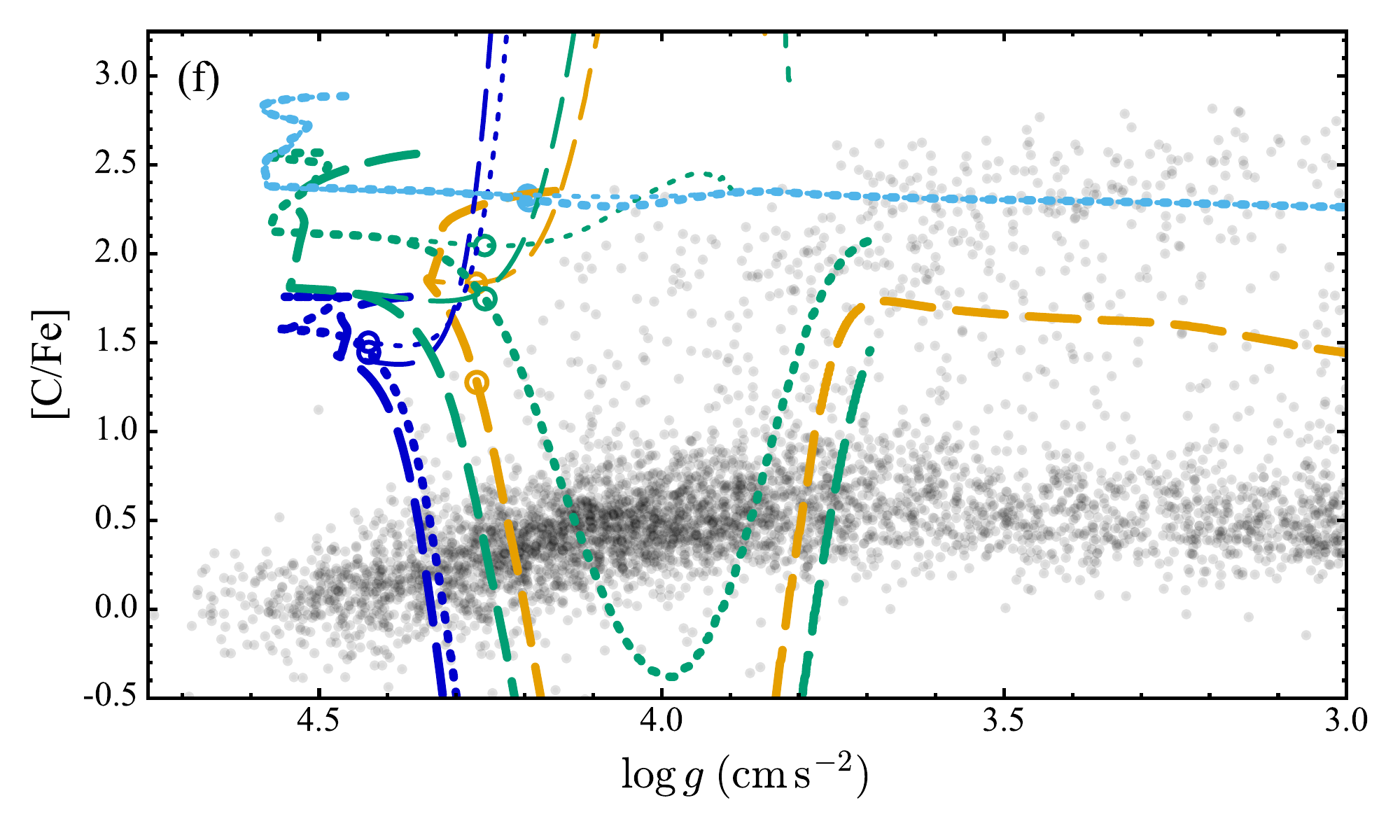}\label{fig:cfe_gsra_0.85}}

\caption{Evolution of {[}Fe/H{]} (upper panels), {[}C/H{]} (middle panels) and {[}C/Fe{]} (lower panels) in CEMP-\emph{s} models of $0.8~\mathrm{M}_{\sun}$ (left panels) and $0.85~\mathrm{M}_{\sun}$ (right panels) with diffusion. Thick lines are models with radiative levitation, whereas thin lines are those without. Empty circles mark an age of 10~Gyr and filled circles mark an age of 13.8~Gyr, i.e. the part of the track between the circles covers the expected age range of CEMP-\emph{s} stars. The small, grey circles show the metal-poor stars observed in the Sloan Digital Sky Survey in the metallicity range $-2.5\leq\text{[Fe/H]}\leq-2.0$ \citep{2013AJ....146..132L}.\label{fig:feh_ch_cfe_gsra_0.8_0.85}}
\end{figure*}

Comparing the observed {[}C/Fe{]} abundances with models must be done with care. Use of a fixed metallicity ({[}Fe/H{]}) range might not be adequate because of the diffusion of iron (Figs.~\ref{fig:feh_gsra_0.8} and \ref{fig:feh_gsra_0.85}). Only the $0.8~\mathrm{M}_{\sun}$ models with levitation generally predict {[}Fe/H{]} to remain within a factor of about two from the initial value ($\text{[Fe/H]}=-2.14$) throughout the evolution, whereas the models without levitation have $[\mathrm{Fe/H}]\lesssim-2.5$ near the turn-off. Meanwhile, most of the $0.85~\mathrm{M}_{\sun}$ models have $[\mathrm{Fe/H}]<-3$ (without levitation) or $[\mathrm{Fe/H}]\gtrsim-1$ (with levitation). 

It is safer to first consider {[}C/H{]} because the largest {[}C/H{]} values should be close to 0.5~dex, independent of metallicity. In the metallicity range typical of CEMP-\emph{s} stars ($-2.5\leq\text{[Fe/H]}\leq-2.0$), this is indeed the case for subgiants ($\log g\lesssim3.7$) but very few turn-off stars ($\log g\approx4$) seem to have $\text{[C/H]}>0$ (Figs.~\ref{fig:ch_gsra_0.8} and \ref{fig:ch_gsra_0.85}). Is this difference in the maximum {[}C/H{]} values between the two groups in the observational data ($\Delta\text{[C/H]}\approx0.5$) evidence of gravitational settling of carbon in CEMP stars? And does the similar 0.5~dex difference seen in the {[}C/Fe{]} data (Figs.~\ref{fig:cfe_gsra_0.8} and \ref{fig:cfe_gsra_0.85}) then imply that iron is levitated just enough that its abundance stays roughly constant throughout the evolution? This seems rather unlikely because then the carbon-normal population should plausibly also have lower {[}C/H{]} (and {[}C/Fe{]}) values near the turn-off which is not the case. On the contrary, the observations suggest that the carbon abundance in carbon-normal metal-poor stars is increasing on the main sequence until they reach the turn-off. This is exactly the opposite behaviour that atomic diffusion predicts! Moreover, there is no obvious candidate for a physical process that could cause the surface carbon abundance to increase on the main sequence.

The carbon-normal metal-poor stars listed in the Stellar Abundances for Galactic Archeology database \citep[SAGA;][]{2008PASJ...60.1159S,2011MNRAS.412..843S} do not seem to show a similar behaviour of increasing carbon abundance on the main sequence (C.~Abate, priv. comm.), although the relatively small number statistics (23 stars with $-2.5\leq\text{[Fe/H]}\leq-2.0$, $\text{[C/Fe]}<1$ and $\log g>4.0$) and the heterogeneity of the data could perhaps hide such a trend. Unfortunately, whether there is a real difference in the upper {[}C/H{]} and {[}C/Fe{]} values between turn-off stars and subgiants remains unclear.

Most of the models are at odds with the \citet{2013AJ....146..132L} data, regardless of whether the abundance differences between turn-off stars and subgiants are caused by atomic diffusion. For example, while some of the $0.8~\mathrm{M}_{\sun}$ models predict abundances that are consistent with the observations, they only do so at very late times, i.e. at ages exceeding the age of the Universe (13.8~Gyr; \citealt{2013ApJS..208...19H}). At earlier times stars with low-mass AGB companions (which thus accreted mass later) and/or with low initial masses (which were thus less evolved at the point of mass transfer) are still relatively unevolved. They should be observable as carbon-rich low-luminosity ($\log g\gtrsim4.1$) objects. But such objects are conspicuous by their absence in the \citet{2013AJ....146..132L} results (at all metallicities; their figure~6). Since there are plenty of carbon-normal low-luminosity stars in the data, it is difficult to imagine how this could be a selection effect, particularly since a few carbon-rich dwarfs have been found in the SDSS data in detailed abundance studies \citep{2008ApJ...678.1351A,2010A&A...513A..72B}.\footnote{Two examples are the CEMP-\emph{no} (or -\emph{r}) star SDSS0036-10 ($\text{[Fe/H]}=-2.4$, $\log g=4.5$, $\text{[C/Fe]}=2.3$, $\text{[Ba/Fe]}=0.3$) and the CEMP-\emph{s} (or -\emph{r}/\emph{s}) star SDSS2047+10 ($\text{[Fe/H]}=-2.1$, $\log g=4.5$, $\text{[C/Fe]}=2.0$, $\text{[Ba/Fe]}=1.5$) from \citet{2008ApJ...678.1351A}. \citet{2010A&A...513A..72B} present the CEMP-\emph{r}/\emph{s} star SDSSJ0912+0216 ($\text{[Fe/H]}=-2.5$, $\log g=4.5$, $\text{[C/Fe]}\approx1.5$, $\text{[Ba/Fe]}=1.5$, $\text{[Eu/Fe]}=1.2$), which may, however, be more evolved with $\log g\approx4.0$ \citep{2013AJ....145...13A}.} By contrast, the $0.85~\mathrm{M}_{\sun}$ CEMP-\emph{s} models are sufficiently evolved but diffusion is so efficient that unrealistic abundances (e.g. $[\mathrm{C/Fe}]<-0.5$ with levitation or $[\mathrm{C/Fe}]>3.5$ without levitation) are predicted in nearly all such stars around the turn-off (Fig.~\ref{fig:cfe_gsra_0.85}). Clearly, in these more massive CEMP-\emph{s} stars some physical process must be countering atomic diffusion, at least near their surface.

The disagreement between observations and models concerning the existence of low-luminosity carbon-rich stars has little to do with atomic diffusion -- even if we identify a process that neatly counteracts diffusion near the surface, models will still predict many carbon-rich unevolved objects. In fact, if this process were to counteract diffusion throughout the star, the tension with observations would increase because diffusion starves the core of fuel and accelerates the evolution, making the star spend less time on the main sequence. Perhaps, the SDSS sample indicates that the mass ratio ($q\equiv M_{2,\mathrm{i}}/M_{1}$) in CEMP-\emph{s} progenitor systems is biased towards unity, contrary to the common assumption of a flat distribution. If $q$ is close to one, the mass transfer occurs relatively late, giving the secondary more time to evolve before it becomes a CEMP-\emph{s} star.

\begin{figure}
\includegraphics[width=1\columnwidth]{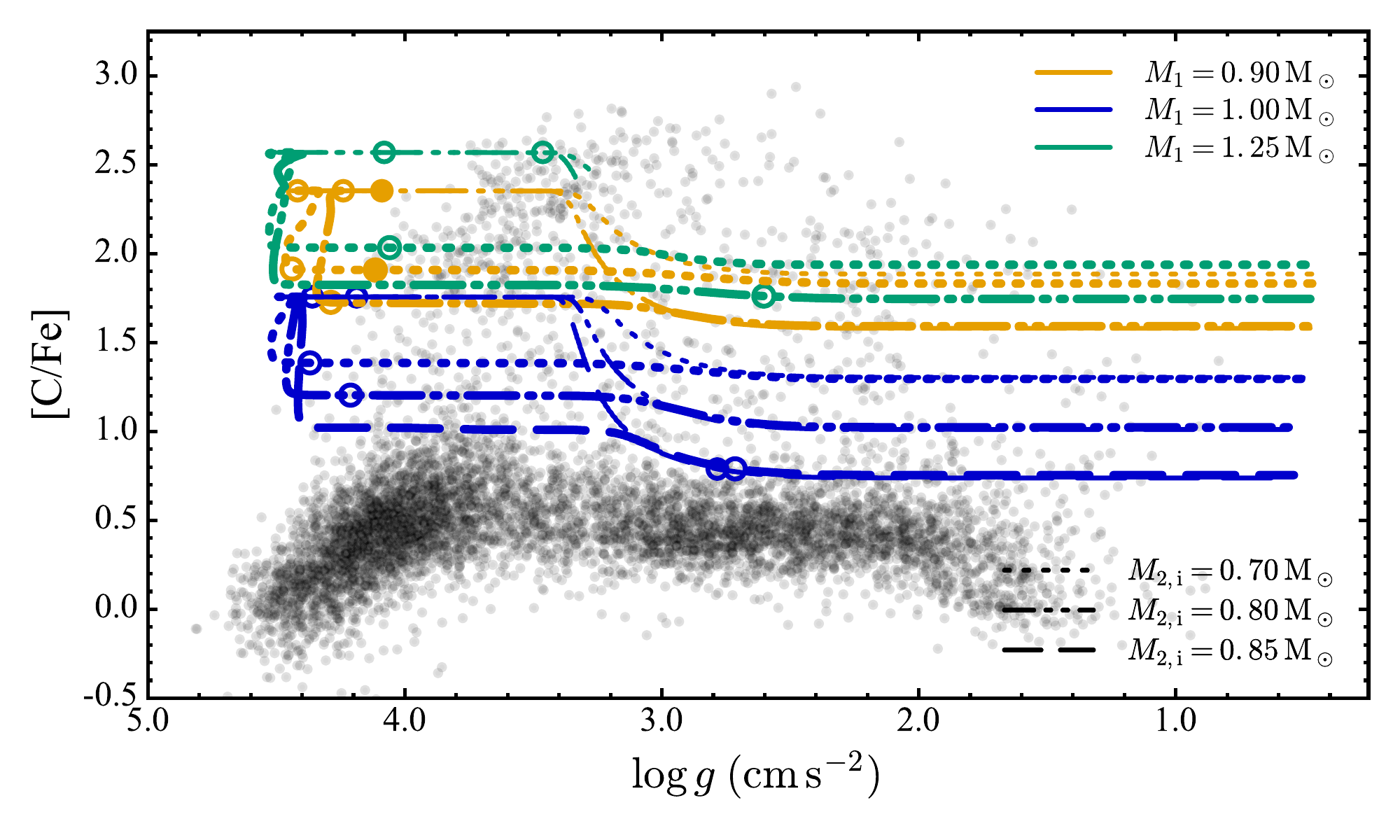}

\caption{Similar to Figs.~\ref{fig:cfe_gsra_0.8} and \ref{fig:cfe_gsra_0.85} but here the lines correspond to CEMP-\emph{s} models of $0.9~\mathrm{M}_{\sun}$ without diffusion. Thick lines are models with thermohaline mixing, whereas thin lines are those without.\label{fig:cfe_notm_0.9_0.95}}
\end{figure}

Overall, models without diffusion are better able to envelop the observational data (Fig.~\ref{fig:cfe_notm_0.9_0.95}). For example, CEMP-\emph{s} models that have accreted less carbon-rich material (here coming from $1~\mathrm{M}_{\sun}$ primaries) have $[\mathrm{C/Fe}]<1.5$ with {[}C/Fe{]} increasing with accreted mass. Models that have accreted more carbon-rich material (from primaries with masses $1.25~\mathrm{M}_{\sun}$, $1.5~\mathrm{M}_{\sun}$) can have $[\mathrm{C/Fe}]\gtrsim2$ throughout evolution. Models with thermohaline mixing seem to be preferred because no sharp change in {[}C/Fe{]} at FDU is evident from the data. Lower-mass models ($M_{2,\text{f}}\approx0.8~\mathrm{M}_{\sun}$) without diffusion would predict similar abundance evolution \citep{2007A&A...464L..57S,2008MNRAS.389.1828S}. However, most of them, coming from systems with relatively low-mass AGB companions compared to earlier works, would be consistent with the data only for $t>13.8~\text{Gyr}$.

\begin{longtab} 
\begin{longtable}{ll>{\raggedright}p{1.2cm}l>{\raggedright}p{1.4cm}lllllll>{\raggedright}p{1.4cm}}
\caption{Results from simulations including atomic diffusion. The columns list
the initial mass of the secondary ($M_{2,\text{i}}$); accreted mass
($\Delta M$); whether levitation was included; the deepest mass coordinate
reached by \foreignlanguage{british}{thermohaline} mixing ($m_{\text{thm}}$);
{[}C/Fe{]} after \foreignlanguage{british}{thermohaline} mixing ends;
the age ($t$), luminosity ($L$), effective temperature ($T_{\text{eff}}$),
surface gravity ($g$), envelope mass ($M_{\text{env}}$), metallicity
({[}Fe/H{]}), and {[}C/Fe{]} when the envelope mass reaches a minimum;
{[}C/Fe{]} after first dredge-up. The
table is sectioned according to the initial primary mass, $M_{1}$.
\label{tab:Results_main}} \\
\hline
\hline
\multirow{2}{*}{{\tiny{}$M_{2,\mathrm{i}}$}} & \multirow{2}{*}{{\tiny{}$\Delta M$}} & \multirow{2}{1.2cm}{{\tiny{}Levitation?}} & \multirow{2}{*}{{\tiny{}$m_{\mathrm{\text{thm}}}$}} & \multirow{2}{1.6cm}{{\tiny{}{[}C/Fe{]} post-th.mix.}} & \multicolumn{7}{l}{{\tiny{}At the time when envelope mass is smallest}} & \multirow{2}{1.4cm}{{\tiny{}$\text{[C/Fe]}$\tablefootmark{b} post-FDU}}\tabularnewline 
\cline{6-12}   
 &  &  &  &  & {\tiny{}$t$ (Gyr)} & {\tiny{}$\log(L/\mathrm{L}_{\sun})$} & {\tiny{}$T_{\mathrm{eff}}$} & {\tiny{}$\log g$} & {\tiny{}$M_{\mathrm{env}}$} & {\tiny{}$\text{[Fe/H]}$\tablefootmark{a}} & {\tiny{}$\text{[C/Fe]}$\tablefootmark{a}} & \tabularnewline
\hline
\endfirsthead 
\caption{continued.}\\ 
\hline
\hline
\multirow{2}{*}{{\tiny{}$M_{2,\mathrm{i}}$}} & \multirow{2}{*}{{\tiny{}$\Delta M$}} & \multirow{2}{1.2cm}{{\tiny{}Levitation?}} & \multirow{2}{*}{{\tiny{}$m_{\mathrm{\text{thm}}}$}} & \multirow{2}{1.6cm}{{\tiny{}{[}C/Fe{]} post-th.mix.}} & \multicolumn{7}{l}{{\tiny{}At the time when envelope mass is smallest}} & \multirow{2}{1.4cm}{{\tiny{}$\text{[C/Fe]}$\tablefootmark{b} post-FDU}}\tabularnewline 
\cline{6-12}  
 &  &  &  &  & {\tiny{}$t$ (Gyr)} & {\tiny{}$\log(L/\mathrm{L}_{\sun})$} & {\tiny{}$T_{\mathrm{eff}}$} & {\tiny{}$\log g$} & {\tiny{}$M_{\mathrm{env}}$} & {\tiny{}$\text{[Fe/H]}$\tablefootmark{a}} & {\tiny{}$\text{[C/Fe]}$\tablefootmark{a}} & \tabularnewline
\hline
\endhead 
\hline
\endfoot 
\multicolumn{13}{c}{{\tiny{}$M_{1}=0.9\thinspace\mathrm{M}_{\sun}$}}\tabularnewline
{\tiny{}$0.700$} & {\tiny{}$0.100$} & {\tiny{}no } & {\tiny{}$0.411$} & {\tiny{}$ 1.88$} & {\tiny{}$15.84$}       & {\tiny{}$ 0.4461$} & {\tiny{}$6366$} & {\tiny{}$4.06$} & {\tiny{}$2.79(-5)$} & {\tiny{}$-2.54$} & {\tiny{}$ 1.82$} & {\tiny{}  \ldots   }  \tabularnewline
{\tiny{}       } & {\tiny{}       } & {\tiny{}yes} & {\tiny{}$0.410$} & {\tiny{}$ 1.88$} & {\tiny{}$15.83$}       & {\tiny{}$ 0.4462$} & {\tiny{}$6365$} & {\tiny{}$4.06$} & {\tiny{}$2.90(-5)$} & {\tiny{}$-2.25$} & {\tiny{}$ 1.56$} & {\tiny{}  \ldots   }  \tabularnewline
{\tiny{}$0.750$} & {\tiny{}$0.050$} & {\tiny{}no } & {\tiny{}$0.526$} & {\tiny{}$ 1.77$} & {\tiny{}$14.02$}       & {\tiny{}$ 0.4573$} & {\tiny{}$6400$} & {\tiny{}$4.06$} & {\tiny{}$1.59(-5)$} & {\tiny{}$-2.61$} & {\tiny{}$ 1.71$} & {\tiny{}  \ldots   }  \tabularnewline
{\tiny{}       } & {\tiny{}       } & {\tiny{}yes} & {\tiny{}$0.525$} & {\tiny{}$ 1.78$} & {\tiny{}$14.00$}       & {\tiny{}$ 0.4544$} & {\tiny{}$6402$} & {\tiny{}$4.06$} & {\tiny{}$1.69(-5)$} & {\tiny{}$-2.13$} & {\tiny{}$ 1.27$} & {\tiny{}$ 1.38$}  \tabularnewline
{\tiny{}$0.800$} & {\tiny{}$0.001$} & {\tiny{}no } & {\tiny{}$0.797$} & {\tiny{}$ 1.85$} & {\tiny{}$11.94$}       & {\tiny{}$ 0.4986$} & {\tiny{}$6471$} & {\tiny{}$4.04$} & {\tiny{}$3.36(-6)$} & {\tiny{}$-3.05$} & {\tiny{}$ 1.86$} & {\tiny{}  \ldots   }  \tabularnewline
{\tiny{}       } & {\tiny{}       } & {\tiny{}yes} & {\tiny{}$0.797$} & {\tiny{}$ 1.83$} & {\tiny{}$11.93$}       & {\tiny{}$ 0.4959$} & {\tiny{}$6465$} & {\tiny{}$4.04$} & {\tiny{}$3.98(-6)$} & {\tiny{}$-1.60$} & {\tiny{}$ 0.46$} & {\tiny{}$ 0.15$}  \tabularnewline
{\tiny{}$0.800$\tablefootmark{\textasteriskcentered}} & {\tiny{}$0.010$} & {\tiny{}no } & {\tiny{}$0.760$} & {\tiny{}$ 1.76$} & {\tiny{}$11.79$}       & {\tiny{}$ 0.4922$} & {\tiny{}$6464$} & {\tiny{}$4.05$} & {\tiny{}$5.28(-6)$} & {\tiny{}$-2.82$} & {\tiny{}$ 1.73$} & {\tiny{}  \ldots   }  \tabularnewline
{\tiny{}       } & {\tiny{}       } & {\tiny{}yes} & {\tiny{}$0.760$} & {\tiny{}$ 1.75$} & {\tiny{}$11.80$}       & {\tiny{}$ 0.4930$} & {\tiny{}$6457$} & {\tiny{}$4.05$} & {\tiny{}$6.04(-6)$} & {\tiny{}$-1.73$} & {\tiny{}$ 0.67$} & {\tiny{}$ 0.72$}  \tabularnewline
{\tiny{}$0.750$} & {\tiny{}$0.100$} & {\tiny{}no } & {\tiny{}$0.462$} & {\tiny{}$ 1.91$} & {\tiny{}$13.24$}       & {\tiny{}$ 0.5482$} & {\tiny{}$6581$} & {\tiny{}$4.05$} & {\tiny{}$3.94(-7)$} & {\tiny{}$-5.32$} & {\tiny{}$ 2.72$} & {\tiny{}  \ldots   }  \tabularnewline
{\tiny{}       } & {\tiny{}       } & {\tiny{}yes} & {\tiny{}$0.461$} & {\tiny{}$ 1.91$} & {\tiny{}$13.25$}       & {\tiny{}$ 0.5490$} & {\tiny{}$6555$} & {\tiny{}$4.04$} & {\tiny{}$4.66(-7)$} & {\tiny{}$-0.63$} & {\tiny{}$-1.67$} & {\tiny{}  \ldots   }  \tabularnewline
{\tiny{}$0.800$} & {\tiny{}$0.050$} & {\tiny{}no } & {\tiny{}$0.576$} & {\tiny{}$ 1.84$} & {\tiny{}$11.50$}       & {\tiny{}$ 0.5567$} & {\tiny{}$6647$} & {\tiny{}$4.05$} & {\tiny{}$1.44(-7)$} & {\tiny{}$-7.71$} & {\tiny{}$ 3.95$} & {\tiny{}  \ldots   }  \tabularnewline
{\tiny{}       } & {\tiny{}       } & {\tiny{}yes} & {\tiny{}$0.578$} & {\tiny{}$ 1.76$} & {\tiny{}$11.48$}       & {\tiny{}$ 0.5522$} & {\tiny{}$6616$} & {\tiny{}$4.05$} & {\tiny{}$1.56(-7)$} & {\tiny{}$-0.35$} & {\tiny{}$-2.60$} & {\tiny{}$ 1.34$}  \tabularnewline
{\tiny{}$0.700$} & {\tiny{}$0.200$} & {\tiny{}no } & {\tiny{}$0.361$} & {\tiny{}$ 2.06$} & {\tiny{}$12.76$\tablefootmark{\dag}}  & {\tiny{}$ 0.4587$} & {\tiny{}$6798$} & {\tiny{}$4.22$} & {\tiny{}$5.97(-8)$} & {\tiny{}$-$inf } & {\tiny{}inf    } & {\tiny{}  \ldots   }  \tabularnewline
{\tiny{}       } & {\tiny{}       } & {\tiny{}yes} & {\tiny{}$0.360$} & {\tiny{}$ 2.05$} & {\tiny{}$12.99$\tablefootmark{\dag}}  & {\tiny{}$ 0.4907$} & {\tiny{}$6781$} & {\tiny{}$4.18$} & {\tiny{}$4.14(-8)$} & {\tiny{}$-0.44$} & {\tiny{}$-3.90$} & {\tiny{}  \ldots   }  \tabularnewline
{\tiny{}$0.800$} & {\tiny{}$0.100$} & {\tiny{}no } & {\tiny{}$0.515$} & {\tiny{}$ 2.06$} & {\tiny{}$10.13$\tablefootmark{\dag}}  & {\tiny{}$ 0.4557$} & {\tiny{}$6858$} & {\tiny{}$4.23$} & {\tiny{}$2.91(-8)$} & {\tiny{}$-$inf } & {\tiny{}inf    } & {\tiny{}  \ldots   }  \tabularnewline
{\tiny{}       } & {\tiny{}       } & {\tiny{}yes} & {\tiny{}$0.518$} & {\tiny{}$ 1.49$} & {\tiny{}$10.54$\tablefootmark{\dag}}  & {\tiny{}$ 0.5170$} & {\tiny{}$6874$} & {\tiny{}$4.18$} & {\tiny{}$1.03(-8)$} & {\tiny{}$-0.42$} & {\tiny{}$-3.17$} & {\tiny{}  \ldots   }  \tabularnewline
\multicolumn{13}{c}{{\tiny{}$M_{1}=1.0\thinspace\mathrm{M}_{\sun}$}}\tabularnewline
{\tiny{}$0.600$} & {\tiny{}$0.200$} & {\tiny{}no } & {\tiny{}$0.342$} & {\tiny{}$ 1.55$} & {\tiny{}$16.00$\tablefootmark{\ddag}} & {\tiny{}$ 0.4132$} & {\tiny{}$6495$} & {\tiny{}$4.13$} & {\tiny{}$4.03(-6)$} & {\tiny{}$-3.05$} & {\tiny{}$ 1.50$} & {\tiny{}  \ldots   }  \tabularnewline
{\tiny{}       } & {\tiny{}       } & {\tiny{}yes} & {\tiny{}$0.341$} & {\tiny{}$ 1.55$} & {\tiny{}$16.00$\tablefootmark{\ddag}} & {\tiny{}$ 0.4120$} & {\tiny{}$6486$} & {\tiny{}$4.13$} & {\tiny{}$4.80(-6)$} & {\tiny{}$-1.84$} & {\tiny{}$ 0.38$} & {\tiny{}  \ldots   }  \tabularnewline
{\tiny{}$0.700$} & {\tiny{}$0.100$} & {\tiny{}no } & {\tiny{}$0.477$} & {\tiny{}$ 1.44$} & {\tiny{}$14.66$}       & {\tiny{}$ 0.4815$} & {\tiny{}$6489$} & {\tiny{}$4.06$} & {\tiny{}$2.21(-6)$} & {\tiny{}$-3.44$} & {\tiny{}$ 1.48$} & {\tiny{}  \ldots   }  \tabularnewline
{\tiny{}       } & {\tiny{}       } & {\tiny{}yes} & {\tiny{}$0.479$} & {\tiny{}$ 1.44$} & {\tiny{}$14.67$}       & {\tiny{}$ 0.4854$} & {\tiny{}$6476$} & {\tiny{}$4.05$} & {\tiny{}$2.66(-6)$} & {\tiny{}$-1.48$} & {\tiny{}$-0.37$} & {\tiny{}$ 1.11$}  \tabularnewline
{\tiny{}$0.750$} & {\tiny{}$0.050$} & {\tiny{}no } & {\tiny{}$0.602$} & {\tiny{}$ 1.36$} & {\tiny{}$13.45$}       & {\tiny{}$ 0.4875$} & {\tiny{}$6498$} & {\tiny{}$4.06$} & {\tiny{}$1.82(-6)$} & {\tiny{}$-3.56$} & {\tiny{}$ 1.45$} & {\tiny{}  \ldots   }  \tabularnewline
{\tiny{}       } & {\tiny{}       } & {\tiny{}yes} & {\tiny{}$0.601$} & {\tiny{}$ 1.36$} & {\tiny{}$13.43$}       & {\tiny{}$ 0.4871$} & {\tiny{}$6488$} & {\tiny{}$4.06$} & {\tiny{}$2.16(-6)$} & {\tiny{}$-1.37$} & {\tiny{}$-0.62$} & {\tiny{}$ 0.83$}  \tabularnewline
{\tiny{}$0.800$} & {\tiny{}$0.001$} & {\tiny{}no } & {\tiny{}$0.797$} & {\tiny{}$ 1.28$} & {\tiny{}$11.93$}       & {\tiny{}$ 0.4992$} & {\tiny{}$6531$} & {\tiny{}$4.06$} & {\tiny{}$9.59(-7)$} & {\tiny{}$-4.11$} & {\tiny{}$ 1.56$} & {\tiny{}  \ldots   }  \tabularnewline
{\tiny{}       } & {\tiny{}       } & {\tiny{}yes} & {\tiny{}$0.797$} & {\tiny{}$ 1.28$} & {\tiny{}$11.92$}       & {\tiny{}$ 0.4987$} & {\tiny{}$6515$} & {\tiny{}$4.05$} & {\tiny{}$1.18(-6)$} & {\tiny{}$-1.10$} & {\tiny{}$-1.25$} & {\tiny{}$ 0.03$}  \tabularnewline
{\tiny{}$0.800$} & {\tiny{}$0.010$} & {\tiny{}no } & {\tiny{}$0.778$} & {\tiny{}$ 1.34$} & {\tiny{}$11.73$}       & {\tiny{}$ 0.5113$} & {\tiny{}$6558$} & {\tiny{}$4.06$} & {\tiny{}$5.99(-7)$} & {\tiny{}$-4.69$} & {\tiny{}$ 1.85$} & {\tiny{}  \ldots   }  \tabularnewline
{\tiny{}       } & {\tiny{}       } & {\tiny{}yes} & {\tiny{}$0.777$} & {\tiny{}$ 1.33$} & {\tiny{}$11.73$}       & {\tiny{}$ 0.5109$} & {\tiny{}$6540$} & {\tiny{}$4.05$} & {\tiny{}$7.05(-7)$} & {\tiny{}$-0.85$} & {\tiny{}$-1.74$} & {\tiny{}$ 0.31$}  \tabularnewline
{\tiny{}$0.650$} & {\tiny{}$0.200$} & {\tiny{}no } & {\tiny{}$0.380$} & {\tiny{}$ 1.56$} & {\tiny{}$13.16$\tablefootmark{\dag}}  & {\tiny{}$ 0.4455$} & {\tiny{}$6764$} & {\tiny{}$4.20$} & {\tiny{}$7.90(-8)$} & {\tiny{}$-$inf } & {\tiny{}inf    } & {\tiny{}  \ldots   }  \tabularnewline
{\tiny{}       } & {\tiny{}       } & {\tiny{}yes} & {\tiny{}$0.382$} & {\tiny{}$ 1.58$} & {\tiny{}$13.51$\tablefootmark{\dag}}  & {\tiny{}$ 0.5007$} & {\tiny{}$6759$} & {\tiny{}$4.14$} & {\tiny{}$3.76(-8)$} & {\tiny{}$-0.42$} & {\tiny{}$-4.47$} & {\tiny{}  \ldots   }  \tabularnewline
{\tiny{}$0.750$} & {\tiny{}$0.100$} & {\tiny{}no } & {\tiny{}$0.528$} & {\tiny{}$ 1.48$} & {\tiny{}$11.33$\tablefootmark{\dag}}  & {\tiny{}$ 0.4298$} & {\tiny{}$6776$} & {\tiny{}$4.21$} & {\tiny{}$7.68(-8)$} & {\tiny{}$-$inf } & {\tiny{}inf    } & {\tiny{}  \ldots   }  \tabularnewline
{\tiny{}       } & {\tiny{}       } & {\tiny{}yes} & {\tiny{}$0.528$} & {\tiny{}$ 1.48$} & {\tiny{}$11.78$\tablefootmark{\dag}}  & {\tiny{}$ 0.5005$} & {\tiny{}$6783$} & {\tiny{}$4.15$} & {\tiny{}$2.78(-8)$} & {\tiny{}$-0.42$} & {\tiny{}$-4.82$} & {\tiny{}  \ldots   }  \tabularnewline
{\tiny{}$0.800$\tablefootmark{\textasteriskcentered}} & {\tiny{}$0.050$} & {\tiny{}no } & {\tiny{}$0.655$} & {\tiny{}$ 1.42$} & {\tiny{}$10.32$\tablefootmark{\dag}}  & {\tiny{}$ 0.4584$} & {\tiny{}$6812$} & {\tiny{}$4.20$} & {\tiny{}$3.87(-8)$} & {\tiny{}$-$inf } & {\tiny{}inf    } & {\tiny{}  \ldots   }  \tabularnewline
{\tiny{}       } & {\tiny{}       } & {\tiny{}yes} & {\tiny{}$0.654$} & {\tiny{}$ 1.39$} & {\tiny{}$10.31$\tablefootmark{\dag}}  & {\tiny{}$ 0.4566$} & {\tiny{}$6781$} & {\tiny{}$4.19$} & {\tiny{}$4.25(-8)$} & {\tiny{}$-0.44$} & {\tiny{}$-4.46$} & {\tiny{}  \ldots   }  \tabularnewline
{\tiny{}$0.700$} & {\tiny{}$0.200$} & {\tiny{}no } & {\tiny{}$0.427$} & {\tiny{}$ 1.62$} & {\tiny{}$ 9.88$\tablefootmark{\dag}}  & {\tiny{}$ 0.3482$} & {\tiny{}$6879$} & {\tiny{}$4.35$} & {\tiny{}$5.60(-8)$} & {\tiny{}$-$inf } & {\tiny{}inf    } & {\tiny{}  \ldots   }  \tabularnewline
{\tiny{}       } & {\tiny{}       } & {\tiny{}yes} & {\tiny{}$0.428$} & {\tiny{}$ 1.58$} & {\tiny{}$10.35$\tablefootmark{\dag}}  & {\tiny{}$ 0.3998$} & {\tiny{}$6918$} & {\tiny{}$4.31$} & {\tiny{}$1.68(-8)$} & {\tiny{}$-0.56$} & {\tiny{}$-7.12$} & {\tiny{}  \ldots   }  \tabularnewline
\multicolumn{13}{c}{{\tiny{}$M_{1}=1.25\thinspace\mathrm{M}_{\sun}$}}\tabularnewline
{\tiny{}$0.600$} & {\tiny{}$0.200$} & {\tiny{}no } & {\tiny{}$0.158$} & {\tiny{}$ 2.21$} & {\tiny{}$14.35$}       & {\tiny{}$ 0.4497$} & {\tiny{}$6302$} & {\tiny{}$4.04$} & {\tiny{}$6.44(-5)$} & {\tiny{}$-2.46$} & {\tiny{}$ 2.15$} & {\tiny{}$ 2.04$}  \tabularnewline
{\tiny{}       } & {\tiny{}       } & {\tiny{}yes} & {\tiny{}$0.156$} & {\tiny{}$ 2.21$} & {\tiny{}$14.34$}       & {\tiny{}$ 0.4487$} & {\tiny{}$6302$} & {\tiny{}$4.04$} & {\tiny{}$6.63(-5)$} & {\tiny{}$-2.34$} & {\tiny{}$ 2.04$} & {\tiny{}$ 2.04$}  \tabularnewline
{\tiny{}$0.700$\tablefootmark{\textasteriskcentered}} & {\tiny{}$0.100$} & {\tiny{}no } & {\tiny{}$0.263$} & {\tiny{}$ 1.94$} & {\tiny{}$13.31$}       & {\tiny{}$ 0.4596$} & {\tiny{}$6358$} & {\tiny{}$4.05$} & {\tiny{}$2.42(-5)$} & {\tiny{}$-2.60$} & {\tiny{}$ 1.87$} & {\tiny{}$ 1.80$}  \tabularnewline
{\tiny{}       } & {\tiny{}       } & {\tiny{}yes} & {\tiny{}$0.264$} & {\tiny{}$ 1.95$} & {\tiny{}$13.28$}       & {\tiny{}$ 0.4534$} & {\tiny{}$6359$} & {\tiny{}$4.05$} & {\tiny{}$2.59(-5)$} & {\tiny{}$-2.26$} & {\tiny{}$ 1.57$} & {\tiny{}$ 1.80$}  \tabularnewline
{\tiny{}$0.750$} & {\tiny{}$0.050$} & {\tiny{}no } & {\tiny{}$0.370$} & {\tiny{}$ 1.78$} & {\tiny{}$12.68$}       & {\tiny{}$ 0.4628$} & {\tiny{}$6400$} & {\tiny{}$4.06$} & {\tiny{}$1.19(-5)$} & {\tiny{}$-2.74$} & {\tiny{}$ 1.71$} & {\tiny{}  \ldots   }  \tabularnewline
{\tiny{}       } & {\tiny{}       } & {\tiny{}yes} & {\tiny{}$0.370$} & {\tiny{}$ 1.78$} & {\tiny{}$12.66$}       & {\tiny{}$ 0.4598$} & {\tiny{}$6397$} & {\tiny{}$4.06$} & {\tiny{}$1.32(-5)$} & {\tiny{}$-2.11$} & {\tiny{}$ 1.12$} & {\tiny{}$ 1.58$}  \tabularnewline
{\tiny{}$0.800$} & {\tiny{}$0.001$} & {\tiny{}no } & {\tiny{}$0.796$} & {\tiny{}$ 1.90$} & {\tiny{}$11.90$}       & {\tiny{}$ 0.4925$} & {\tiny{}$6457$} & {\tiny{}$4.04$} & {\tiny{}$3.56(-6)$} & {\tiny{}$-3.15$} & {\tiny{}$ 1.88$} & {\tiny{}  \ldots   }  \tabularnewline
{\tiny{}       } & {\tiny{}       } & {\tiny{}yes} & {\tiny{}$0.796$} & {\tiny{}$ 1.90$} & {\tiny{}$11.89$}       & {\tiny{}$ 0.4914$} & {\tiny{}$6449$} & {\tiny{}$4.04$} & {\tiny{}$4.40(-6)$} & {\tiny{}$-1.69$} & {\tiny{}$ 0.50$} & {\tiny{}  \ldots   }  \tabularnewline
{\tiny{}$0.800$} & {\tiny{}$0.010$} & {\tiny{}no } & {\tiny{}$0.600$} & {\tiny{}$ 1.52$} & {\tiny{}$11.58$}       & {\tiny{}$ 0.5006$} & {\tiny{}$6507$} & {\tiny{}$4.05$} & {\tiny{}$1.46(-6)$} & {\tiny{}$-3.74$} & {\tiny{}$ 1.64$} & {\tiny{}  \ldots   }  \tabularnewline
{\tiny{}       } & {\tiny{}       } & {\tiny{}yes} & {\tiny{}$0.600$} & {\tiny{}$ 1.51$} & {\tiny{}$11.57$}       & {\tiny{}$ 0.4979$} & {\tiny{}$6495$} & {\tiny{}$4.05$} & {\tiny{}$1.88(-6)$} & {\tiny{}$-1.30$} & {\tiny{}$-0.62$} & {\tiny{}$ 0.92$}  \tabularnewline
{\tiny{}$0.650$} & {\tiny{}$0.200$} & {\tiny{}no } & {\tiny{}$0.149$} & {\tiny{}$ 2.12$} & {\tiny{}$11.65$}       & {\tiny{}$ 0.5482$} & {\tiny{}$6513$} & {\tiny{}$4.03$} & {\tiny{}$1.13(-6)$} & {\tiny{}$-3.96$} & {\tiny{}$ 2.32$} & {\tiny{}  \ldots   }  \tabularnewline
{\tiny{}       } & {\tiny{}       } & {\tiny{}yes} & {\tiny{}$0.150$} & {\tiny{}$ 2.12$} & {\tiny{}$11.62$}       & {\tiny{}$ 0.5426$} & {\tiny{}$6501$} & {\tiny{}$4.03$} & {\tiny{}$1.47(-6)$} & {\tiny{}$-1.17$} & {\tiny{}$-0.29$} & {\tiny{}  \ldots   }  \tabularnewline
{\tiny{}$0.750$} & {\tiny{}$0.100$} & {\tiny{}no } & {\tiny{}$0.309$} & {\tiny{}$ 1.96$} & {\tiny{}$10.90$}       & {\tiny{}$ 0.5598$} & {\tiny{}$6589$} & {\tiny{}$4.04$} & {\tiny{}$3.02(-7)$} & {\tiny{}$-6.13$} & {\tiny{}$ 3.10$} & {\tiny{}  \ldots   }  \tabularnewline
{\tiny{}       } & {\tiny{}       } & {\tiny{}yes} & {\tiny{}$0.308$} & {\tiny{}$ 1.97$} & {\tiny{}$10.88$}       & {\tiny{}$ 0.5578$} & {\tiny{}$6561$} & {\tiny{}$4.03$} & {\tiny{}$3.60(-7)$} & {\tiny{}$-0.56$} & {\tiny{}$-1.98$} & {\tiny{}$ 1.78$}  \tabularnewline
{\tiny{}$0.800$} & {\tiny{}$0.050$} & {\tiny{}no } & {\tiny{}$0.419$} & {\tiny{}$ 1.81$} & {\tiny{}$10.28$}       & {\tiny{}$ 0.5552$} & {\tiny{}$6662$} & {\tiny{}$4.06$} & {\tiny{}$1.15(-7)$} & {\tiny{}$-$inf } & {\tiny{}inf    } & {\tiny{}  \ldots   }  \tabularnewline
{\tiny{}       } & {\tiny{}       } & {\tiny{}yes} & {\tiny{}$0.421$} & {\tiny{}$ 1.81$} & {\tiny{}$10.30$}       & {\tiny{}$ 0.5597$} & {\tiny{}$6625$} & {\tiny{}$4.05$} & {\tiny{}$1.27(-7)$} & {\tiny{}$-0.36$} & {\tiny{}$-3.19$} & {\tiny{}  \ldots   }  \tabularnewline
\multicolumn{13}{c}{{\tiny{}$M_{1}=1.5\thinspace\mathrm{M}_{\sun}$}}\tabularnewline
{\tiny{}$0.600$} & {\tiny{}$0.200$} & {\tiny{}no } & {\tiny{}$0.000$} & {\tiny{}$ 2.37$} & {\tiny{}$12.70$}       & {\tiny{}$ 0.2746$} & {\tiny{}$6166$} & {\tiny{}$4.18$} & {\tiny{}$1.12(-3)$} & {\tiny{}$-2.29$} & {\tiny{}$ 2.35$} & {\tiny{}$ 2.28$}  \tabularnewline
{\tiny{}       } & {\tiny{}       } & {\tiny{}yes} & {\tiny{}$0.000$} & {\tiny{}$ 2.40$} & {\tiny{}$12.72$}       & {\tiny{}$ 0.2751$} & {\tiny{}$6166$} & {\tiny{}$4.18$} & {\tiny{}$1.13(-3)$} & {\tiny{}$-2.28$} & {\tiny{}$ 2.34$} & {\tiny{}$ 2.28$}  \tabularnewline
{\tiny{}$0.800$} & {\tiny{}$0.010$} & {\tiny{}no } & {\tiny{}$0.495$} & {\tiny{}$ 1.67$} & {\tiny{}$11.56$}       & {\tiny{}$ 0.4913$} & {\tiny{}$6449$} & {\tiny{}$4.05$} & {\tiny{}$4.15(-6)$} & {\tiny{}$-3.11$} & {\tiny{}$ 1.64$} & {\tiny{}$ 1.22$}  \tabularnewline
{\tiny{}       } & {\tiny{}       } & {\tiny{}yes} & {\tiny{}$0.493$} & {\tiny{}$ 1.67$} & {\tiny{}$11.54$}       & {\tiny{}$ 0.4886$} & {\tiny{}$6445$} & {\tiny{}$4.05$} & {\tiny{}$5.01(-6)$} & {\tiny{}$-1.73$} & {\tiny{}$ 0.35$} & {\tiny{}$ 1.22$}  \tabularnewline
{\tiny{}$0.650$} & {\tiny{}$0.200$} & {\tiny{}no } & {\tiny{}$0.000$} & {\tiny{}$ 2.37$} & {\tiny{}$10.44$}       & {\tiny{}$ 0.3817$} & {\tiny{}$6307$} & {\tiny{}$4.14$} & {\tiny{}$1.30(-4)$} & {\tiny{}$-2.41$} & {\tiny{}$ 2.32$} & {\tiny{}$ 2.25$}  \tabularnewline
{\tiny{}       } & {\tiny{}       } & {\tiny{}yes} & {\tiny{}$0.000$} & {\tiny{}$ 2.38$} & {\tiny{}$10.42$}       & {\tiny{}$ 0.3804$} & {\tiny{}$6305$} & {\tiny{}$4.14$} & {\tiny{}$1.34(-4)$} & {\tiny{}$-2.35$} & {\tiny{}$ 2.28$} & {\tiny{}$ 2.25$}  \tabularnewline
\multicolumn{13}{c}{{\tiny{}no accretion}}\tabularnewline
{\tiny{}$0.750$} & {\tiny{}$0.000$} & {\tiny{}no } & {\tiny{}$0.000$} & {\tiny{}$ 0.00$} & {\tiny{}$15.06$}       & {\tiny{}$ 0.3914$} & {\tiny{}$6345$} & {\tiny{}$4.08$} & {\tiny{}$2.59(-5)$} & {\tiny{}$-2.63$} & {\tiny{}$-0.09$} & {\tiny{}  \ldots   }  \tabularnewline
{\tiny{}       } & {\tiny{}       } & {\tiny{}yes} & {\tiny{}$0.000$} & {\tiny{}$ 0.00$} & {\tiny{}$15.06$}       & {\tiny{}$ 0.3901$} & {\tiny{}$6345$} & {\tiny{}$4.09$} & {\tiny{}$2.75(-5)$} & {\tiny{}$-2.33$} & {\tiny{}$-0.24$} & {\tiny{}  \ldots   }  \tabularnewline
{\tiny{}$0.800$} & {\tiny{}$0.000$} & {\tiny{}no } & {\tiny{}$0.000$} & {\tiny{}$ 0.00$} & {\tiny{}$11.93$}       & {\tiny{}$ 0.4951$} & {\tiny{}$6580$} & {\tiny{}$4.07$} & {\tiny{}$4.48(-7)$} & {\tiny{}$-5.38$} & {\tiny{}$ 0.73$} & {\tiny{}  \ldots   }  \tabularnewline
{\tiny{}       } & {\tiny{}       } & {\tiny{}yes} & {\tiny{}$0.000$} & {\tiny{}$ 0.00$} & {\tiny{}$11.94$}       & {\tiny{}$ 0.4961$} & {\tiny{}$6558$} & {\tiny{}$4.07$} & {\tiny{}$5.27(-7)$} & {\tiny{}$-0.80$} & {\tiny{}$-3.33$} & {\tiny{}$-0.02$}  \tabularnewline
{\tiny{}$0.850$} & {\tiny{}$0.000$} & {\tiny{}no } & {\tiny{}$0.000$} & {\tiny{}$ 0.00$} & {\tiny{}$ 8.35$\tablefootmark{\dag}}  & {\tiny{}$ 0.3818$} & {\tiny{}$6819$} & {\tiny{}$4.27$} & {\tiny{}$6.75(-8)$} & {\tiny{}$-$inf } & {\tiny{}inf    } & {\tiny{}  \ldots   }  \tabularnewline
{\tiny{}       } & {\tiny{}       } & {\tiny{}yes} & {\tiny{}$0.000$} & {\tiny{}$ 0.00$} & {\tiny{}$ 8.51$\tablefootmark{\dag}}  & {\tiny{}$ 0.4027$} & {\tiny{}$6814$} & {\tiny{}$4.25$} & {\tiny{}$5.28(-8)$} & {\tiny{}$-1.07$} & {\tiny{}$-4.25$} & {\tiny{}  \ldots   }  \tabularnewline
\end{longtable} 
\tablefoot{
All masses are in solar masses; other quantities are in cgs units unless indicated otherwise. Values of $M_\text{env}$ are given in the format $n(m)=n\times10^m$ for concision.
\tablefoottext{a} {An `inf' indicates that one of the mass fractions is below $10^{-12}$.}
\tablefoottext{b} {Most of the models stop earlier.}
\tablefoottext{\textasteriskcentered} {Systems (with levitation) used in resolution tests (see Section~\ref{subsec:uncertainties}).}
\tablefoottext{\dag} {Models stop before reaching the minimum of $M_{\text{env}}$. The listed values are from the last converged model.}
\tablefoottext{\ddag} {Models reach $t=16\thinspace\text{Gyr}$ before reaching the minimum of $M_{\text{env}}$. The listed values are for the final model.}
}
\end{longtab}

\begin{longtab} 
\begin{longtable}{ll>{\raggedright}p{1cm}l>{\raggedright}p{1.4cm}lllllll>{\raggedright}p{1.4cm}}
\caption{Results from simulations without atomic diffusion. Columns have the
same meaning as in Table~\ref{tab:Results_main} except for the third
column which here indicates whether \foreignlanguage{british}{thermohaline}
mixing is included.\label{tab:Results_massive}} \\
\hline
\hline
\multirow{2}{*}{{\tiny{}$M_{2,\mathrm{i}}$}} & \multirow{2}{*}{{\tiny{}$\Delta M$}} & \multirow{2}{1cm}{\foreignlanguage{british}{{\tiny{}Th.}\foreignlanguage{english}{{\tiny{} mixing?}}}} & \multirow{2}{*}{{\tiny{}$m_{\text{thm}}$\tablefootmark{a}}} & \multirow{2}{1.6cm}{{\tiny{}$\text{[C/Fe]}$\tablefootmark{a} post-th.mix.}} & \multicolumn{7}{l}{{\tiny{}At the time when envelope mass is smallest}} & \multirow{2}{1.4cm}{{\tiny{}$\text{[C/Fe]}$\tablefootmark{b} post-FDU}}\tabularnewline
\cline{6-12} 
 &  &  &  &  & {\tiny{}$t$ (Gyr)} & {\tiny{}$\log(L/\mathrm{L}_{\sun})$} & {\tiny{}$T_{\mathrm{eff}}$} & {\tiny{}$\log g$} & {\tiny{}$M_{\mathrm{env}}$} & {\tiny{}{[}Fe/H{]}} & {\tiny{}{[}C/Fe{]}} & \tabularnewline
\hline 
\endfirsthead 
\caption{continued.}\\ 
\hline
\hline
\multirow{2}{*}{{\tiny{}$M_{2,\mathrm{i}}$}} & \multirow{2}{*}{{\tiny{}$\Delta M$}} & \multirow{2}{1cm}{\foreignlanguage{british}{{\tiny{}Th.}\foreignlanguage{english}{{\tiny{} mixing?}}}} & \multirow{2}{*}{{\tiny{}$m_{\text{thm}}$\tablefootmark{a}}} & \multirow{2}{1.6cm}{{\tiny{}$\text{[C/Fe]}$\tablefootmark{a} post-th.mix.}} & \multicolumn{7}{l}{{\tiny{}At the time when envelope mass is smallest}} & \multirow{2}{1.4cm}{{\tiny{}$\text{[C/Fe]}$\tablefootmark{b} post-FDU}}\tabularnewline
\cline{6-12} 
 &  &  &  &  & {\tiny{}$t$ (Gyr)} & {\tiny{}$\log(L/\mathrm{L}_{\sun})$} & {\tiny{}$T_{\mathrm{eff}}$} & {\tiny{}$\log g$} & {\tiny{}$M_{\mathrm{env}}$} & {\tiny{}{[}Fe/H{]}} & {\tiny{}{[}C/Fe{]}} & \tabularnewline
\hline
\endhead 
\hline
\endfoot 
\multicolumn{13}{c}{{\tiny{}$M_{1}=0.9\thinspace\mathrm{M}_{\sun}$}}\tabularnewline
{\tiny{}$0.700$} & {\tiny{}$0.200$} & {\tiny{}no } & {\tiny{}  \ldots   } & {\tiny{}  \ldots   } & {\tiny{}$13.98$}       & {\tiny{}$ 0.6032$} & {\tiny{}$6719$} & {\tiny{}$4.05$} & {\tiny{}$1.59(-6)$} & {\tiny{}$-2.16$} & {\tiny{}$ 2.35$} & {\tiny{}$ 1.89$}  \tabularnewline
{\tiny{}       } & {\tiny{}       } & {\tiny{}yes} & {\tiny{}$0.329$} & {\tiny{}$ 1.91$} & {\tiny{}$13.93$}       & {\tiny{}$ 0.6405$} & {\tiny{}$6990$} & {\tiny{}$4.08$} & {\tiny{}$1.37(-8)$} & {\tiny{}$-2.14$} & {\tiny{}$ 1.91$} & {\tiny{}$ 1.83$}  \tabularnewline
{\tiny{}$0.800$} & {\tiny{}$0.100$} & {\tiny{}no } & {\tiny{}  \ldots   } & {\tiny{}  \ldots   } & {\tiny{}$11.29$}       & {\tiny{}$ 0.6136$} & {\tiny{}$6734$} & {\tiny{}$4.05$} & {\tiny{}$1.15(-6)$} & {\tiny{}$-2.16$} & {\tiny{}$ 2.35$} & {\tiny{}$ 1.58$}  \tabularnewline
{\tiny{}       } & {\tiny{}       } & {\tiny{}yes} & {\tiny{}$0.461$} & {\tiny{}$ 1.72$} & {\tiny{}$11.23$}       & {\tiny{}$ 0.6221$} & {\tiny{}$7079$} & {\tiny{}$4.12$} & {\tiny{}$6.39(-9)$} & {\tiny{}$-2.14$} & {\tiny{}$ 1.72$} & {\tiny{}$ 1.59$}  \tabularnewline
{\tiny{}$0.750$} & {\tiny{}$0.200$} & {\tiny{}no } & {\tiny{}  \ldots   } & {\tiny{}  \ldots   } & {\tiny{}$12.20$}       & {\tiny{}$ 0.6829$} & {\tiny{}$7019$} & {\tiny{}$4.07$} & {\tiny{}$1.07(-8)$} & {\tiny{}$-2.16$} & {\tiny{}$ 2.35$} & {\tiny{}$ 1.85$}  \tabularnewline
{\tiny{}       } & {\tiny{}       } & {\tiny{}yes} & {\tiny{}$0.372$} & {\tiny{}$ 1.91$} & {\tiny{}$12.05$}       & {\tiny{}$ 0.6799$} & {\tiny{}$7359$} & {\tiny{}$4.16$} & {\tiny{}$2.84(-10)$} & {\tiny{}$-2.14$} & {\tiny{}$ 1.91$} & {\tiny{}$ 1.82$}  \tabularnewline
\multicolumn{13}{c}{{\tiny{}$M_{1}=1.0\thinspace\mathrm{M}_{\sun}$}}\tabularnewline
{\tiny{}$0.650$} & {\tiny{}$0.200$} & {\tiny{}no } & {\tiny{}  \ldots   } & {\tiny{}  \ldots   } & {\tiny{}$14.18$}       & {\tiny{}$ 0.5809$} & {\tiny{}$6841$} & {\tiny{}$4.08$} & {\tiny{}$1.59(-7)$} & {\tiny{}$-2.17$} & {\tiny{}$ 1.76$} & {\tiny{}$ 1.34$}  \tabularnewline
{\tiny{}       } & {\tiny{}       } & {\tiny{}yes} & {\tiny{}$0.343$} & {\tiny{}$ 1.39$} & {\tiny{}$14.14$}       & {\tiny{}$ 0.5877$} & {\tiny{}$6928$} & {\tiny{}$4.10$} & {\tiny{}$3.49(-8)$} & {\tiny{}$-2.15$} & {\tiny{}$ 1.39$} & {\tiny{}$ 1.32$}  \tabularnewline
{\tiny{}$0.750$} & {\tiny{}$0.100$} & {\tiny{}no } & {\tiny{}  \ldots   } & {\tiny{}  \ldots   } & {\tiny{}$12.39$}       & {\tiny{}$ 0.5819$} & {\tiny{}$6836$} & {\tiny{}$4.08$} & {\tiny{}$1.73(-7)$} & {\tiny{}$-2.17$} & {\tiny{}$ 1.76$} & {\tiny{}$ 1.05$}  \tabularnewline
{\tiny{}       } & {\tiny{}       } & {\tiny{}yes} & {\tiny{}$0.455$} & {\tiny{}$ 1.21$} & {\tiny{}$12.35$}       & {\tiny{}$ 0.5838$} & {\tiny{}$6953$} & {\tiny{}$4.11$} & {\tiny{}$2.53(-8)$} & {\tiny{}$-2.15$} & {\tiny{}$ 1.21$} & {\tiny{}$ 1.06$}  \tabularnewline
{\tiny{}$0.800$} & {\tiny{}$0.050$} & {\tiny{}no } & {\tiny{}  \ldots   } & {\tiny{}  \ldots   } & {\tiny{}$11.17$}       & {\tiny{}$ 0.5820$} & {\tiny{}$6849$} & {\tiny{}$4.08$} & {\tiny{}$1.38(-7)$} & {\tiny{}$-2.17$} & {\tiny{}$ 1.76$} & {\tiny{}$ 0.78$}  \tabularnewline
{\tiny{}       } & {\tiny{}       } & {\tiny{}yes} & {\tiny{}$0.543$} & {\tiny{}$ 1.02$} & {\tiny{}$11.16$}       & {\tiny{}$ 0.5861$} & {\tiny{}$6979$} & {\tiny{}$4.11$} & {\tiny{}$1.73(-8)$} & {\tiny{}$-2.14$} & {\tiny{}$ 1.01$} & {\tiny{}$ 0.79$}  \tabularnewline
{\tiny{}$0.700$} & {\tiny{}$0.200$} & {\tiny{}no } & {\tiny{}  \ldots   } & {\tiny{}  \ldots   } & {\tiny{}$12.04$}       & {\tiny{}$ 0.6436$} & {\tiny{}$7198$} & {\tiny{}$4.13$} & {\tiny{}$6.53(-10)$} & {\tiny{}$-2.17$} & {\tiny{}$ 1.76$} & {\tiny{}$ 1.30$}  \tabularnewline
{\tiny{}       } & {\tiny{}       } & {\tiny{}yes} & {\tiny{}$0.381$} & {\tiny{}$ 1.39$} & {\tiny{}$11.98$}       & {\tiny{}$ 0.6391$} & {\tiny{}$7291$} & {\tiny{}$4.16$} & {\tiny{}$3.65(-10)$} & {\tiny{}$-2.15$} & {\tiny{}$ 1.39$} & {\tiny{}$ 1.30$}  \tabularnewline
{\tiny{}$0.800$} & {\tiny{}$0.100$} & {\tiny{}no } & {\tiny{}  \ldots   } & {\tiny{}  \ldots   } & {\tiny{}$10.31$}       & {\tiny{}$ 0.6459$} & {\tiny{}$7190$} & {\tiny{}$4.13$} & {\tiny{}$6.91(-10)$} & {\tiny{}$-2.17$} & {\tiny{}$ 1.76$} & {\tiny{}$ 1.01$}  \tabularnewline
{\tiny{}       } & {\tiny{}       } & {\tiny{}yes} & {\tiny{}$0.511$} & {\tiny{}$ 1.20$} & {\tiny{}$10.24$}       & {\tiny{}$ 0.6369$} & {\tiny{}$7321$} & {\tiny{}$4.17$} & {\tiny{}$3.14(-10)$} & {\tiny{}$-2.15$} & {\tiny{}$ 1.20$} & {\tiny{}$ 1.02$}  \tabularnewline
{\tiny{}$0.850$} & {\tiny{}$0.050$} & {\tiny{}no } & {\tiny{}  \ldots   } & {\tiny{}  \ldots   } & {\tiny{}$ 9.17$}       & {\tiny{}$ 0.6493$} & {\tiny{}$7193$} & {\tiny{}$4.12$} & {\tiny{}$6.71(-10)$} & {\tiny{}$-2.17$} & {\tiny{}$ 1.76$} & {\tiny{}$ 0.74$}  \tabularnewline
{\tiny{}       } & {\tiny{}       } & {\tiny{}yes} & {\tiny{}$0.593$} & {\tiny{}$ 1.02$} & {\tiny{}$ 9.10$}       & {\tiny{}$ 0.6384$} & {\tiny{}$7350$} & {\tiny{}$4.17$} & {\tiny{}$2.74(-10)$} & {\tiny{}$-2.14$} & {\tiny{}$ 1.02$} & {\tiny{}$ 0.75$}  \tabularnewline
{\tiny{}$0.750$} & {\tiny{}$0.200$} & {\tiny{}no } & {\tiny{}  \ldots   } & {\tiny{}  \ldots   } & {\tiny{}$10.20$}       & {\tiny{}$ 0.6613$} & {\tiny{}$7544$} & {\tiny{}$4.22$} & {\tiny{}$1.52(-10)$} & {\tiny{}$-2.17$} & {\tiny{}$ 1.76$} & {\tiny{}$ 1.27$}  \tabularnewline
{\tiny{}       } & {\tiny{}       } & {\tiny{}yes} & {\tiny{}$0.433$} & {\tiny{}$ 1.38$} & {\tiny{}$10.13$}       & {\tiny{}$ 0.6538$} & {\tiny{}$7631$} & {\tiny{}$4.25$} & {\tiny{}$1.21(-10)$} & {\tiny{}$-2.15$} & {\tiny{}$ 1.38$} & {\tiny{}$ 1.27$}  \tabularnewline
\multicolumn{13}{c}{{\tiny{}$M_{1}=1.25\thinspace\mathrm{M}_{\sun}$}}\tabularnewline
{\tiny{}$0.700$} & {\tiny{}$0.200$} & {\tiny{}no } & {\tiny{}  \ldots   } & {\tiny{}  \ldots   } & {\tiny{}$10.18$}       & {\tiny{}$ 0.6008$} & {\tiny{}$6677$} & {\tiny{}$4.04$} & {\tiny{}$3.31(-6)$} & {\tiny{}$-2.15$} & {\tiny{}$ 2.57$} & {\tiny{}\ldots   }  \tabularnewline
{\tiny{}       } & {\tiny{}       } & {\tiny{}yes} & {\tiny{}$0.194$} & {\tiny{}$ 2.03$} & {\tiny{}$10.08$}       & {\tiny{}$ 0.6829$} & {\tiny{}$6963$} & {\tiny{}$4.03$} & {\tiny{}$2.76(-9)$} & {\tiny{}$-2.14$} & {\tiny{}$ 2.03$} & {\tiny{}$ 1.94$}  \tabularnewline
{\tiny{}$0.800$} & {\tiny{}$0.100$} & {\tiny{}no } & {\tiny{}  \ldots   } & {\tiny{}  \ldots   } & {\tiny{}$ 9.24$}       & {\tiny{}$ 0.6103$} & {\tiny{}$6688$} & {\tiny{}$4.04$} & {\tiny{}$2.67(-6)$} & {\tiny{}$-2.15$} & {\tiny{}$ 2.57$} & {\tiny{}\ldots   }  \tabularnewline
{\tiny{}       } & {\tiny{}       } & {\tiny{}yes} & {\tiny{}$0.324$} & {\tiny{}$ 1.83$} & {\tiny{}$ 9.02$}       & {\tiny{}$ 0.6225$} & {\tiny{}$7091$} & {\tiny{}$4.13$} & {\tiny{}$6.12(-9)$} & {\tiny{}$-2.14$} & {\tiny{}$ 1.82$} & {\tiny{}$ 1.75$}  \tabularnewline
\multicolumn{13}{c}{{\tiny{}no accretion}}\tabularnewline
{\tiny{}$0.750$} & {\tiny{}$0.000$} & {\tiny{}no } & {\tiny{}  \ldots   } & {\tiny{}  \ldots   } & {\tiny{}$15.44$}       & {\tiny{}$ 0.4033$} & {\tiny{}$6509$} & {\tiny{}$4.12$} & {\tiny{}$2.31(-5)$} & {\tiny{}$-2.14$} & {\tiny{}$ 0.00$} & {\tiny{}\ldots   }  \tabularnewline
{\tiny{}$0.800$} & {\tiny{}$0.000$} & {\tiny{}no } & {\tiny{}  \ldots   } & {\tiny{}  \ldots   } & {\tiny{}$12.17$}       & {\tiny{}$ 0.5013$} & {\tiny{}$6741$} & {\tiny{}$4.11$} & {\tiny{}$1.16(-6)$} & {\tiny{}$-2.14$} & {\tiny{}$ 0.00$} & {\tiny{}$-0.01$}  \tabularnewline
{\tiny{}$0.850$} & {\tiny{}$0.000$} & {\tiny{}no } & {\tiny{}  \ldots   } & {\tiny{}  \ldots   } & {\tiny{}$ 9.75$}       & {\tiny{}$ 0.5897$} & {\tiny{}$7047$} & {\tiny{}$4.12$} & {\tiny{}$7.61(-9)$} & {\tiny{}$-2.14$} & {\tiny{}$ 0.00$} & {\tiny{}$-0.03$}  \tabularnewline
\end{longtable} 
\tablefoot{
Values of $M_\text{env}$ are given in the format $n(m)=n\times10^m$ for concision.
\tablefoottext{a} {Undefined for models without thermohaline mixing.}
\tablefoottext{b} {Some of the models stop earlier.}
}
\end{longtab}

Last, there are some objects with $[\mathrm{C/Fe}]\gtrsim2.5$ whose surface gravities ($\log g\lesssim3$) imply that they are close to the end of FDU if not past it. How might we explain such objects, assuming that they were polluted by an AGB companion? Since carbon is not produced in the star, it is hard to imagine how the surface carbon abundance could be above the one in the accreted material. This limits the possible primaries to those that are able to produce at least this much carbon. From the models of \citet{2012ApJ...747....2L} these are AGB stars with $1.25\lesssim M_{1}/\mathrm{M}_{\sun}\lesssim3$. Lower-mass AGB stars do not produce enough carbon; and higher-mass stars convert the carbon into nitrogen in the lower part of their convective envelope (a process known as hot-bottom-burning). Moreover, a large amount of mass must be transferred because the combined carbon reduction from thermohaline mixing and FDU can only be about $0.5~\text{dex}$ at most \citep[the maximum $\text{[C/Fe]}$ given by the AGB models of][is about $+3.2$]{2012ApJ...747....2L}. This implies an accreted-to-initial mass ratio, $\Delta M/M_{2,\text{i}}$, between about $0.25$, if all the mixing occurs during FDU (where the envelope mass grows to about $0.5~\mathrm{M}_{\sun}$), and $0.5$, if thermohaline mixing is efficient (as it will be in the absence of some inhibitory process because of how unevolved the progenitor of the CEMP-\emph{s} star must be). The respective accreted-to-final mass ratios, $\Delta M/M_{2,\text{f}}$, are between $0.2$ and $0.35$. Hence, the progenitor systems of the most carbon-rich evolved stars must be small mass-ratio binaries in which a lot of mass has been transferred to the secondary. It may be difficult to account for such stars without also predicting too many low-luminosity carbon-rich stars from cases where less mass has been transferred.

\section{Discussion\label{subsec:uncertainties}}

The results presented in this paper may lead one to wonder whether the abundance anomalies predicted by our diffusion models are overestimated. We have run multiple tests to address this concern. First, we have tried to reproduce the results of \citet{2002ApJ...568..979R}. In particular, we have compared the abundance evolution in a $M=0.8~\mathrm{M}_{\sun}$ model with an initial composition taken from their table~1 ($Z=1.7\times10^{-4}$; $[\mathrm{Fe/H}]=-2.31$). We obtain good agreement in terms of the temperature and luminosity at the turn-off although our model is longer-lived by about $0.5~\text{Gyr}$. The abundance anomalies from settling and/or levitation of He, C, N, and Fe agree within 0.1~dex. For other elements the abundances differ by $0.3\text{--}0.4~\text{dex}$. Given that in our model the convective envelope mass is smaller by about a factor of two throughout most of the evolution (and the minimum size of the convective envelope in our model is only about $3.7\times10^{-6}~\mathrm{M}_{\sun}$ whereas \citet{2002ApJ...568..979R} get about $2.5\times10^{-5}~\mathrm{M}_{\sun}$), these differences are plausible. Judging from their figure~2, a smaller envelope mass in their model would lead to greater over-abundances of O, Ne, and Mg, and a smaller over-abundance of Si. All of these changes would reduce the discrepancies between their model and ours.

Second, we have tested whether the large abundance anomalies predicted by our models stem specifically from our simplified treatment of diffusion. For this purpose we have ported into our code the relevant parts from the code used by \citet[priv. comm.]{2010A&A...511A..87H}\footnote{In their code the full set of Burgers flow equations \citep{1969fecg.book.....B} is solved and their treatment of diffusion is thus valid for arbitrary compositions.} and run a $M=0.8~\mathrm{M}_{\sun}$, $Z=10^{-4}$ diffusive model without radiative levitation (with the ZAMS chemical composition from Table~\ref{tab:xinp}). In this model diffusion reduces the helium and metal abundances in the envelope on much shorter timescales. The same $M=0.8~\mathrm{M}_{\sun}$ model run with the MESA code \citep{2011ApJS..192....3P,2015ApJS..220...15P} yields similar results, which is reassuring given that the treatment of diffusion is based on the work of \citet{1969fecg.book.....B} and \citet{1994ApJ...421..828T} in both MESA and \citet{2010A&A...511A..87H}. The abundances in the MESA model are depleted to a much greater degree: 5~dex for helium (compared to our 2.5~dex) and 5--6~dex (3--4~dex) for metals, even though the envelope masses are in a good agreement (the minimum envelope mass is $4.5\times10^{-7}~\mathrm{M}_{\sun}$ in STARS and $6.0\times10^{-7}~\mathrm{M}_{\sun}$ in MESA). The conclusion from all three of these tests is the same -- if anything, the diffusion models presented here underestimate the amount of diffusion that we would get from a more rigorous treatment.

We have performed spatial resolution tests in three systems (denoted by an asterisk in Table~\ref{tab:Results_main}) by varying the default number of meshpoints (999) by a factor of two. All models give consistent results (within a couple of percent) in terms of the global properties, depth of thermohaline mixing, abundance anomalies after turn-off, and post-FDU abundances. The size of the convective envelope at minimum is consistent within ten percent. We thus conclude that the models are sufficiently resolved.

Our approach of interpolating the opacities and accelerations from tables computed during the run time necessitates the introduction of some numerical parameters. These parameters control mainly the amount by which some species has to change to warrant the computation of a new table. We have done extensive tests to make sure that our results do not depend on the choice of these parameters, i.e. the tables are computed often enough. As stated earlier, we set the temperature above which we use the old opacity tables that include conduction \citep{2004MNRAS.348..201E} to $\log T=7.3$. We have since included the conductive opacities from \citet{2007ApJ...661.1094C} in our code and made sure that use of OP opacities above $\log T=7.3$ would have virtually no effect on any of our results. 

The size of the convective envelope throughout the evolution depends somewhat on the choice of the mixing-length parameter with larger values resulting in more massive envelopes. Our value, $\alpha_{\mathrm{MLT}}=2.0$, is based on a calibration between the radius, effective temperature, and luminosity of a $Z=0.0142$, $M=1~\mathrm{M}_{\sun}$ diffusive model with OP opacities at an age of 4.56~Gyr and the Sun \citep[our $\alpha_\text{MLT}$ value is slightly smaller than the value of 2.025 presented by][because of the different opacities]{2016A&A...586A.119S}. Stars of masses, metallicities, and evolutionary stages different from the Sun should have other values of $\alpha_{\text{MLT}}$ \citep[e.g. ][]{2014MNRAS.445.4366T} but meaningful quantitative predictions are virtually impossible. In a $0.8~\mathrm{M}_{\sun}$, $Z=10^{-4}$ model increasing or decreasing $\alpha_{\mathrm{MLT}}$ by 5\% accordingly changes the envelope mass by about 50\%, which, given that $M_{\mathrm{env}}<10^{-6}~\mathrm{M}_{\sun}$, translates into substantial changes in the surface abundances (Fig.~\ref{fig:ab_vs_menv}). Since theoretical models suggest that at lower metallicities one should use lower $\alpha_{\mathrm{MLT}}$ values \citep[at least for $-0.6<\text{[Fe/H]}<+0.3$;][]{2012ApJ...755L..12B}, it is unlikely that we have overestimated the importance of diffusion by underestimating the value of $\alpha_{\mathrm{MLT}}$.

\subsection{Missing mixing processes\label{subsec:missing-phys}}

The strong abundance anomalies predicted by diffusive models are not observed in CEMP stars. This suggests that in real stars atomic diffusion is inhibited by some physical process that we have not included in our models. While we leave a more in-depth investigation of possible culprits to future work, we examine a simple test case here. We add a ``turbulent'' diffusion term to $D_{\mathrm{mix}}$ in Eq.~\eqref{eq:dxdt} as proposed by \citet{2000ApJ...529..338R,2005ApJ...619..538R}:
\begin{equation}
D_{\mathrm{T}}=D_{0}D_{\mathrm{He}}\left(T_{0}\right)\left[\frac{\rho}{\rho\left(T_{0}\right)}\right]^{-3}.\label{eq:dturb}
\end{equation}
This type of parametrization extends the surface mixing region down to where the local temperature is somewhat larger than $T_{0}$. \citet{2005ApJ...619..538R} find that observations of lithium abundances in population~II stars require $T_{0}\approx10^{6}~\text{K}$. Similar values have been found to reproduce the small systematic abundance differences between turn-off and giant stars in old globular clusters \citep{2006Natur.442..657K,2012ApJ...753...48N,2014A&A...567A..72G}. Nevertheless, the ad-hoc nature of this prescription should be kept in mind. With this prescription we can primarily constrain the depth to which some form of mixing must occur to reconcile the models with observations, but not the physical processes responsible for this mixing.

\begin{figure}
\subfloat{\includegraphics[width=1\columnwidth]{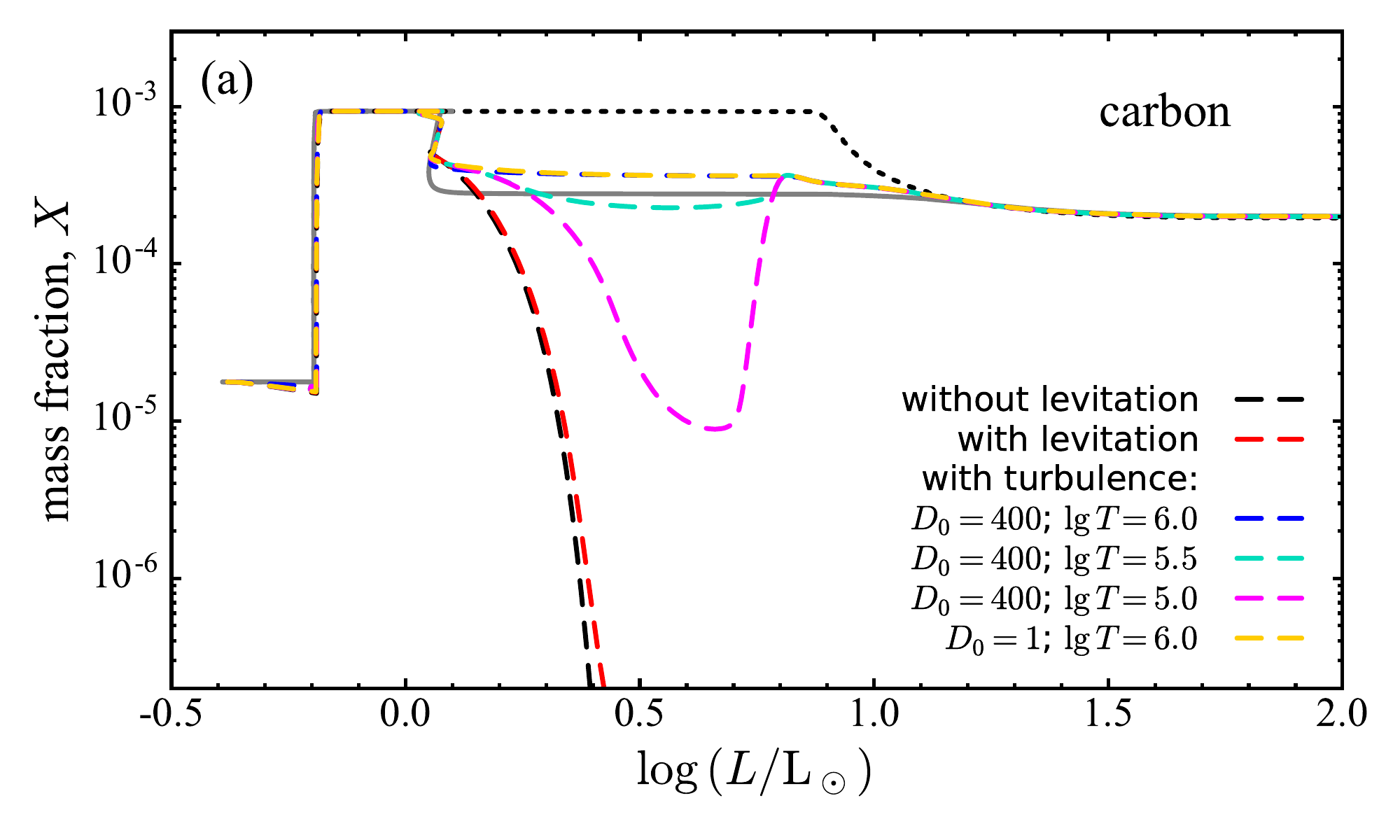}\label{fig:turb-test_C}}

\subfloat{\includegraphics[width=1\columnwidth]{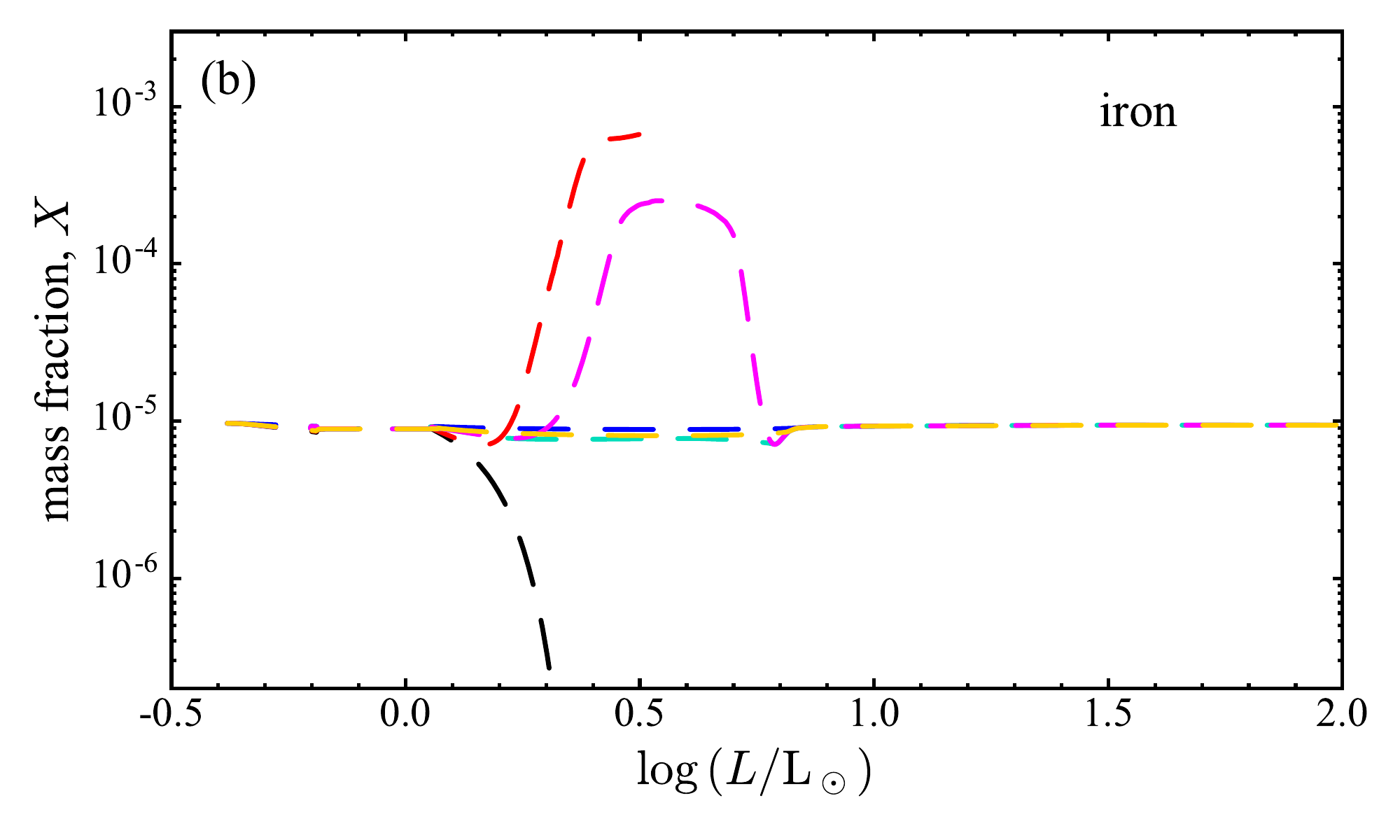}\label{fig:turb-test_Fe}}

\subfloat{\includegraphics[width=1\columnwidth]{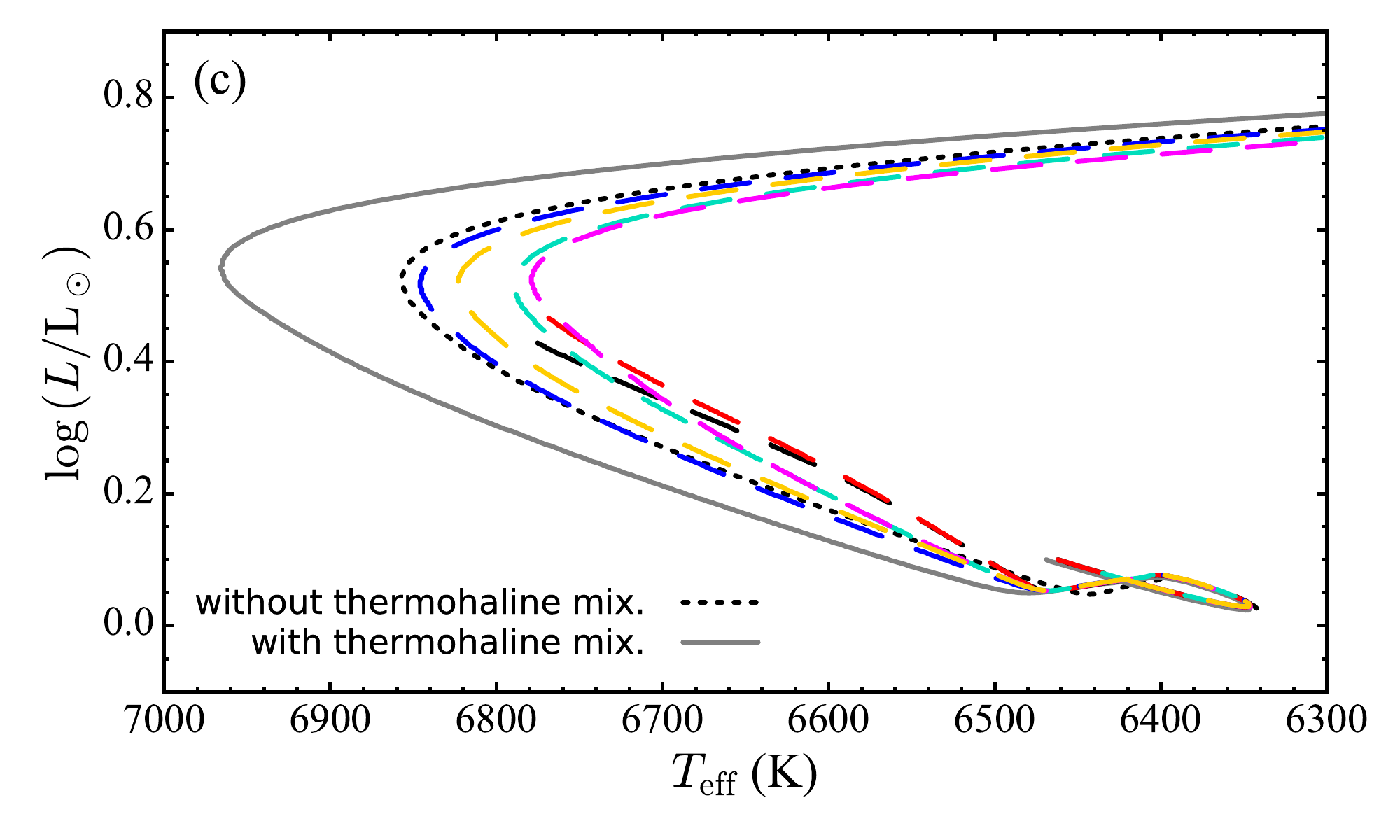}\label{fig:turb-test_hrd}}

\caption{Effect of turbulence on the evolution of carbon (a) and iron (b) mass fractions, and the HRD (c) of a $0.75\mathrm{~M}_{\sun}$ star accreting $0.1~\mathrm{M}_{\sun}$ of material from a $1~\mathrm{M}_{\sun}$ primary. Large abundance variations are expected after accretion in absence of turbulence (black and red dashes). Inclusion of turbulence as in Eq.~\eqref{eq:dturb} erases all abundance signatures of atomic diffusion on the post-mass-transfer main sequence (dark blue) even for small turbulent diffusion coefficients (orange). Only when the temperature parameter is reduced to $\log T_{0}\lesssim5.5$ do the abundance variations start to reappear (light blue and magenta). The thermohaline-mixing-only model (solid grey) is hotter than the models with turbulence because in the latter diffusion still modifies the layers with $T\gtrsim T_{0}$ and thermohaline mixing is not as deep. For clarity, the HRD only shows the post-mass-transfer part of the evolutionary tracks.\label{fig:turb-test}}
\end{figure}

We test the effect of turbulent diffusion on the evolution of a $0.75\mathrm{~M}_{\sun}$ star accreting $0.1~\mathrm{M}_{\sun}$ of material from a $1~\mathrm{M}_{\sun}$ primary. While in the absence of turbulence the resulting $0.85~\mathrm{M}_{\sun}$ model shows extremely large abundance anomalies ($[\mathrm{Fe/H}]>-0.5$ with levitation and $[\mathrm{Fe/H}]<-9.2$ without levitation), turbulence with $D_{0}=400$ \citep[as used by][]{2005ApJ...619..538R} and $\log T_{0}=6.0$ completely negates them (Fig.~\ref{fig:turb-test}). Even much smaller turbulent diffusion coefficients (e.g. $D_{0}=1$) suffice to erase the anomalies. Indeed, the key parameter here is $T_{0}$ -- as long as the mixing region remains large enough ($\log T_{0}\gtrsim5.5$ or $M_{\mathrm{env}}\gtrsim10^{-4}~\mathrm{M}_{\sun}$), atomic diffusion is strongly suppressed. In terms of global properties, models with more pervasive turbulence are hotter and therefore more closely resemble models without diffusion. This can be seen from comparing the turbulent models with the model with thermohaline mixing only (solid grey line) in Fig.~\ref{fig:turb-test_hrd}.\footnote{The model with $D_{0}=400$ and $\log T_{0}=6.0$ is quite different from the basic model with no diffusion or thermohaline mixing, although their tracks almost coincide. In the basic model the accreted material remains on the surface of the star. In the turbulent model the material is diluted but not as much as in a model without diffusion because of the stabilizing $\mu$-gradient in layers where $T\gtrsim T_{0}$.}

Figure~\ref{fig:turb-test} also shows that while the turbulent diffusion prescription of \citet{2005ApJ...619..538R} can inhibit diffusion in the outer layers of a star, it has almost no influence on thermohaline mixing. This is not to say, however, that some form of turbulence \citep[e.g. rotationally driven horizontal turbulence;][]{2008ApJ...684..626D} could not inhibit thermohaline mixing as well. Rather, investigating this requires treating the different processes together instead of considering them as independent and simply adding the individual diffusion coefficients \citep{2013A&A...553A...1M}. Such a treatment is beyond the scope of this paper.

\subsection{Mass loss\label{subsec:Mass-loss}}

So far we have ignored mass loss. Simple estimates imply that it may be too important to neglect. With a mass-loss rate comparable to the current solar value, $2\text{--}3\times10^{-14}~\mathrm{M}_{\sun}\thinspace\text{yr}^{-1}$ \citep{1998ASPC..154..131W}, a star will lose more than $10^{-4}~\mathrm{M}_{\sun}$ over the roughly $10^{10}$ years it spends on the main sequence. This is a very large amount compared to the envelope masses of our models (Fig.~\ref{fig:ab_vs_menv}) and could greatly interfere with atomic diffusion.

In the absence of mass loss the convective envelope of a star moves outwards in mass until the beginning of FDU. But when mass loss erodes the surface, this outward movement is halted and eventually reversed while the star is still on the main sequence. As the envelope now moves inwards, the surface abundances reflect the composition of the progressively deeper layers that get exposed. Qualitatively, if the mass-loss rate is sufficiently high, the removal of the outer layers is so fast that diffusion has not had enough time to modify the newly exposed layers and only small abundance anomalies can develop \citep{1995ApJ...438L..87S}. On the other hand, mass-loss rates below some limit must be negligible and have essentially no effect on the surface abundances.

\begin{figure*}
\subfloat{\includegraphics[width=1\columnwidth]{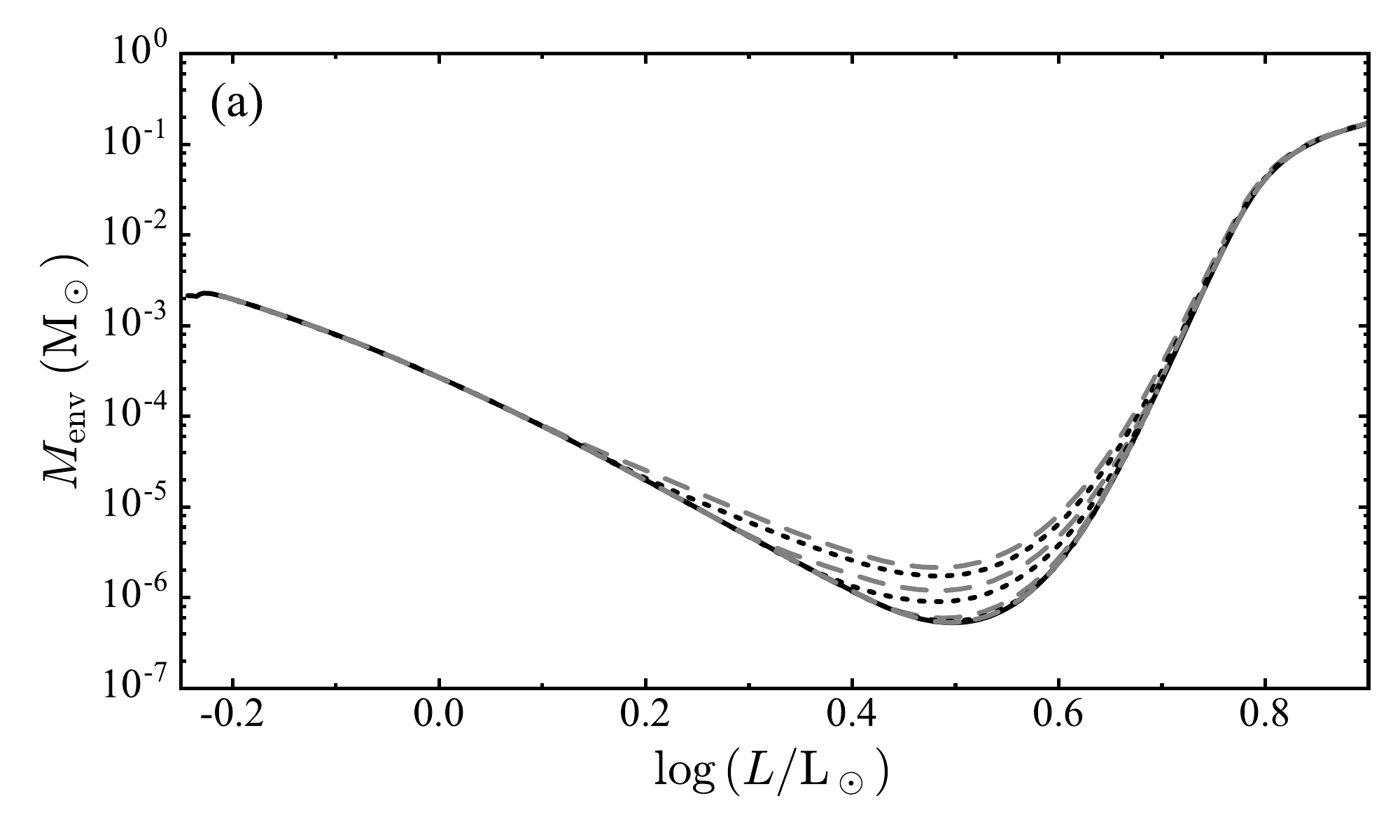}\label{fig:mass-loss_0.80_Menv-vs-L}}\hspace{\columnsep}\subfloat{\includegraphics[width=1\columnwidth]{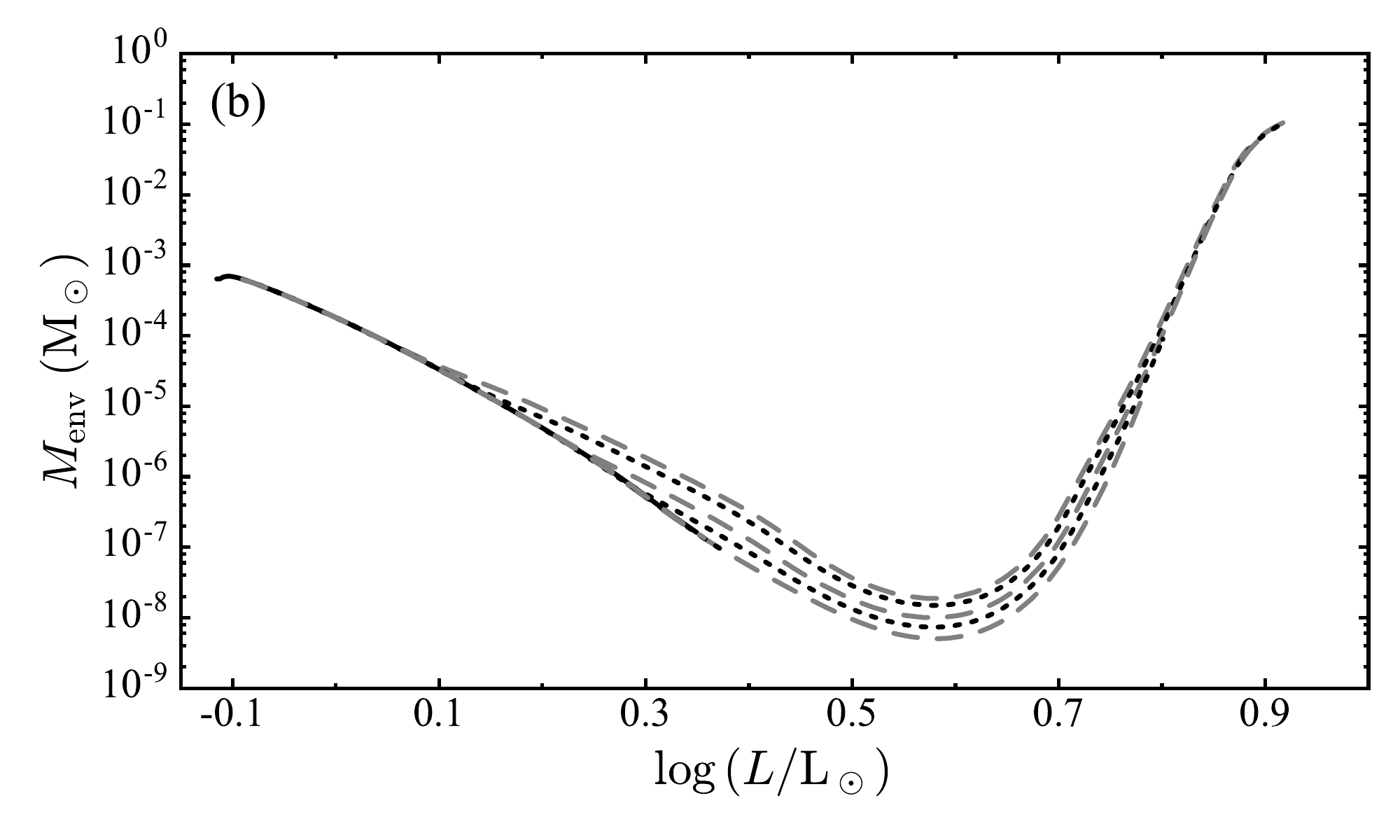}\label{fig:mass-loss_0.85_Menv-vs-L}}

\subfloat{\includegraphics[width=1\columnwidth]{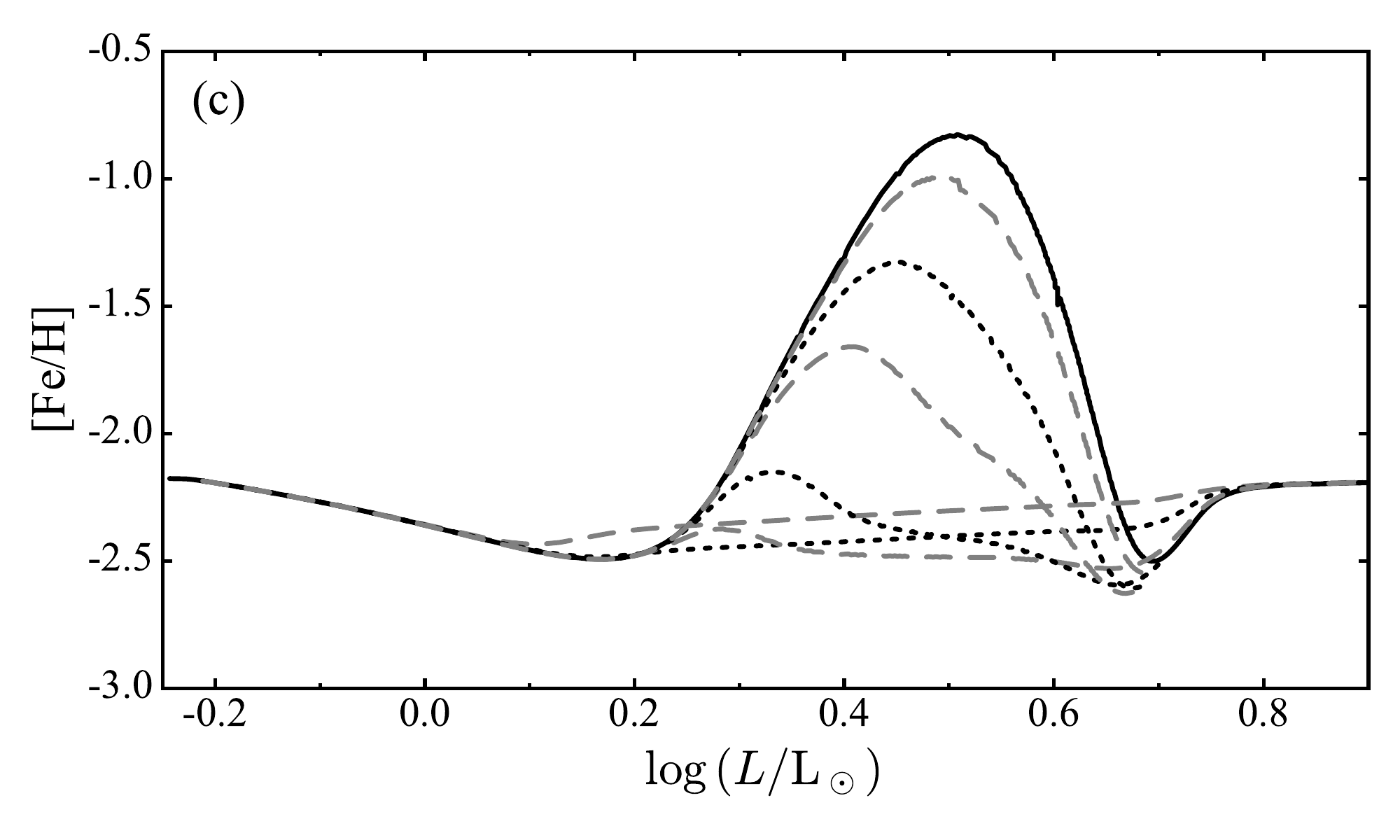}\label{fig:mass-loss_0.80_FeH-vs-L}}\hspace{\columnsep}\subfloat{\includegraphics[width=1\columnwidth]{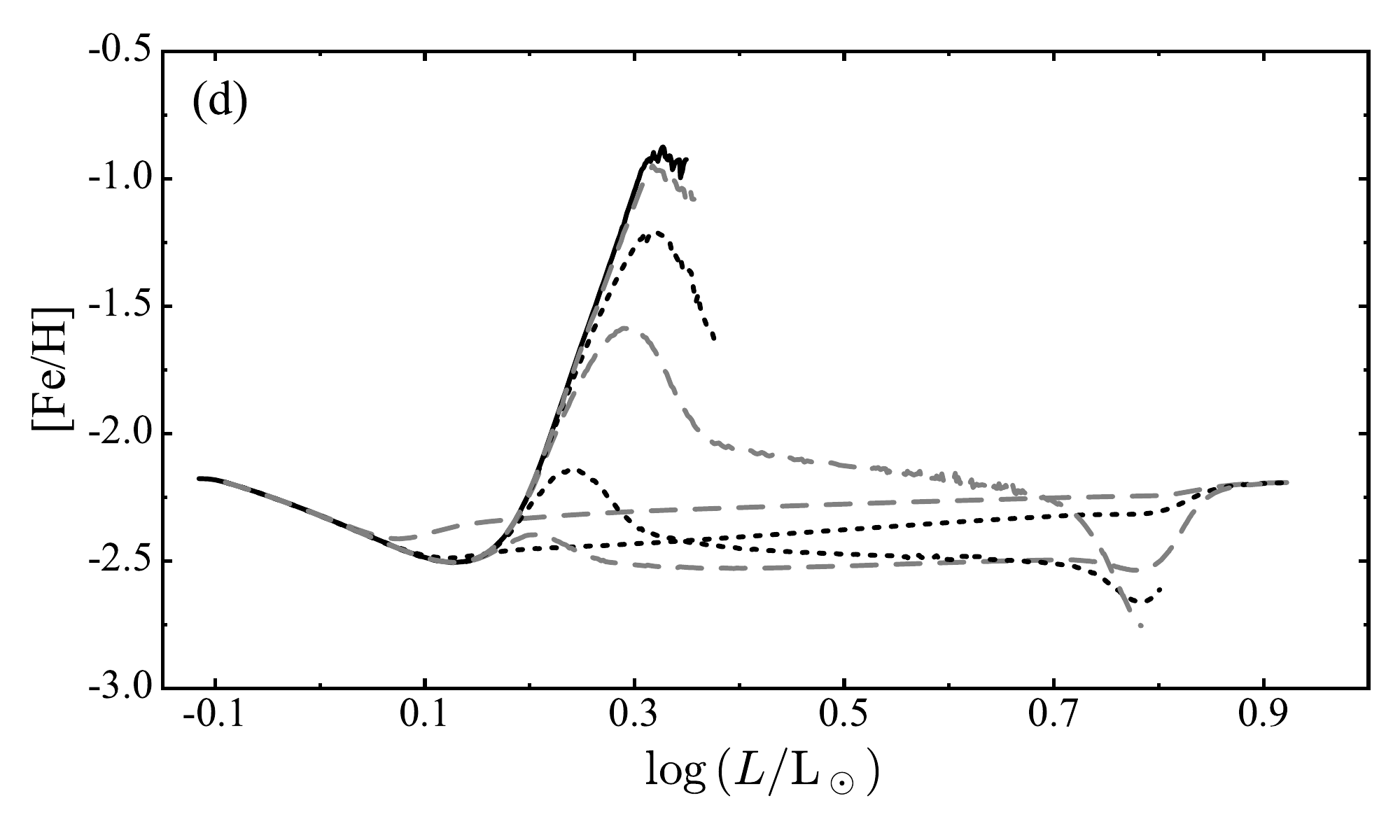}\label{fig:mass-loss_0.85_FeH-vs-L}}

\subfloat{\includegraphics[width=1\columnwidth]{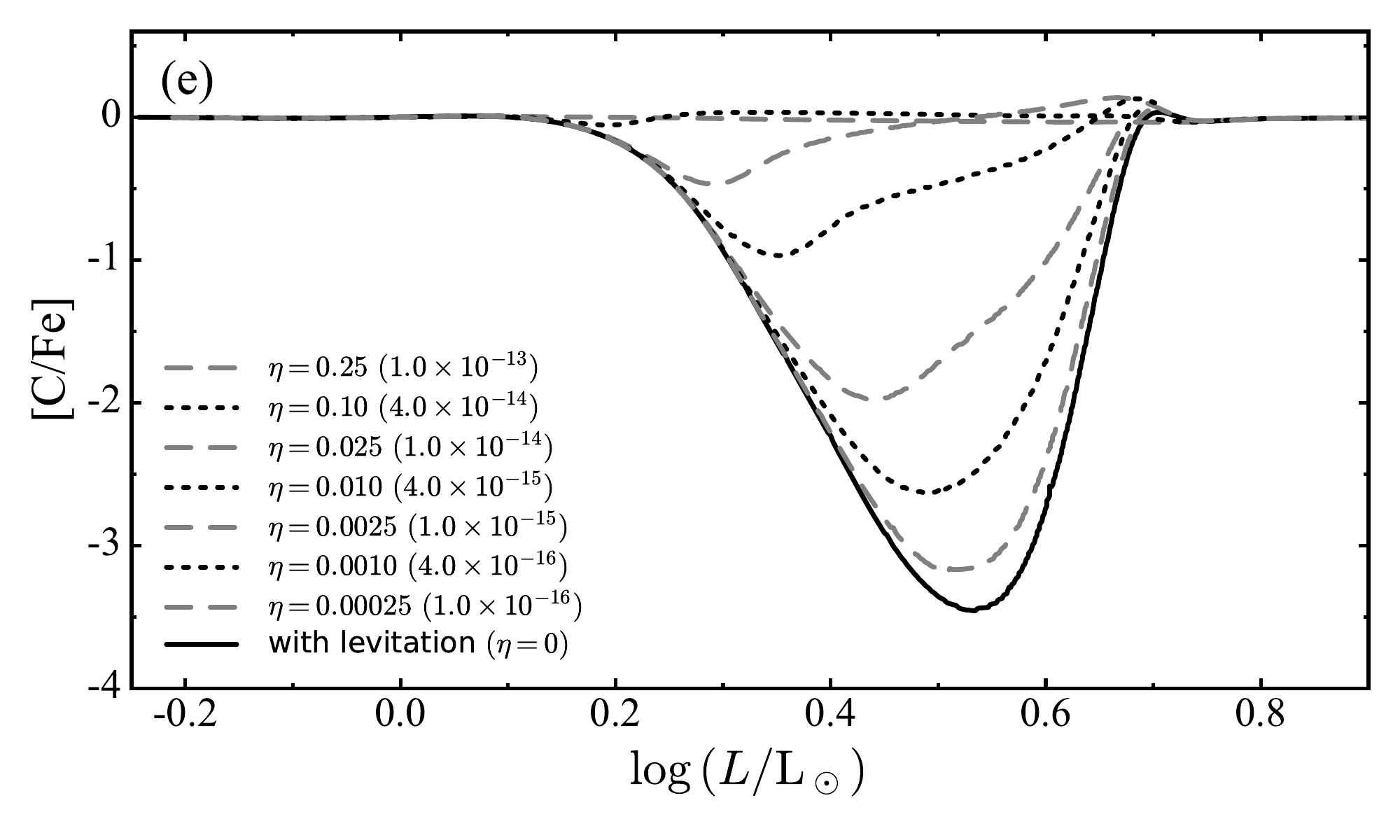}\label{fig:mass-loss_0.80_CFe-vs-L}}\hspace{\columnsep}\subfloat{\includegraphics[width=1\columnwidth]{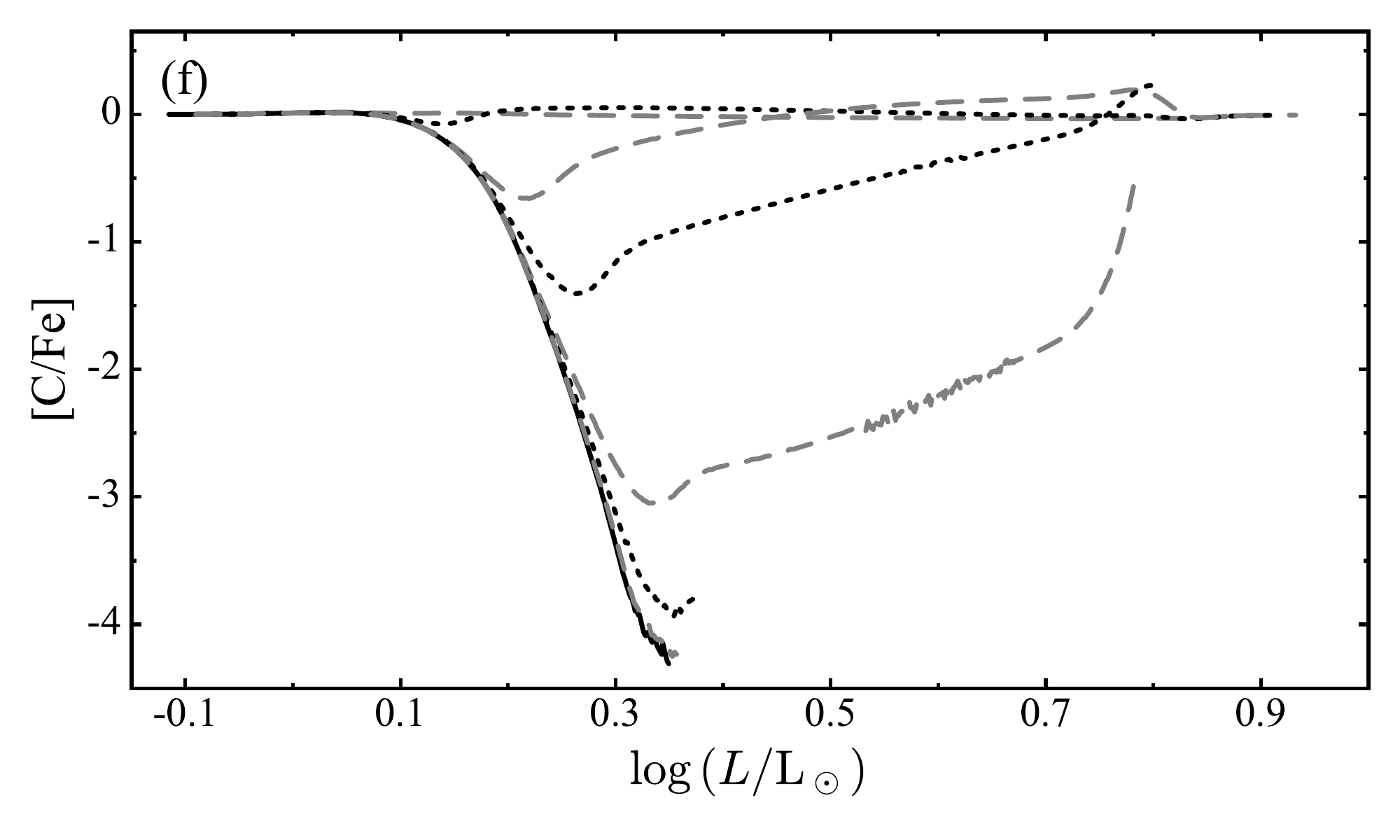}\label{fig:mass-loss_0.85_CFe-vs-L}}

\caption{Envelope mass and abundance evolution in metal-poor $0.8~\mathrm{M}_{\sun}$ (left panels) and $0.85~\mathrm{M}_{\sun}$ (right panels) models with different Reimers-type mass-loss rates. The values in the parentheses are the mass-loss rates at $\log L\approx0$; the average mass-loss rate during the main sequence is about 50\% higher as a result of the mass, radius, and luminosity scaling in Reimers' law.\label{fig:mass-loss}}
\end{figure*}

We now estimate what mass-loss rates are necessary to prevent the development of abundance anomalies and what mass-loss rates are negligible. For simplicity, we use \citet{1975MSRSL...8..369R} mass-loss formula with different factors $\eta$:

\begin{equation}
\dot{M}=-4\times10^{-13}\eta\frac{LR}{M}\left(\frac{LR}{M}\right)_{\sun}^{-1}~\mathrm{M}_{\sun}\thinspace\text{yr}^{-1}.\label{eq:RML}
\end{equation}
Here we only consider metal-poor $0.8~\mathrm{M}_{\sun}$ and $0.85~\mathrm{M}_{\sun}$ models without accretion. We find that if $\eta\gtrsim0.1$, the effects of atomic diffusion are almost entirely erased in both models (Fig.~\ref{fig:mass-loss}). This translates to mass-loss rates of a few times $10^{-14}\text{--}10^{-13}~\mathrm{M}_{\sun}\thinspace\text{yr}^{-1}$ throughout the main sequence. In contrast, when the mass-loss rate falls below about $10^{-16}~\mathrm{M}_{\sun}\thinspace\text{yr}^{-1}$, the abundance evolution proceeds as in models without mass loss. Intermediate mass-loss rates result in less extreme but non-negligible abundance variations. Note that the lost material is assumed to have the same composition as the surface at that time. Depending on the mass-loss rate and mechanism, some elements may be lost more readily than others, leading to more complicated abundance variations \citep[e.g.][]{1987ApJ...322..302M,2010A&A...521A..62V}.

Is it reasonable to expect somewhat super-solar mass-loss rates from CEMP-\emph{s} stars on the main sequence? That is difficult to say. The form of mass loss in these stars is presumably the same as in normal low-mass main sequence stars -- as magnetized winds originating in a corona that is heated by turbulent dissipation of Alfv\'en waves \citep[e.g.][]{2007ApJ...659.1592S,2011ApJ...741...54C}. While mass-loss rates in these stars are too small to be directly observable, indirect measurements based on the interaction of the wind with the interstellar medium yield values within about an order of magnitude of the solar mass-loss rate \citep{2002ApJ...574..412W,2005ApJ...628L.143W}. Various theoretical models also normally predict mass-loss rates in this range \citep[e.g.][]{2007A&A...463...11H,2011ApJ...741...54C,2015A&A...577A..28J}. While most of these models concern stars with near solar metallicity, to first order the mass-loss rates are not expected to depend on metallicity. Note that the mass-loss rate in our test models increases over time because of the $LR/M$ scaling. This is because there is no rotational dependence in the Reimers' mass-loss law which is reasonable given that it was derived from observations of red giants (i.e. slow rotators). But CEMP-\emph{s} stars could have high rotation rates after the mass transfer phase, if the transferred material carries with it some angular momentum. Over most of the post-mass-transfer main sequence evolution their mass-loss rate would then be decreasing as the wind carried away the excess angular momentum \citep[e.g. for solar-type stars][give $\dot{M}\sim\Omega^{1.3}\sim t^{-0.75}$]{2015A&A...577A..28J}, which should result in higher average mass-loss rates. This scenario may even have further complications aside from mass loss, because rapid rotation could directly lead to enhanced chemical mixing in the star.

In any case, the mass-loss rates of CEMP-\emph{s} dwarfs are likely at least a few times $10^{-15}~\mathrm{M}_{\sun}\text{yr}^{-1}$ throughout their evolution. Such mass-loss rates should at least moderate the effects of atomic diffusion and help explain why turn-off stars with extreme abundance anomalies are not observed. Additionally, or alternatively, some form of turbulence might play a role. As discussed by \citet{2010A&A...521A..62V}, two models, one with mass loss and one with turbulent diffusion, that predict the same surface abundances do not necessarily have the same internal abundance profiles. In a model with turbulence the abundance profiles are flat down to some depth (e.g. determined by $T_{0}$ in Eq.~\eqref{eq:dturb}). In a model with mass loss no mixing is enforced outside the convective region so the abundance profiles will not be flat unless the mass-loss rate is large ($\dot{M}\gtrsim10^{-13}~\mathrm{M}_{\sun}\thinspace\text{yr}^{-1}$). Asteroseismic measurements sensitive to the internal structure of a star might in principle be able to distinguish between the two types of models \citep[e.g.][]{2012ApJ...746...16V}. However, in practice the difference in the internal structure might be too small at these low metallicities for such measurements to be possible.

\section{\label{sec:Conclusions}Conclusions}

In this paper we present stellar evolution models of \emph{s}-process-rich carbon-enhanced metal-poor (CEMP-\emph{s}) stars under the assumption that they form when a low-mass metal-poor star accretes material from an AGB companion. Motivated by results from binary population synthesis calculations of \citet{2015A&A...581A..62A}, our models cover current CEMP-\emph{s} star masses between $0.8$ and $0.95~\mathrm{M}_{\sun}$ deriving from initial secondary masses between $0.6$ and $0.8~\mathrm{M}_{\sun}$, and initial primary masses between $0.9$ and $1.5~\mathrm{M}_{\sun}$. Our main focus is the post-mass-transfer evolution of the surface abundances of carbon and iron driven by thermohaline mixing and atomic diffusion, including radiative levitation.

Our simulations with atomic diffusion indicate that CEMP-\emph{s} stars should show large surface abundance variations on the main sequence, particularly as they approach the turn-off. This is because they have very shallow convective envelopes ($M_{\mathrm{env}}\lesssim10^{-4}~\mathrm{M}_{\sun}$ throughout most of the evolution and perhaps as little as $10^{-8}~\mathrm{M}_{\sun}$ near the turn-off) and, therefore, short diffusion timescales. In stars whose envelope masses fall below about $10^{-5}~\mathrm{M}_{\sun}$ (which happens in most of our models, including nearly all those with $M>0.8~\mathrm{M}_{\sun}$) the abundances should vary by a factor of about ten. This factor rapidly increases with decreasing envelope mass resulting in unrealistic abundances (Fig.~\ref{fig:ab_vs_menv}). But even though our treatment of diffusion is not as detailed as in some other works, we do not find evidence that the surface abundance variations predicted by our models are exaggerated.

Radiative levitation has only a minor influence on carbon but a large one on iron. Whereas in diffusive models without levitation the metallicity ({[}Fe/H{]}) of the star decreases until first dredge-up, in models with levitation the metallicity can increase as the star evolves along the main sequence. Consequently, models with levitation predict reduced carbon enhancements ({[}C/Fe{]}) around the turn-off. This implies a systematic difference between the {[}C/Fe{]} values of stars near the turn-off and those at the beginning of FDU. Unfortunately, whether there is any such difference is difficult to establish even from the largest homogeneous sample of observational data \citep[from SDSS;][]{2013AJ....146..132L}. Any such difference, however, would clearly be smaller than predicted by most of our models (Fig.~\ref{fig:feh_ch_cfe_gsra_0.8_0.85}). And while some of our $0.8~\mathrm{M}_{\sun}$ models do predict only a small variation in {[}C/Fe{]}, at ages typical of metal-poor halo stars most of them are still relatively unevolved and should be visible as carbon-rich low-luminosity objects. Very few such objects have been observed.

Although they too would predict many low-luminosity carbon-rich stars, models without atomic diffusion are generally much more successful at covering the range of observations (Fig.~\ref{fig:cfe_notm_0.9_0.95}). We thus conclude that atomic diffusion cannot be acting alone near the surface convection zone of real CEMP-\emph{s} stars and needs to be largely counteracted by some other physical process(es). For example, a turbulent diffusive process like proposed by \citet{2002ApJ...568..979R} can suppress surface abundance variations almost entirely by extending the mixing region to depths where the temperature exceeds about $10^{6}~\text{K}$ (about $10^{-4}~\mathrm{M}_{\sun}$ from the surface; Fig.~\ref{fig:turb-test}). Additionally, at least the most extreme abundance variations (corresponding to stars with the smallest envelopes) should also be moderated by mass loss. In fact, a mass-loss rate of a few times the current solar value sustained throughout the evolution could on its own prevent substantial abundance anomalies from developing (Fig.~\ref{fig:mass-loss}).

While this work has primarily dealt with carbon and iron, given how divergent their abundance evolution is expected to be, these conclusions should extend to other elements, including those produced by neutron capture. The common assumption that the material coming from the AGB companion has simply been diluted by some factor after accretion onto the CEMP-\emph{s} star is likely not too far from the truth.

\begin{acknowledgements}
We thank the anonymous referee for comments that have helped improve the clarity of the paper. We thank Carlo Abate for help with the model grid selection, constructive comments on this manuscript, and many useful discussions. We also thank Young Sun Lee and Timothy Beers for sharing the SDSS CEMP data, Haili Hu for sharing her code, and Evert Glebbeek and Olivier Richard for useful discussions. RJS is the recipient of a Sofja Kovalevskaja Award from the Alexander von Humboldt Foundation.
\end{acknowledgements}

\bibliographystyle{aa}

\bibliography{bibliography.bib}

\begin{thebibliography}{107}
\expandafter\ifx\csname natexlab\endcsname\relax\def\natexlab#1{#1}\fi

\bibitem[{{Abate} {et~al.}(2015){Abate}, {Pols}, {Stancliffe}, {Izzard},
  {Karakas}, {Beers}, \& {Lee}}]{2015A&A...581A..62A}
{Abate}, C., {Pols}, O.~R., {Stancliffe}, R.~J., {et~al.} 2015, \aap, 581, A62

\bibitem[{{Ahn} {et~al.}(2014){Ahn}, {Alexandroff}, {Allende Prieto}, {Anders},
  {Anderson}, {Anderton}, {Andrews}, {Aubourg}, {Bailey}, {Bastien}, \&
  et~al.}]{2014ApJS..211...17A}
{Ahn}, C.~P., {Alexandroff}, R., {Allende Prieto}, C., {et~al.} 2014, \apjs,
  211, 17

\bibitem[{{Ahn} {et~al.}(2012){Ahn}, {Alexandroff}, {Allende Prieto},
  {Anderson}, {Anderton}, {Andrews}, {Aubourg}, {Bailey}, {Balbinot}, {Barnes},
  \& et~al.}]{2012ApJS..203...21A}
{Ahn}, C.~P., {Alexandroff}, R., {Allende Prieto}, C., {et~al.} 2012, \apjs,
  203, 21

\bibitem[{{Alexander} \& {Ferguson}(1994)}]{1994ApJ...437..879A}
{Alexander}, D.~R. \& {Ferguson}, J.~W. 1994, \apj, 437, 879

\bibitem[{{Allen} {et~al.}(2012){Allen}, {Ryan}, {Rossi}, {Beers}, \&
  {Tsangarides}}]{2012A&A...548A..34A}
{Allen}, D.~M., {Ryan}, S.~G., {Rossi}, S., {Beers}, T.~C., \& {Tsangarides},
  S.~A. 2012, \aap, 548, A34

\bibitem[{{Aoki} {et~al.}(2013){Aoki}, {Beers}, {Lee}, {Honda}, {Ito},
  {Takada-Hidai}, {Frebel}, {Suda}, {Fujimoto}, {Carollo}, \&
  {Sivarani}}]{2013AJ....145...13A}
{Aoki}, W., {Beers}, T.~C., {Lee}, Y.~S., {et~al.} 2013, \aj, 145, 13

\bibitem[{{Aoki} {et~al.}(2008){Aoki}, {Beers}, {Sivarani}, {Marsteller},
  {Lee}, {Honda}, {Norris}, {Ryan}, \& {Carollo}}]{2008ApJ...678.1351A}
{Aoki}, W., {Beers}, T.~C., {Sivarani}, T., {et~al.} 2008, \apj, 678, 1351

\bibitem[{{Asplund} {et~al.}(2009){Asplund}, {Grevesse}, {Sauval}, \&
  {Scott}}]{2009ARA&A..47..481A}
{Asplund}, M., {Grevesse}, N., {Sauval}, A.~J., \& {Scott}, P. 2009, \araa, 47,
  481

\bibitem[{{Badnell} {et~al.}(2005){Badnell}, {Bautista}, {Butler}, {Delahaye},
  {Mendoza}, {Palmeri}, {Zeippen}, \& {Seaton}}]{2005MNRAS.360..458B}
{Badnell}, N.~R., {Bautista}, M.~A., {Butler}, K., {et~al.} 2005, \mnras, 360,
  458

\bibitem[{{Beers} \& {Christlieb}(2005)}]{2005ARA&A..43..531B}
{Beers}, T.~C. \& {Christlieb}, N. 2005, \araa, 43, 531

\bibitem[{{Beers} {et~al.}(1985){Beers}, {Preston}, \&
  {Shectman}}]{1985AJ.....90.2089B}
{Beers}, T.~C., {Preston}, G.~W., \& {Shectman}, S.~A. 1985, \aj, 90, 2089

\bibitem[{{Beers} {et~al.}(1992){Beers}, {Preston}, \&
  {Shectman}}]{1992AJ....103.1987B}
{Beers}, T.~C., {Preston}, G.~W., \& {Shectman}, S.~A. 1992, \aj, 103, 1987

\bibitem[{{Behara} {et~al.}(2010){Behara}, {Bonifacio}, {Ludwig}, {Sbordone},
  {Gonz{\'a}lez Hern{\'a}ndez}, \& {Caffau}}]{2010A&A...513A..72B}
{Behara}, N.~T., {Bonifacio}, P., {Ludwig}, H.-G., {et~al.} 2010, \aap, 513,
  A72

\bibitem[{{Bisterzo} {et~al.}(2010){Bisterzo}, {Gallino}, {Straniero},
  {Cristallo}, \& {K{\"a}ppeler}}]{2010MNRAS.404.1529B}
{Bisterzo}, S., {Gallino}, R., {Straniero}, O., {Cristallo}, S., \&
  {K{\"a}ppeler}, F. 2010, \mnras, 404, 1529

\bibitem[{{Bisterzo} {et~al.}(2011){Bisterzo}, {Gallino}, {Straniero},
  {Cristallo}, \& {K{\"a}ppeler}}]{2011MNRAS.418..284B}
{Bisterzo}, S., {Gallino}, R., {Straniero}, O., {Cristallo}, S., \&
  {K{\"a}ppeler}, F. 2011, \mnras, 418, 284

\bibitem[{{Bisterzo} {et~al.}(2012){Bisterzo}, {Gallino}, {Straniero},
  {Cristallo}, \& {K{\"a}ppeler}}]{2012MNRAS.422..849B}
{Bisterzo}, S., {Gallino}, R., {Straniero}, O., {Cristallo}, S., \&
  {K{\"a}ppeler}, F. 2012, \mnras, 422, 849

\bibitem[{{Bisterzo} {et~al.}(2014){Bisterzo}, {Travaglio}, {Gallino},
  {Wiescher}, \& {K{\"a}ppeler}}]{2014ApJ...787...10B}
{Bisterzo}, S., {Travaglio}, C., {Gallino}, R., {Wiescher}, M., \&
  {K{\"a}ppeler}, F. 2014, \apj, 787, 10

\bibitem[{{B{\"o}hm-Vitense}(1958)}]{1958ZA.....46..108B}
{B{\"o}hm-Vitense}, E. 1958, \zap, 46, 108

\bibitem[{{Bonaca} {et~al.}(2012){Bonaca}, {Tanner}, {Basu}, {Chaplin},
  {Metcalfe}, {Monteiro}, {Ballot}, {Bedding}, {Bonanno}, {Broomhall},
  {Bruntt}, {Campante}, {Christensen-Dalsgaard}, {Corsaro}, {Elsworth},
  {Garc{\'{\i}}a}, {Hekker}, {Karoff}, {Kjeldsen}, {Mathur}, {R{\'e}gulo},
  {Roxburgh}, {Stello}, {Trampedach}, {Barclay}, {Burke}, \&
  {Caldwell}}]{2012ApJ...755L..12B}
{Bonaca}, A., {Tanner}, J.~D., {Basu}, S., {et~al.} 2012, \apjl, 755, L12

\bibitem[{{Bressan} {et~al.}(2012){Bressan}, {Marigo}, {Girardi}, {Salasnich},
  {Dal Cero}, {Rubele}, \& {Nanni}}]{2012MNRAS.427..127B}
{Bressan}, A., {Marigo}, P., {Girardi}, L., {et~al.} 2012, \mnras, 427, 127

\bibitem[{{Burgers}(1969)}]{1969fecg.book.....B}
{Burgers}, J.~M. 1969, {Flow Equations for Composite Gases}

\bibitem[{{Canuto}(1970)}]{1970ApJ...159..641C}
{Canuto}, V. 1970, \apj, 159, 641

\bibitem[{{Carollo} {et~al.}(2012){Carollo}, {Beers}, {Bovy}, {Sivarani},
  {Norris}, {Freeman}, {Aoki}, {Lee}, \& {Kennedy}}]{2012ApJ...744..195C}
{Carollo}, D., {Beers}, T.~C., {Bovy}, J., {et~al.} 2012, \apj, 744, 195

\bibitem[{{Cassisi} {et~al.}(2007){Cassisi}, {Potekhin}, {Pietrinferni},
  {Catelan}, \& {Salaris}}]{2007ApJ...661.1094C}
{Cassisi}, S., {Potekhin}, A.~Y., {Pietrinferni}, A., {Catelan}, M., \&
  {Salaris}, M. 2007, \apj, 661, 1094

\bibitem[{{Castellani} \& {degl'Innocenti}(1999)}]{1999A&A...344...97C}
{Castellani}, V. \& {degl'Innocenti}, S. 1999, \aap, 344, 97

\bibitem[{{Christlieb} {et~al.}(2001){Christlieb}, {Green}, {Wisotzki}, \&
  {Reimers}}]{2001A&A...375..366C}
{Christlieb}, N., {Green}, P.~J., {Wisotzki}, L., \& {Reimers}, D. 2001, \aap,
  375, 366

\bibitem[{{Christlieb} {et~al.}(2008){Christlieb}, {Sch{\"o}rck}, {Frebel},
  {Beers}, {Wisotzki}, \& {Reimers}}]{2008A&A...484..721C}
{Christlieb}, N., {Sch{\"o}rck}, T., {Frebel}, A., {et~al.} 2008, \aap, 484,
  721

\bibitem[{{Cranmer} \& {Saar}(2011)}]{2011ApJ...741...54C}
{Cranmer}, S.~R. \& {Saar}, S.~H. 2011, \apj, 741, 54

\bibitem[{{Denissenkov}(2010)}]{2010ApJ...723..563D}
{Denissenkov}, P.~A. 2010, \apj, 723, 563

\bibitem[{{Denissenkov} \& {Pinsonneault}(2008)}]{2008ApJ...684..626D}
{Denissenkov}, P.~A. \& {Pinsonneault}, M. 2008, \apj, 684, 626

\bibitem[{{D'Ercole} {et~al.}(2010){D'Ercole}, {D'Antona}, {Ventura},
  {Vesperini}, \& {McMillan}}]{2010MNRAS.407..854D}
{D'Ercole}, A., {D'Antona}, F., {Ventura}, P., {Vesperini}, E., \& {McMillan},
  S.~L.~W. 2010, \mnras, 407, 854

\bibitem[{{Eggleton}(1971)}]{1971MNRAS.151..351E}
{Eggleton}, P.~P. 1971, \mnras, 151, 351

\bibitem[{{Eggleton}(1972)}]{1972MNRAS.156..361E}
{Eggleton}, P.~P. 1972, \mnras, 156, 361

\bibitem[{{Eggleton} {et~al.}(1973){Eggleton}, {Faulkner}, \&
  {Flannery}}]{1973A&A....23..325E}
{Eggleton}, P.~P., {Faulkner}, J., \& {Flannery}, B.~P. 1973, \aap, 23, 325

\bibitem[{{Eldridge} \& {Tout}(2004)}]{2004MNRAS.348..201E}
{Eldridge}, J.~J. \& {Tout}, C.~A. 2004, \mnras, 348, 201

\bibitem[{{Gonzalez} {et~al.}(1995){Gonzalez}, {LeBlanc}, {Artru}, \&
  {Michaud}}]{1995A&A...297..223G}
{Gonzalez}, J.-F., {LeBlanc}, F., {Artru}, M.-C., \& {Michaud}, G. 1995, \aap,
  297, 223

\bibitem[{{Gruyters} {et~al.}(2014){Gruyters}, {Nordlander}, \&
  {Korn}}]{2014A&A...567A..72G}
{Gruyters}, P., {Nordlander}, T., \& {Korn}, A.~J. 2014, \aap, 567, A72

\bibitem[{{Hansen} {et~al.}(2016{\natexlab{a}}){Hansen}, {Andersen},
  {Nordstr{\"o}m}, {Beers}, {Placco}, {Yoon}, \&
  {Buchhave}}]{2016A&A...586A.160H}
{Hansen}, T.~T., {Andersen}, J., {Nordstr{\"o}m}, B., {et~al.}
  2016{\natexlab{a}}, \aap, 586, A160

\bibitem[{{Hansen} {et~al.}(2016{\natexlab{b}}){Hansen}, {Andersen},
  {Nordstr{\"o}m}, {Beers}, {Placco}, {Yoon}, \&
  {Buchhave}}]{2016A&A...588A...3H}
{Hansen}, T.~T., {Andersen}, J., {Nordstr{\"o}m}, B., {et~al.}
  2016{\natexlab{b}}, \aap, 588, A3

\bibitem[{{Hansen} {et~al.}(2015){Hansen}, {Andersen}, {Nordstr{\"o}m},
  {Beers}, {Yoon}, \& {Buchhave}}]{2015A&A...583A..49H}
{Hansen}, T.~T., {Andersen}, J., {Nordstr{\"o}m}, B., {et~al.} 2015, \aap, 583,
  A49

\bibitem[{{Herwig}(2005)}]{2005ARA&A..43..435H}
{Herwig}, F. 2005, \araa, 43, 435

\bibitem[{{Hinshaw} {et~al.}(2013){Hinshaw}, {Larson}, {Komatsu}, {Spergel},
  {Bennett}, {Dunkley}, {Nolta}, {Halpern}, {Hill}, {Odegard}, {Page}, {Smith},
  {Weiland}, {Gold}, {Jarosik}, {Kogut}, {Limon}, {Meyer}, {Tucker}, {Wollack},
  \& {Wright}}]{2013ApJS..208...19H}
{Hinshaw}, G., {Larson}, D., {Komatsu}, E., {et~al.} 2013, \apjs, 208, 19

\bibitem[{{Holzwarth} \& {Jardine}(2007)}]{2007A&A...463...11H}
{Holzwarth}, V. \& {Jardine}, M. 2007, \aap, 463, 11

\bibitem[{{Hu} {et~al.}(2010){Hu}, {Glebbeek}, {Thoul}, {Dupret}, {Stancliffe},
  {Nelemans}, \& {Aerts}}]{2010A&A...511A..87H}
{Hu}, H., {Glebbeek}, E., {Thoul}, A.~A., {et~al.} 2010, \aap, 511, A87

\bibitem[{{Hu} {et~al.}(2011){Hu}, {Tout}, {Glebbeek}, \&
  {Dupret}}]{2011MNRAS.418..195H}
{Hu}, H., {Tout}, C.~A., {Glebbeek}, E., \& {Dupret}, M.-A. 2011, \mnras, 418,
  195

\bibitem[{{Hubbard} \& {Lampe}(1969)}]{1969ApJS...18..297H}
{Hubbard}, W.~B. \& {Lampe}, M. 1969, \apjs, 18, 297

\bibitem[{{Iglesias} \& {Rogers}(1996)}]{1996ApJ...464..943I}
{Iglesias}, C.~A. \& {Rogers}, F.~J. 1996, \apj, 464, 943

\bibitem[{{Johnstone} {et~al.}(2015){Johnstone}, {G{\"u}del}, {Brott}, \&
  {L{\"u}ftinger}}]{2015A&A...577A..28J}
{Johnstone}, C.~P., {G{\"u}del}, M., {Brott}, I., \& {L{\"u}ftinger}, T. 2015,
  \aap, 577, A28

\bibitem[{{Jonsell} {et~al.}(2006){Jonsell}, {Barklem}, {Gustafsson},
  {Christlieb}, {Hill}, {Beers}, \& {Holmberg}}]{2006A&A...451..651J}
{Jonsell}, K., {Barklem}, P.~S., {Gustafsson}, B., {et~al.} 2006, \aap, 451,
  651

\bibitem[{{Jorissen} {et~al.}(2016){Jorissen}, {Van Eck}, {Van Winckel},
  {Merle}, {Boffin}, {Andersen}, {Nordstr{\"o}m}, {Udry}, {Masseron},
  {Lenaerts}, \& {Waelkens}}]{2016A&A...586A.158J}
{Jorissen}, A., {Van Eck}, S., {Van Winckel}, H., {et~al.} 2016, \aap, 586,
  A158

\bibitem[{{Kippenhahn} {et~al.}(1980){Kippenhahn}, {Ruschenplatt}, \&
  {Thomas}}]{1980A&A....91..175K}
{Kippenhahn}, R., {Ruschenplatt}, G., \& {Thomas}, H.-C. 1980, \aap, 91, 175

\bibitem[{{Kobayashi} {et~al.}(2011){Kobayashi}, {Karakas}, \&
  {Umeda}}]{2011MNRAS.414.3231K}
{Kobayashi}, C., {Karakas}, A.~I., \& {Umeda}, H. 2011, \mnras, 414, 3231

\bibitem[{{Korn} {et~al.}(2006){Korn}, {Grundahl}, {Richard}, {Barklem},
  {Mashonkina}, {Collet}, {Piskunov}, \& {Gustafsson}}]{2006Natur.442..657K}
{Korn}, A.~J., {Grundahl}, F., {Richard}, O., {et~al.} 2006, \nat, 442, 657

\bibitem[{{Lee} {et~al.}(2013){Lee}, {Beers}, {Masseron}, {Plez}, {Rockosi},
  {Sobeck}, {Yanny}, {Lucatello}, {Sivarani}, {Placco}, \&
  {Carollo}}]{2013AJ....146..132L}
{Lee}, Y.~S., {Beers}, T.~C., {Masseron}, T., {et~al.} 2013, \aj, 146, 132

\bibitem[{{Lucatello} {et~al.}(2006){Lucatello}, {Beers}, {Christlieb},
  {Barklem}, {Rossi}, {Marsteller}, {Sivarani}, \& {Lee}}]{2006ApJ...652L..37L}
{Lucatello}, S., {Beers}, T.~C., {Christlieb}, N., {et~al.} 2006, \apjl, 652,
  L37

\bibitem[{{Lucatello} {et~al.}(2005){Lucatello}, {Tsangarides}, {Beers},
  {Carretta}, {Gratton}, \& {Ryan}}]{2005ApJ...625..825L}
{Lucatello}, S., {Tsangarides}, S., {Beers}, T.~C., {et~al.} 2005, \apj, 625,
  825

\bibitem[{{Lugaro} {et~al.}(2012){Lugaro}, {Karakas}, {Stancliffe}, \&
  {Rijs}}]{2012ApJ...747....2L}
{Lugaro}, M., {Karakas}, A.~I., {Stancliffe}, R.~J., \& {Rijs}, C. 2012, \apj,
  747, 2

\bibitem[{{Maeder} {et~al.}(2013){Maeder}, {Meynet}, {Lagarde}, \&
  {Charbonnel}}]{2013A&A...553A...1M}
{Maeder}, A., {Meynet}, G., {Lagarde}, N., \& {Charbonnel}, C. 2013, \aap, 553,
  A1

\bibitem[{{Masseron} {et~al.}(2010){Masseron}, {Johnson}, {Plez}, {van Eck},
  {Primas}, {Goriely}, \& {Jorissen}}]{2010A&A...509A..93M}
{Masseron}, T., {Johnson}, J.~A., {Plez}, B., {et~al.} 2010, \aap, 509, A93

\bibitem[{{McClure} \& {Woodsworth}(1990)}]{1990ApJ...352..709M}
{McClure}, R.~D. \& {Woodsworth}, A.~W. 1990, \apj, 352, 709

\bibitem[{{Mendoza} {et~al.}(2007){Mendoza}, {Seaton}, {Buerger},
  {Bellor{\'{\i}}n}, {Mel{\'e}ndez}, {Gonz{\'a}lez}, {Rodr{\'{\i}}guez},
  {Delahaye}, {Palacios}, {Pradhan}, \& {Zeippen}}]{2007MNRAS.378.1031M}
{Mendoza}, C., {Seaton}, M.~J., {Buerger}, P., {et~al.} 2007, \mnras, 378, 1031

\bibitem[{{Michaud}(1977)}]{1977Natur.266..433M}
{Michaud}, G. 1977, \nat, 266, 433

\bibitem[{{Michaud} {et~al.}(1987){Michaud}, {Dupuis}, {Fontaine}, \&
  {Montmerle}}]{1987ApJ...322..302M}
{Michaud}, G., {Dupuis}, J., {Fontaine}, G., \& {Montmerle}, T. 1987, \apj,
  322, 302

\bibitem[{{Michaud} {et~al.}(2010){Michaud}, {Richer}, \&
  {Richard}}]{2010A&A...510A.104M}
{Michaud}, G., {Richer}, J., \& {Richard}, O. 2010, \aap, 510, A104

\bibitem[{{Nordlander} {et~al.}(2012){Nordlander}, {Korn}, {Richard}, \&
  {Lind}}]{2012ApJ...753...48N}
{Nordlander}, T., {Korn}, A.~J., {Richard}, O., \& {Lind}, K. 2012, \apj, 753,
  48

\bibitem[{{Paquette} {et~al.}(1986){Paquette}, {Pelletier}, {Fontaine}, \&
  {Michaud}}]{1986ApJS...61..177P}
{Paquette}, C., {Pelletier}, C., {Fontaine}, G., \& {Michaud}, G. 1986, \apjs,
  61, 177

\bibitem[{{Paxton} {et~al.}(2011){Paxton}, {Bildsten}, {Dotter}, {Herwig},
  {Lesaffre}, \& {Timmes}}]{2011ApJS..192....3P}
{Paxton}, B., {Bildsten}, L., {Dotter}, A., {et~al.} 2011, \apjs, 192, 3

\bibitem[{{Paxton} {et~al.}(2015){Paxton}, {Marchant}, {Schwab}, {Bauer},
  {Bildsten}, {Cantiello}, {Dessart}, {Farmer}, {Hu}, {Langer}, {Townsend},
  {Townsley}, \& {Timmes}}]{2015ApJS..220...15P}
{Paxton}, B., {Marchant}, P., {Schwab}, J., {et~al.} 2015, \apjs, 220, 15

\bibitem[{{Placco} {et~al.}(2014){Placco}, {Frebel}, {Beers}, \&
  {Stancliffe}}]{2014ApJ...797...21P}
{Placco}, V.~M., {Frebel}, A., {Beers}, T.~C., \& {Stancliffe}, R.~J. 2014,
  \apj, 797, 21

\bibitem[{{Pols} {et~al.}(1995){Pols}, {Tout}, {Eggleton}, \&
  {Han}}]{1995MNRAS.274..964P}
{Pols}, O.~R., {Tout}, C.~A., {Eggleton}, P.~P., \& {Han}, Z. 1995, \mnras,
  274, 964

\bibitem[{{Reimers}(1975)}]{1975MSRSL...8..369R}
{Reimers}, D. 1975, Memoires of the Societe Royale des Sciences de Liege, 8,
  369

\bibitem[{{Richard} {et~al.}(2002{\natexlab{a}}){Richard}, {Michaud}, \&
  {Richer}}]{2002ApJ...580.1100R}
{Richard}, O., {Michaud}, G., \& {Richer}, J. 2002{\natexlab{a}}, \apj, 580,
  1100

\bibitem[{{Richard} {et~al.}(2005){Richard}, {Michaud}, \&
  {Richer}}]{2005ApJ...619..538R}
{Richard}, O., {Michaud}, G., \& {Richer}, J. 2005, \apj, 619, 538

\bibitem[{{Richard} {et~al.}(2002{\natexlab{b}}){Richard}, {Michaud}, {Richer},
  {Turcotte}, {Turck-Chi{\`e}ze}, \& {VandenBerg}}]{2002ApJ...568..979R}
{Richard}, O., {Michaud}, G., {Richer}, J., {et~al.} 2002{\natexlab{b}}, \apj,
  568, 979

\bibitem[{{Richer} {et~al.}(2000){Richer}, {Michaud}, \&
  {Turcotte}}]{2000ApJ...529..338R}
{Richer}, J., {Michaud}, G., \& {Turcotte}, S. 2000, \apj, 529, 338

\bibitem[{{Seaton}(1997)}]{1997MNRAS.289..700S}
{Seaton}, M.~J. 1997, \mnras, 289, 700

\bibitem[{{Seaton}(2007)}]{2007MNRAS.382..245S}
{Seaton}, M.~J. 2007, \mnras, 382, 245

\bibitem[{{Sneden} {et~al.}(2008){Sneden}, {Cowan}, \&
  {Gallino}}]{2008ARA&A..46..241S}
{Sneden}, C., {Cowan}, J.~J., \& {Gallino}, R. 2008, \araa, 46, 241

\bibitem[{{Stancliffe}(2009)}]{2009MNRAS.394.1051S}
{Stancliffe}, R.~J. 2009, \mnras, 394, 1051

\bibitem[{{Stancliffe} {et~al.}(2009){Stancliffe}, {Church}, {Angelou}, \&
  {Lattanzio}}]{2009MNRAS.396.2313S}
{Stancliffe}, R.~J., {Church}, R.~P., {Angelou}, G.~C., \& {Lattanzio}, J.~C.
  2009, \mnras, 396, 2313

\bibitem[{{Stancliffe} \& {Eldridge}(2009)}]{2009MNRAS.396.1699S}
{Stancliffe}, R.~J. \& {Eldridge}, J.~J. 2009, \mnras, 396, 1699

\bibitem[{{Stancliffe} {et~al.}(2016){Stancliffe}, {Fossati}, {Passy}, \&
  {Schneider}}]{2016A&A...586A.119S}
{Stancliffe}, R.~J., {Fossati}, L., {Passy}, J.-C., \& {Schneider}, F.~R.~N.
  2016, \aap, 586, A119

\bibitem[{{Stancliffe} \& {Glebbeek}(2008)}]{2008MNRAS.389.1828S}
{Stancliffe}, R.~J. \& {Glebbeek}, E. 2008, \mnras, 389, 1828

\bibitem[{{Stancliffe} {et~al.}(2007){Stancliffe}, {Glebbeek}, {Izzard}, \&
  {Pols}}]{2007A&A...464L..57S}
{Stancliffe}, R.~J., {Glebbeek}, E., {Izzard}, R.~G., \& {Pols}, O.~R. 2007,
  \aap, 464, L57

\bibitem[{{Stancliffe} {et~al.}(2013){Stancliffe}, {Kennedy}, {Lau}, \&
  {Beers}}]{2013MNRAS.435..698S}
{Stancliffe}, R.~J., {Kennedy}, C.~R., {Lau}, H.~H.~B., \& {Beers}, T.~C. 2013,
  \mnras, 435, 698

\bibitem[{{Starkenburg} {et~al.}(2014){Starkenburg}, {Shetrone}, {McConnachie},
  \& {Venn}}]{2014MNRAS.441.1217S}
{Starkenburg}, E., {Shetrone}, M.~D., {McConnachie}, A.~W., \& {Venn}, K.~A.
  2014, \mnras, 441, 1217

\bibitem[{{Suda} {et~al.}(2008){Suda}, {Katsuta}, {Yamada}, {Suwa}, {Ishizuka},
  {Komiya}, {Sorai}, {Aikawa}, \& {Fujimoto}}]{2008PASJ...60.1159S}
{Suda}, T., {Katsuta}, Y., {Yamada}, S., {et~al.} 2008, \pasj, 60, 1159

\bibitem[{{Suda} {et~al.}(2011){Suda}, {Yamada}, {Katsuta}, {Komiya},
  {Ishizuka}, {Aoki}, \& {Fujimoto}}]{2011MNRAS.412..843S}
{Suda}, T., {Yamada}, S., {Katsuta}, Y., {et~al.} 2011, \mnras, 412, 843

\bibitem[{{Suzuki}(2007)}]{2007ApJ...659.1592S}
{Suzuki}, T.~K. 2007, \apj, 659, 1592

\bibitem[{{Swenson}(1995)}]{1995ApJ...438L..87S}
{Swenson}, F.~J. 1995, \apjl, 438, L87

\bibitem[{{Talon}(2008)}]{2008EAS....32...81T}
{Talon}, S. 2008, in EAS Publications Series, Vol.~32, EAS Publications Series,
  ed. C.~{Charbonnel} \& J.-P. {Zahn}, 81--130

\bibitem[{{Thompson} {et~al.}(2008){Thompson}, {Ivans}, {Bisterzo}, {Sneden},
  {Gallino}, {Vauclair}, {Burley}, {Shectman}, \&
  {Preston}}]{2008ApJ...677..556T}
{Thompson}, I.~B., {Ivans}, I.~I., {Bisterzo}, S., {et~al.} 2008, \apj, 677,
  556

\bibitem[{{Thoul} {et~al.}(1994){Thoul}, {Bahcall}, \&
  {Loeb}}]{1994ApJ...421..828T}
{Thoul}, A.~A., {Bahcall}, J.~N., \& {Loeb}, A. 1994, \apj, 421, 828

\bibitem[{{Trampedach} {et~al.}(2014){Trampedach}, {Stein},
  {Christensen-Dalsgaard}, {Nordlund}, \& {Asplund}}]{2014MNRAS.445.4366T}
{Trampedach}, R., {Stein}, R.~F., {Christensen-Dalsgaard}, J., {Nordlund},
  {\AA}., \& {Asplund}, M. 2014, \mnras, 445, 4366

\bibitem[{{Travaglio} {et~al.}(1999){Travaglio}, {Galli}, {Gallino}, {Busso},
  {Ferrini}, \& {Straniero}}]{1999ApJ...521..691T}
{Travaglio}, C., {Galli}, D., {Gallino}, R., {et~al.} 1999, \apj, 521, 691

\bibitem[{{Travaglio} {et~al.}(2001){Travaglio}, {Gallino}, {Busso}, \&
  {Gratton}}]{2001ApJ...549..346T}
{Travaglio}, C., {Gallino}, R., {Busso}, M., \& {Gratton}, R. 2001, \apj, 549,
  346

\bibitem[{{Ulrich}(1972)}]{1972ApJ...172..165U}
{Ulrich}, R.~K. 1972, \apj, 172, 165

\bibitem[{{Valiante} {et~al.}(2009){Valiante}, {Schneider}, {Bianchi}, \&
  {Andersen}}]{2009MNRAS.397.1661V}
{Valiante}, R., {Schneider}, R., {Bianchi}, S., \& {Andersen}, A.~C. 2009,
  \mnras, 397, 1661

\bibitem[{{van Saders} \& {Pinsonneault}(2012)}]{2012ApJ...746...16V}
{van Saders}, J.~L. \& {Pinsonneault}, M.~H. 2012, \apj, 746, 16

\bibitem[{{VandenBerg} {et~al.}(2002){VandenBerg}, {Richard}, {Michaud}, \&
  {Richer}}]{2002ApJ...571..487V}
{VandenBerg}, D.~A., {Richard}, O., {Michaud}, G., \& {Richer}, J. 2002, \apj,
  571, 487

\bibitem[{{Ventura} {et~al.}(2014){Ventura}, {Criscienzo}, {D'Antona},
  {Vesperini}, {Tailo}, {Dell'Agli}, \& {D'Ercole}}]{2014MNRAS.437.3274V}
{Ventura}, P., {Criscienzo}, M.~D., {D'Antona}, F., {et~al.} 2014, \mnras, 437,
  3274

\bibitem[{{Vick} {et~al.}(2010){Vick}, {Michaud}, {Richer}, \&
  {Richard}}]{2010A&A...521A..62V}
{Vick}, M., {Michaud}, G., {Richer}, J., \& {Richard}, O. 2010, \aap, 521, A62

\bibitem[{{Wang}(1998)}]{1998ASPC..154..131W}
{Wang}, Y.-M. 1998, in Astronomical Society of the Pacific Conference Series,
  Vol. 154, Cool Stars, Stellar Systems, and the Sun, ed. R.~A. {Donahue} \&
  J.~A. {Bookbinder}, 131

\bibitem[{{Wood} {et~al.}(2002){Wood}, {M{\"u}ller}, {Zank}, \&
  {Linsky}}]{2002ApJ...574..412W}
{Wood}, B.~E., {M{\"u}ller}, H.-R., {Zank}, G.~P., \& {Linsky}, J.~L. 2002,
  \apj, 574, 412

\bibitem[{{Wood} {et~al.}(2005){Wood}, {M{\"u}ller}, {Zank}, {Linsky}, \&
  {Redfield}}]{2005ApJ...628L.143W}
{Wood}, B.~E., {M{\"u}ller}, H.-R., {Zank}, G.~P., {Linsky}, J.~L., \&
  {Redfield}, S. 2005, \apjl, 628, L143

\bibitem[{{Yanny} {et~al.}(2009){Yanny}, {Rockosi}, {Newberg}, {Knapp},
  {Adelman-McCarthy}, {Alcorn}, {Allam}, {Allende Prieto}, {An}, {Anderson},
  {Anderson}, {Bailer-Jones}, {Bastian}, {Beers}, {Bell}, {Belokurov},
  {Bizyaev}, {Blythe}, {Bochanski}, {Boroski}, {Brinchmann}, {Brinkmann},
  {Brewington}, {Carey}, {Cudworth}, {Evans}, {Evans}, {Gates}, {G{\"a}nsicke},
  {Gillespie}, {Gilmore}, {Nebot Gomez-Moran}, {Grebel}, {Greenwell}, {Gunn},
  {Jordan}, {Jordan}, {Harding}, {Harris}, {Hendry}, {Holder}, {Ivans},
  {Ivezi{\v c}}, {Jester}, {Johnson}, {Kent}, {Kleinman}, {Kniazev},
  {Krzesinski}, {Kron}, {Kuropatkin}, {Lebedeva}, {Lee}, {French Leger},
  {L{\'e}pine}, {Levine}, {Lin}, {Long}, {Loomis}, {Lupton}, {Malanushenko},
  {Malanushenko}, {Margon}, {Martinez-Delgado}, {McGehee}, {Monet}, {Morrison},
  {Munn}, {Neilsen}, {Nitta}, {Norris}, {Oravetz}, {Owen}, {Padmanabhan},
  {Pan}, {Peterson}, {Pier}, {Platson}, {Re Fiorentin}, {Richards}, {Rix},
  {Schlegel}, {Schneider}, {Schreiber}, {Schwope}, {Sibley}, {Simmons},
  {Snedden}, {Allyn Smith}, {Stark}, {Stauffer}, {Steinmetz}, {Stoughton},
  {SubbaRao}, {Szalay}, {Szkody}, {Thakar}, {Sivarani}, {Tucker}, {Uomoto},
  {Vanden Berk}, {Vidrih}, {Wadadekar}, {Watters}, {Wilhelm}, {Wyse}, {Yarger},
  \& {Zucker}}]{2009AJ....137.4377Y}
{Yanny}, B., {Rockosi}, C., {Newberg}, H.~J., {et~al.} 2009, \aj, 137, 4377

\bibitem[{{York} {et~al.}(2000){York}, {Adelman}, {Anderson}, {Anderson},
  {Annis}, {Bahcall}, {Bakken}, {Barkhouser}, {Bastian}, {Berman}, {Boroski},
  {Bracker}, {Briegel}, {Briggs}, {Brinkmann}, {Brunner}, {Burles}, {Carey},
  {Carr}, {Castander}, {Chen}, {Colestock}, {Connolly}, {Crocker}, {Csabai},
  {Czarapata}, {Davis}, {Doi}, {Dombeck}, {Eisenstein}, {Ellman}, {Elms},
  {Evans}, {Fan}, {Federwitz}, {Fiscelli}, {Friedman}, {Frieman}, {Fukugita},
  {Gillespie}, {Gunn}, {Gurbani}, {de Haas}, {Haldeman}, {Harris}, {Hayes},
  {Heckman}, {Hennessy}, {Hindsley}, {Holm}, {Holmgren}, {Huang}, {Hull},
  {Husby}, {Ichikawa}, {Ichikawa}, {Ivezi{\'c}}, {Kent}, {Kim}, {Kinney},
  {Klaene}, {Kleinman}, {Kleinman}, {Knapp}, {Korienek}, {Kron}, {Kunszt},
  {Lamb}, {Lee}, {Leger}, {Limmongkol}, {Lindenmeyer}, {Long}, {Loomis},
  {Loveday}, {Lucinio}, {Lupton}, {MacKinnon}, {Mannery}, {Mantsch}, {Margon},
  {McGehee}, {McKay}, {Meiksin}, {Merelli}, {Monet}, {Munn}, {Narayanan},
  {Nash}, {Neilsen}, {Neswold}, {Newberg}, {Nichol}, {Nicinski}, {Nonino},
  {Okada}, {Okamura}, {Ostriker}, {Owen}, {Pauls}, {Peoples}, {Peterson},
  {Petravick}, {Pier}, {Pope}, {Pordes}, {Prosapio}, {Rechenmacher}, {Quinn},
  {Richards}, {Richmond}, {Rivetta}, {Rockosi}, {Ruthmansdorfer}, {Sandford},
  {Schlegel}, {Schneider}, {Sekiguchi}, {Sergey}, {Shimasaku}, {Siegmund},
  {Smee}, {Smith}, {Snedden}, {Stone}, {Stoughton}, {Strauss}, {Stubbs},
  {SubbaRao}, {Szalay}, {Szapudi}, {Szokoly}, {Thakar}, {Tremonti}, {Tucker},
  {Uomoto}, {Vanden Berk}, {Vogeley}, {Waddell}, {Wang}, {Watanabe},
  {Weinberg}, {Yanny}, {Yasuda}, \& {SDSS Collaboration}}]{2000AJ....120.1579Y}
{York}, D.~G., {Adelman}, J., {Anderson}, Jr., J.~E., {et~al.} 2000, \aj, 120,
  1579

\end{thebibliography}

\end{document}